\documentclass[aps,prd,twocolumn,showpacs,floatfix,preprintnumbers,amsfont,amsmath,amssymb,nofootinbib,superscriptaddress]{revtex4-1}

\usepackage{color}
\usepackage{bbold}
\usepackage{hyperref}
\usepackage[normalem]{ulem}
\usepackage{lipsum}
\usepackage{url}
\usepackage{float}
\usepackage[table]{xcolor}
\usepackage{graphicx}
\usepackage[font=small, justification=raggedright, format=plain]{caption}
\usepackage{subcaption}
\usepackage{mathtools}
\usepackage{bm}
\usepackage{siunitx}
\usepackage{multirow}

\usepackage{array,etoolbox}
\preto\tabular{\setcounter{magicrownumbers}{-1}}
\newcounter{magicrownumbers}

\colorlet{RED}{red}

\def\chieff{\chi_{\rm eff}}
\def\nullhyp{\mathcal{H}_0}
\def\althyp{\mathcal{H}_1}
\def\pastro{p_{\rm astro}}
\def\msun{\rm M_{\odot}}
\newcommand{\conv}{\circledast}

\newcommand{\cn}{\mathcal{N}}
\newcommand{\lt}{\left(}
\newcommand{\rt}{\right)}
\newcommand{\abs}[1]{\left\vert #1 \right\vert}
\newcommand{\vabs}[1]{\left\vert\left\vert #1 \right\vert\right\vert}
\newcommand{\mcal}[1]{\mathcal{#1}}
\newcommand{\dd}{\delta_{\rm D}}
\newcommand{\sk}[1]{}

\newcommand{\be}{\begin{equation}}
\newcommand{\ee}{\end{equation}}
\newcommand{\ba}{\begin{eqnarray}}
\newcommand{\ea}{\end{eqnarray}}

\newcommand{\apjl}{Astrophys. J. Lett.}

\newcommand{\mnras}{Mon. Not. R. Astron. Soc.}

\newcommand{\changed}[1]{#1}

\allowdisplaybreaks
\begin{document}

\title{New binary black hole mergers in the LIGO--Virgo O3a data}

\author{Seth Olsen}
\email{srolsen@princeton.edu}
\affiliation{\mbox{Department of Physics, Princeton University, Princeton, NJ 08540, USA}}
\author{Tejaswi Venumadhav}
\affiliation{\mbox{Department of Physics, University of California at Santa Barbara, Santa Barbara, CA 93106, USA}}
\affiliation{\mbox{International Centre for Theoretical Sciences, Tata Institute of Fundamental Research, Bangalore 560089, India}}
\author{Jonathan Mushkin}
\affiliation{\mbox{Department of Particle Physics \& Astrophysics, Weizmann Institute of Science, Rehovot 76100, Israel}}
\author{Javier Roulet}
\affiliation{\mbox{Kavli Institute for Theoretical Physics, University of California at Santa Barbara, Santa Barbara, CA 93106, USA}}
\author{Barak Zackay}
\affiliation{\mbox{Department of Particle Physics \& Astrophysics, Weizmann Institute of Science, Rehovot 76100, Israel}}
\author{Matias Zaldarriaga}
\affiliation{\mbox{School of Natural Sciences, Institute for Advanced Study, 1 Einstein Drive, Princeton, NJ 08540, USA}}

\date{July 12, 2022}

%%%%%%%%%%%%%%%%%%%%%%%%%%%%%%%%%%%%%%%%%%%%%%%%%%%%%%%%%%%%%%%%%%%%%%%%%%%%%%%

\begin{abstract}

We report the detection of ten new binary black hole (BBH) mergers in the publicly released data from the the first half of the third observing run (O3a) of advanced LIGO and advanced Virgo. 
%Candidates are identified using an updated version of the search pipeline described in \citet{ias_pipeline_o1_catalog_new_search_prd2019} (the ``IAS pipeline" \cite{ias_o2_pipeline_new_events_prd2020}), and 
%events are declared according to criteria similar to those in the GWTC-2.1 catalog \cite{lvc_o3a_deep_gwtc2_1_update_2021}. 
\changed{We identify candidates using an updated version of the search pipeline described in \citet{ias_pipeline_o1_catalog_new_search_prd2019} (the ``IAS pipeline" \cite{ias_o2_pipeline_new_events_prd2020}) and compile a catalog of signals that pass a significance threshold of astrophysical probability greater than 0.5 (following the GWTC-2.1 \cite{lvc_o3a_deep_gwtc2_1_update_2021} and 3-OGC \cite{nitz_o3a_3ogc_catalog_2021} catalogs).} 
The updated \changed{IAS pipeline} is sensitive to a larger region of parameter space, applies a template prior that accounts for different search volume as a function of intrinsic parameters, and uses an improved coherent detection statistic that optimally combines the data from the Hanford and Livingston detectors. 
Among the ten new events, we observe interesting astrophysical scenarios including sources with confidently large effective spin parameters in both the positive and negative directions, high-mass black holes that are difficult to form in stellar collapse models due to (pulsational) pair instability, and low-mass mergers that bridge the gap between neutron stars and the lightest observed black holes. 
We \changed{infer source parameters in} the upper and lower black hole mass gaps with both extreme and near-unity mass ratios, and one of the possible neutron star--black hole (NSBH) mergers is well localized for electromagnetic (EM) counterpart searches. 
We 
%\changed{\sout{see a substantial increase in significance for many of the events previously reported by other pipelines, and we}} 
detect all of the GWTC-2.1 BBH mergers with coincident data in Hanford and Livingston except for three loud events that get vetoed, which is compatible with the false-positive rate of our veto procedure, and three that fall below the detection threshold. 
We also return to significance the event GW190909\_114149, which was reduced to a sub-threshold trigger after its initial appearance in GWTC-2 \cite{lvc_o3a_gwtc2_catalog_2021}. 
This amounts to a total of 42 BBH mergers detected by our pipeline's search of the coincident Hanford--Livingston O3a data.

\end{abstract}

\maketitle

\section{Introduction}

The LIGO--Virgo Collaboration (LVC) reported the detection of gravitational waves (GWs) from 38 BBH mergers and one binary neutron star (BNS) merger in the first half of their third observing run \cite{lvc_o3a_gwtc2_catalog_2021}. After the GWTC-2 catalog and O3a data were released, \citet{nitz_o3a_3ogc_catalog_2021} performed an independent analysis to produce the 3-OGC catalog, which recovered the GWTC-2 events and added four new BBH mergers. The LVC later released a deeper catalog of candidates, GWTC-2.1 \cite{lvc_o3a_deep_gwtc2_1_update_2021}, declaring eight BBH detections that were not in GWTC-2 (including the four events first reported in 3-OGC \cite{nitz_o3a_3ogc_catalog_2021}) and revoking three of the previously declared events. This made for a total of 43 declared BBH events in the O3a data, with 37 having coincidence in the Hanford and Livingston detectors.

\begin{figure*}
    \centering
    \includegraphics[width=\linewidth]{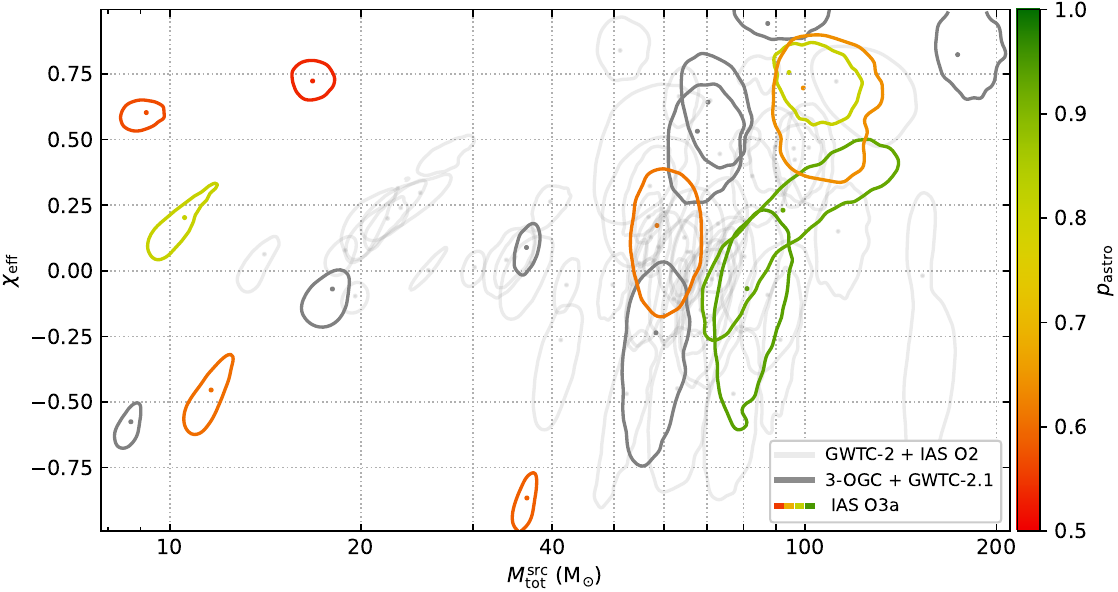}
    \caption{Source-frame total mass and effective spin for the BBH events found over O1, O2, and O3a. The contours enclose \changed{50\% of the probability} and the median is indicated by a dot. PE was done with the IMRPhenomXPHM waveform approximant \cite{xphm_pratten2020} and a prior that is uniform in detector-frame constituent masses, effective spin, and comoving volume-time ($VT$), with redshift computations using Planck15 results \cite{cosmology_planck2015}. Contours for the ten new events are colored by their $\pastro$ values, and the events added in 3-OGC \cite{nitz_o3a_3ogc_catalog_2021} and GWTC-2.1 \cite{lvc_o3a_deep_gwtc2_1_update_2021} are colored solid gray. Events in the LVC or IAS catalogs up to GWTC-2 \cite{O1catalog_LVC2016, ias_pipeline_o1_catalog_new_search_prd2019, gwtc1_o2catalog_LVC2018, ias_o2_pipeline_new_events_prd2020, lvc_o3a_gwtc2_catalog_2021} are transparent gray. GW190909\_114149 is included as transparent gray since it was first announced in GWTC-2, though it was relegated to the sub-threshold list in GWTC-2.1 and therefore only appears in this plot because our pipeline recovered is with a declarable $\pastro$ (candidates with marginal scores in both our pipeline and the GWTC-2.1 analysis are left out of this sample). Our search only covers events with Hanford--Livingston coincident triggers, but in this figure we include all BBH events declared in the LVC catalogs through O3a regardless of detector configuration.}
    \label{fig:population}
\end{figure*}

In this work, we add to these catalogs ten new BBH merger candidates which passed the detection bar in our Hanford--Livingston coincident search of the public O3a data \cite{GWOSC}, as well as one event which was previously declared by the LVC but subsequently revoked. 
Our search was conducted with an improved version of the pipeline developed by \citet{ias_pipeline_o1_catalog_new_search_prd2019}, which also added detections to existing catalogs \cite{O1catalog_LVC2016, gwtc1_o2catalog_LVC2018, NitzCatalog_1-OGC_o1_2018, NitzCatalog_2-OGC_o2_2020} in a reanalysis of previous observing runs \cite{ias_o2_pipeline_new_events_prd2020}. \changed{The full population through O3a is shown in Fig.~\ref{fig:population}.}

At this early stage in GW detection, each new event represents an opportunity to challenge our understanding of BBH formation and dynamics, and possibly even to probe fundamental black hole (BH) physics and cosmology \cite{standard_siren_190814_hubble_measure_Palmese2020, quasinormal_ringdown_190521_nitzCapano2021}. We are, however, limited by the signal-to-noise ratio (SNR) of individual events when attempting to constrain fundamental physics, and an empirical understanding of BBH formation and dynamics naturally requires more than one sample from the astrophysical population. By considering a whole catalog of events, we can improve the accuracy and precision of inferred theoretical constraints \cite{coherent_mode_stacking_BH_spectroscopy_Yang2017pretorius, constraints_on_beyondGR_Perkins2021yunes, measure_hubble_from_population_Abbott2021lvc, unmodeled_mode_stacking_measure_HM_from_population_OBrien2019costa, testing_large_scale_GR_important_for_pop_mass_dist_Ezquiaga2021}, and we can begin to construct a phenomenological picture of the BBH merger population \cite{lvc_o3a_population_properties_2021, ias_o3a_population_analysis_prd2021roulet}.

As the number of detections grows, we reduce the statistical errors in population analysis and refine our estimates for the astrophysical distribution of BBH mergers as a function of the constituent BH masses and spins. The inferred distribution can be used to compute constraints on BBH formation channel models, most broadly divided into dynamical formation in dense environments \cite{o2006binary, Samsing:2013kua, petrovich2017greatly, antonini2016merging}, such as star clusters \cite{star_cluster_dynamical_v_isolated_rates_mapelli2020, metallicity_in_young_star_clusters2020a, young_star_clusters_populating_mass_gap2020a, cluster_hierarchical_metallicity_spin_mapelli2021, young_star_clusters_heavy_remnants2021} and active galactic nuclei (AGN) disks \cite{agn_accretion_disk_merger_population2020a, agn_bbh_population_chieff_q_simulation_mckernan_ford2019, mass_gap_agn_bbh_mergers2021, merger_rates_agn_Tagawa_2020a, hierarchical_mergers_agn_kocsis2019, ishibashi_rate_estimate_merger_agn2020a}); and binary co-evolution in isolation \cite{belczynski2002comprehensive, voss2003galactic, belczynski2008compact, dominik2013double,spera2015mass, stevenson2017formation, giacobbo2018progenitors, mandel2016merging} 
or with external agents \cite{antonini2012secular, stone2016assisted, antonini2014blackbreakdown,liu2018blacktriple, bartos2017rapid, hierarchical_7merger_scenario2020b, hierarchical_from_triples2021}. Refer to \citet{review_formation_channels_stellarBH_Mapelli2021} for a recent review of formation channels.

One prediction of stellar evolution models is the existence of gaps in the distribution of BH masses: 
an ``upper mass gap" (UMG) between $\sim 45\, \msun$ and $\sim$\,$135\, \msun$, due to the impact of the pulsational pair instability and pair instability supernova in massive stars~\cite{umg_pre_pairinstability_early_fowler1964ApJS, umg_Barkat1967PairInstability, umg_another_old_pairinstability_bond1984ApJ, umg_HegerWoosley2002massgap, umg_Woosley2007pairinstabilitySN, umg_Woosley2017ppisn, umg_Farmer2019LowerEdgeBHMassGap, umg_in_collapse_simulations_chen2014, umg_yoshida2016_PPI_simulations, umg_pair_instability_mass_loss_Belczynski2016}; 
and the so-called ``lower mass gap" (LMG) in the range of roughly $2$--$5\, \msun$, between the maximal neutron star mass (constrained to $[2, 2.6]\,\msun$ according to recent work by \citet{max_NS_mass_est_from_EOS_space_alsing2018berti}) and the minimal stellar collapse BH mass \cite{lmg_OBS_EM_stellarBH_Mdist_from_few_xrayNSBH_Bailyn1998, lmg_OBS_EM_BHdistInfer_from_16xrayNSBH_ozel2010, lmg_OBS_EM_BHdist_from_15xrayNSBHrocheOver_Farr2011ilya, lmg_model_3methods_1gap_fastSNtime_2012, lmg_bridging_need_NSBHfeature_Farah2021fishbach_holz, lmg_from_CCSNe_simulations_Liu2020-21sun}. 
BBH mergers that challenge the UMG \cite{fishingIAS19_prd2021, lvc_event_GW190521, lvc_o3a_gwtc2_catalog_2021, lvc_o3a_deep_gwtc2_1_update_2021} 
or the LMG \cite{GW190814ApJ2020_LVC, lvc_o3a_deep_gwtc2_1_update_2021} have been reported in the past, and their inclusion in astrophysical population analysis has a significant impact on the inferred mass distribution \cite{lvc_o3a_population_properties_2021, ias_o3a_population_analysis_prd2021roulet}. The set of new events presented here contains multiple examples in each of these mass gap regions, including possible NSBH mergers and what may be the most distant source detected to date (see Table~\ref{tab:signalsFound}).

Apart from the masses, the best measured intrinsic parameter of BBH events is the effective spin, defined \changed{as the mass-weighted average of the orbit-aligned spin components}:
\begin{align} \label{eq:chieff_def}
\chi_{\rm eff} = \frac{m_1 \chi_{1,z} + m_2 \chi_{2,z}}{m_1 + m_2}\,.
\end{align}
where $m_i$ are the \changed{BH} masses and $\chi_{i,z}$ are the dimensionless spin projections on the orbital angular momentum. 
In addition to being well measured, the sign and magnitude of this parameter are each informative about the source's formation channel \cite{need_dynamical_to_misalign_orbit_and_spin_Rodriguez2016, isotropic_vs_aligned_spin_subpops_prefer_iso_Farr2017ilya, chieff_isolated_field_binary_matias2018, field_binary_no_chieff_pop2020}.
However, the predicted distributions in formation channels and the relative rates between channels can be sensitive to a number of highly uncertain prior assumptions, such as metallicity and the distributions of natal BH masses and spins \cite{rate_hierarchical_cluster_spinningLO_nospinHI_rodriguez2019, hierarchical_mergerFamily_dynamical_mass_dist_matters2021, star_cluster_rates_eccentricity_rigriguez2018samsing, cluster_hierarchical_metallicity_spin_mapelli2021, metallicity_effects_mapelli2017, star_cluster_dynamical_v_isolated_rates_mapelli2020, hierarchical_rate_sensitive_to_natal_spins_Fragione2021kocsis}, as well as unaccounted dynamical factors in the models used to simulate populations \cite{umg_isolated_evolution_dredge-up_CostaMapelli2020, umg_edge_sensitive_to_tdepConvect_starModels_48max_Renzo2020farmer, bbh_evolution_agn_merger_timescale_ishibashi2020a}.

Previous works have attempted to address the fact that prior assumptions about the astrophysical spin distribution can impact not only the Bayesian parameter estimation (PE) for individual events, but also the inferred population properties \cite{impact_of_priors_Vitale2017}. One possibility is to use population-informed priors to reanalyze individual events (see, e.g., \citet{pop_informed_spin_prior_low_chieff_Miller2020farr_callister}), but if the sampling priors led to some regions of parameter space being inadequately explored, then reweighting procedures might fail to converge to the correct distribution. One can attempt to constrain population inference in a prior-agnostic way (see, e.g., \citet{pop_spin_isolated_field_vs_dynamical_NOprior_Talbot2017thrane}), but the effects of prior assumptions in modeled searches are inevitable, especially near the detection threshold where small differences in estimated significance determine which events are included or excluded. In light of our ignorance of the true astrophysical distribution, a good strategy is to choose priors that are uniform (i.e., uninformative) in the best-measured (i.e., most informative) parameters \cite{ias_gw190521_prd2021olsen}. For this reason we adopt the uniform effective spin prior introduced by \citet{flat_chieff_prior_o1event1216_formation_channels2019}, as opposed to the isotropic spin prior used to infer parameters in other catalogs \cite{lvc_o3a_deep_gwtc2_1_update_2021, nitz_o3a_3ogc_catalog_2021}. While the latter is motivated by dynamical formation channels where the constituent masses and spins are all independently distributed, our method more strongly prioritizes the best-measured combination of mass and spin variables when assigning significance to events and estimating their parameters.

The mass distribution is coupled to the spin distribution in many formation channel models (see, e.g., discussion by \citet{signatures_of_dynamical_mapelli2020}) -- especially near mass gap edges \cite{umg_Mapelli2020rotationEffectOnMassGap} -- and correlation between masses and spins has been found in the detected population \cite{mass_chieff_trends_pop_farr2020Safarzadeh}. Indeed, population models which allow for this mass--spin correlation are significantly better at fitting the population than models which do not \cite{who_ordered_that-better_models_have_q_chieff_corr-Callister2021farr}, as are models which allow for independently modeled sub-populations \cite{ias_o3a_population_analysis_prd2021roulet, bimodal_spin_subpops_formation_channels_Galaudage2021ilya}. Included in the new events reported here are examples of well-measured large effective spins in both directions (see Table~\ref{tab:signalsFound}), which will improve statistics in the ongoing empirical analyses of the population's underlying spin distribution.

In this work we report ten new BBH merger events, declarable under the criteria that the signal's probability of astrophysical origin, $\pastro$, is at least one half (following \citet{lvc_o3a_deep_gwtc2_1_update_2021} and \citet{nitz_o3a_3ogc_catalog_2021}). We confirm 
%\changed{\sout{or raise}} 
the significance of all but six of the 37 Hanford--Livingston coincident BBH mergers reported by \citet{lvc_o3a_deep_gwtc2_1_update_2021}, with three LVC candidates vetoed by our pipeline (failing signal consistency or excess power tests), and three LVC candidates falling below the detection threshold (see Table~\ref{tab:lvc events}). 
We also detect GW190909\_114149, which was reduced to sub-threshold between GWTC-2 \cite{lvc_o3a_gwtc2_catalog_2021} and GWTC-2.1 \cite{lvc_o3a_deep_gwtc2_1_update_2021}, and this puts the total at 42 BBH mergers detected in our pipeline's Hanford--Livingston coincident search of the O3a data (to be supplemented by a forthcoming publication of our ``disparate detector response" search based on single-detector triggers, which includes Hanford--Virgo and Livingston--Virgo events).
%\changed{\sout{In many of the previously reported events we also see a improved false alarm rate (FAR) compared with other pipelines.}}

Among the ten new events reported here, several of the inferred sources will make important contributions to constraints on interesting astrophysical scenarios: three events have confidently large positive (aligned) effective spin; two events have confidently negative (anti-aligned) effective spin, one having $\chieff < -0.5$ with over $99\%$ confidence; two events have near-unity mass ratio with primary mass posteriors confidently above $45\, \msun$ (UMG), and third has a likelihood peak at extreme mass ratio corresponding to an intermediate mass black hole (IMBH) primary of $\sim120\, \msun$ merging with a stellar mass companion of $\sim12\, \msun$; and four events have secondary mass posteriors confidently below $5\, \msun$ (LMG), including one at extreme mass ratio and two with secondary mass posteriors whose $90\%$ credible intervals extend below $2.3\, \msun$, which indicates the possibility that the binary contains a neutron star (NS). If we estimate the number of false positives by summing the complements of the reported $\pastro$ values, we find that roughly three events are expected to be noise transients rather than astrophysical signals. \changed{It is important to note that both $\pastro$ estimates and inferred source parameters depend on the choice of prior, with results becoming more sensitive to this choice as SNR decreases. We have made a public GitHub repository (\url{https://github.com/seth-olsen/new_BBH_mergers_O3a_IAS_pipeline}) containing all the information needed for using different astrophysical models to estimate $p_{\rm astro}$ (see, e.g., Ref.~\cite{ias_popO2_Roulet_2020}) and reweight posterior samples (see, e.g., Ref.~\cite{reweighting_importance_sampling_thrane2019}).}
%(compared with roughly two of the eight events added in GWTC-2.1 being attributable to noise based on $\pastro$ maximized over all the pipelines reported by LVC \cite{lvc_o3a_deep_gwtc2_1_update_2021}).

The rest of the paper is organized as follows: in \S \ref{sec:changes} we review changes to the IAS pipeline between the O2 and O3a analyses. In \S \ref{sec:events} we discuss the ten BBH mergers first reported in this work (see Table~\ref{tab:signalsFound}). In \S \ref{sec:previous_events} we report our results for events already included in GWTC-2.1 \cite{lvc_o3a_deep_gwtc2_1_update_2021}, noting differences (see Table~\ref{tab:lvc events}). We summarize the results in \S \ref{sec:conclusions} and discuss the astrophysical implications of the new events. Corner plots of posterior distributions for new events can be found in Appendix \ref{Appendix:posteriors}, with PE samples publicly available at \url{https://github.com/seth-olsen/new_BBH_mergers_O3a_IAS_pipeline}. Our computation of $\pastro$ is described in Appendix \ref{Appendix:pastro}. Our method for weighting regions of our geometric template bank by phase space volume is explained in Appendix \ref{Appendix:templateprior}. A detailed derivation of our method for computing the coherent multi-detector statistic is presented in Appendix \ref{Appendix:collection}.

%%%%%%%%%%%%%%%%%%%%%%%%%%%%%%%%%%%%%%%%%%%%%%%%%%%%%%%%%
%%%%%%%%%%%%%%%%%%%%%%%%%%%%%%%%%%%%%%%%%%%%%%%%%%%%%%%%%
\section{Changes to the O2 analysis pipeline}
\label{sec:changes}

Our analysis pipeline is similar in overall structure to the one we used in the O2 analysis~\cite{ias_pipeline_o1_catalog_new_search_prd2019} but differs in the following aspects:
%%%%%%%%%%%%%%%%%%%%%%%%%%%%%%%%%%%%%%%%%%%%%%%%%%%%%%%%%
%%%%%%%%%%%%%%%%%%%%%%%%%%%%%%%%%%%%%%%%%%%%%%%%%%%%%%%%%

\begin{figure}[H]
    \centering
    \includegraphics[width=\linewidth]{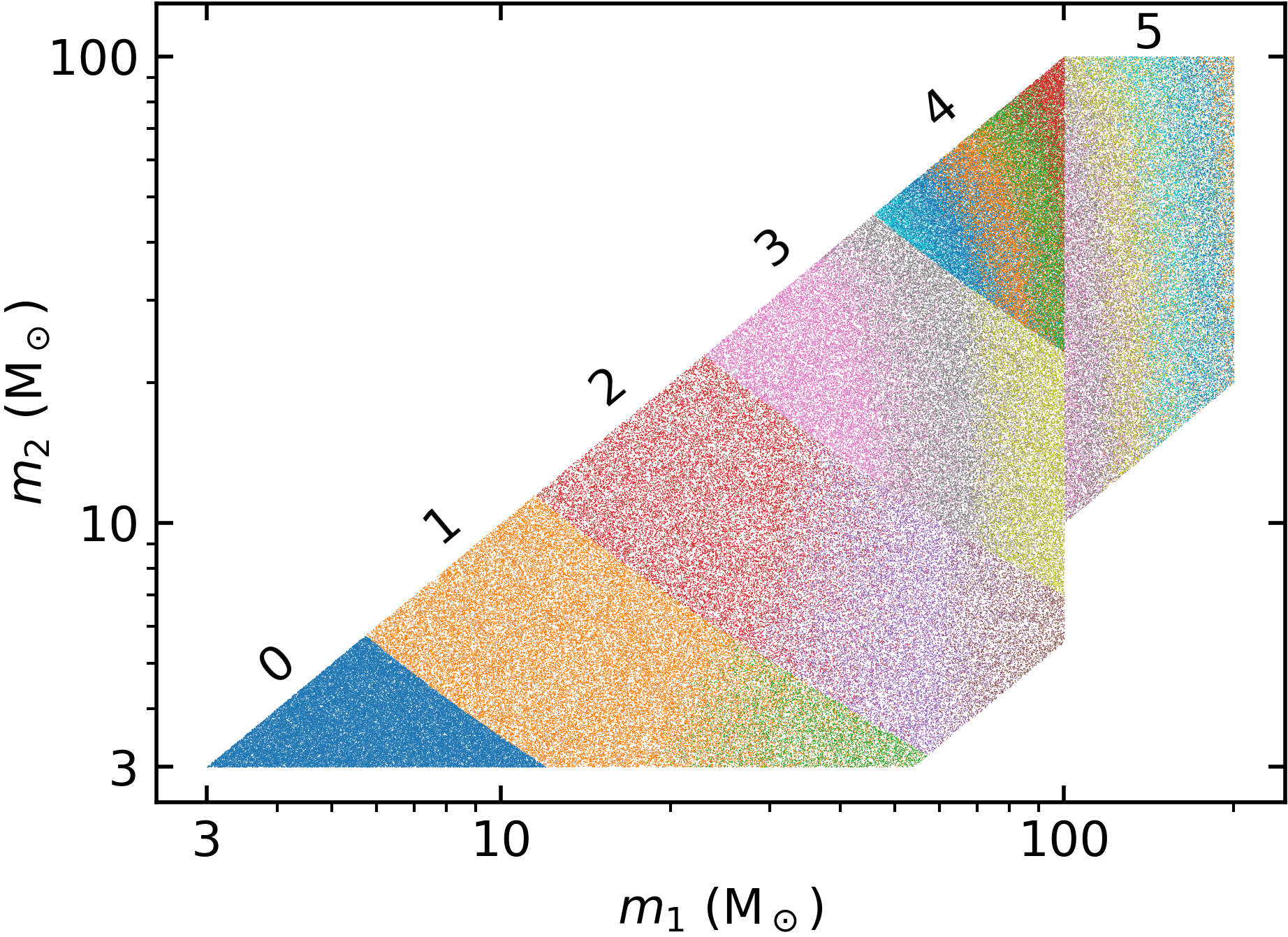}
    \caption{Template banks used in this work to cover the target BBH region. \changed{An independent search is conducted in each of the labeled banks. Colors indicate the (discrete) values of the leading dimension of each bank where we zoom into the tangent plane to construct geometric coordinates from the phase mismatch of nearby templates. Note that waveforms of the same color share the same frequency domain amplitude profile and only differ in their phase, which is the basis for our geometric placement approach (described in Ref.~\cite{ias_template_bank_PSD_roulet2019}).}}
    \label{fig:banks}
\end{figure}

\begin{enumerate}
    \item \textit{Template bank construction:} 
    In this analysis we follow the same method as in \cite{ias_o2_pipeline_new_events_prd2020} to construct template banks, now using a noise power spectral density (PSD) representative of O3a data. Due to the improved sensitivity at high frequencies relative to O2, we expand the range of frequencies to 24--\SI{600}{\hertz} for \texttt{BBH 0-4} in order to satisfy the criterion that we retain $>99\%$ of the matched-filtering SNR over the entire parameter space.
    In the O3a analysis we add a sixth BBH bank (\texttt{BBH 5}) to search for heavier mergers, possibly including IMBH constituents. An IMBH is roughly defined by the range of masses $10^2$--$10^6\,\rm M_\odot$, which we can only just begin to probe given the current low-frequency sensitivity. Our high-mass template bank covers detector frame primary masses in the $[100, 200]\,{\rm M}_\odot$ range and secondary masses within $[10, 100]\,{\rm M}_\odot$, motivated by the fact that our templates contain only the fundamental multipole mode, whose merger frequency moves below the sensitive band for higher-mass binaries. Higher-order multipole modes also become increasingly important at extreme mass ratios, so we limit the mass ratio to a minimum of $1/10$ (in contrast with the lighter banks' limit of $1/18$). The final difference between \texttt{BBH 5} and \texttt{BBH 0-4} is that we construct the high-mass bank using a frequency range of $20$--\SI{512}{\hertz} because a negligible amount of SNR lies outside this band for the masses in \texttt{BBH 5}. \changed{This new bank was responsible for 2 of our 42 detections, as well as the veto of GW190521 and the sub-threshold trigger for the GWTC-2.1 event GW190426\_190642. We expect detection to become more difficult in this region of parameter space because detector noise washes out the low-frequency inspiral of the heavier events, with only a small number of cycles near the merger falling in the sensitive band. We illustrate the template coverage for all banks over the space of detector-frame constituent masses in Fig.~\ref{fig:banks}}

    \item \textit{Preprocessing and flagging the data:}
    We down-sample the public $4096$ Hz data to a sampling rate of $2048$ Hz in the search compared with $1024$ Hz in O2, since our templates now contain frequencies up to $600$ Hz, and hence we need the Nyquist frequency to be above this limit.
    We also updated the method we use to flag frequency ranges containing loud lines, which defines the ranges that are excluded from our excess power tests. Previously, we defined as lines those regions for which the noise amplitude spectral density (ASD, the square root of the PSD) exceeds a smoothed version of the ASD by a fraction that cannot occur due to reasonable measurement noise. 
    We found that some of the lines in the data have a fine frequency structure, with multiple lines occurring in a narrow frequency range (of a few Hz), which can throw off this procedure since the lines bleed to adjacent frequencies when the ASD is smoothed. To address this spectral leakage, we now iterate the line-identification procedure a few times: each time, we use a boxcar filter in the frequency domain (width 1 Hz) to smooth the ASD and then compare with the non-smoothed ASD to flag lines, which define regions that we replace by the smoothed ASD in what we pass to the next iteration as the non-smoothed ASD. 
    In practice, we repeat this procedure three times to achieve convergence. 
    Note that this is only to identify lines, and we still use the full ASD (estimated using the Welch method \cite{welch1967use}) to define the whitening filter throughout the search.
    \changed{These signal processing changes are not expected to have a large effect on the sensitivity of the pipeline but we include them here for completeness.}
    
    \item \textit{Coherent score estimation}:
    We developed a new multi-detector score for ranking candidates that is maximally informative of the signal hypothesis (in the Gaussian noise case): we coherently combine information from the entire matched-filter timeseries in each detector to build an analog of the Bayesian evidence that is commonly used in PE, and we compute it efficiently enough to use it for all search triggers (both with physical detector time shifts and unphysical lags arising from timeslides). As in earlier versions of the pipeline, we apply extra corrections on top of this to account for the non-Gaussian ``glitches" \cite{BlipGlitches, GravitySpy} that produce an excess background. We describe the derivation of the coherent score and the algorithm to compute it in Appendix~\ref{Appendix:collection}. \changed{We expect this to improve sensitivity by moving the ranking statistic closer to the optimal evidence integral.}
    
    \item \textit{Template prior}:
    Previously we assumed a template prior that was uniform in our geometric bank coordinates, but now we apply a template prior that is uniform in the detector-frame constituent masses and the effective spin, as described in Appendix~\ref{Appendix:templateprior}. \changed{We expect this to improve our sensitivity to sources with lower effective spin magnitude and more symmetric masses compared to the prior that is uniform in geometric coordinates, which favors regions of parameter space with extreme values of effective spin and mass ratio (where waveform shape changes most rapidly with respect to changes in physical parameter space).}
    
    \item \textit{Computing $p_{\rm astro}$:} The probability of astrophysical origin for a trigger of ranking score $\Sigma$ is defined in terms of the foreground and background distribution of triggers ${\rm d} N/{\rm d} \Sigma$ as:
    \begin{equation}
        p_{\rm astro}(\Sigma)
        = \frac{\frac{{\rm d} N}{{\rm d} \Sigma}(\Sigma \mid \althyp)}
        {\frac{{\rm d} N}{{\rm d} \Sigma}(\Sigma \mid \nullhyp) + \frac{{\rm d} N}{{\rm d} \Sigma}(\Sigma \mid \althyp)} \, , \label{eq:pastrodef}
    \end{equation}
    where the ranking score is normalized so that all banks are on the same scale, and the null (noise) hypothesis ($\nullhyp$) and alternative (signal) hypothesis ($\althyp$) are that the data was only noise or that it contained an astrophysical BBH merger signal, respectively. We describe our method for estimating the density of triggers ${\rm d} N/{\rm d} \Sigma$ as a function of ranking score in Appendix~\ref{Appendix:pastro}. \changed{The benefits of this new method are improvements in the efficiency and robustness of our $p_{\rm astro}$ estimation, but we do not expect this update to change the pipeline's sensitivity.}
    
\end{enumerate}

%%%%%%%%%%%%%%%%%%%%%%%%%%%%%%%%%%%%%%%%%%%%%%%%%%%%%%%%%
%%%%%%%%%%%%%%%%%%%%%%%%%%%%%%%%%%%%%%%%%%%%%%%%%%%%%%%%%
%%%%%%%%%%%%%%%%%%%%%%%%%%%%%%%%%%%%%%%%%%%%%%%%%%%%%%%%%
%%%%%%%%%%%%%%%%%%%%%%%%%%%%%%% NEW EVENTS
\bgroup
\def\arraystretch{1.5}
\begin{table*}
    \centering
    \caption{New events with $p_{\rm astro} > 0.5$. 
    The parameter ranges are the results of PE with the waveform model \texttt{IMRPhenomXPHM}, which includes higher harmonics and precession, unlike the waveforms used to generate the template bank. The sampling priors are uniform in detector-frame constituent masses, effective spin, and comoving $VT$. The PE also takes into account the data from the Virgo detector when available, unlike the search. Likelihoods marked with an asterisk include the contribution from Virgo\changed{, and the absence of an asterisk means Virgo data was not used in our PE for that event}.}
    \begin{tabular}{|c|c|ccccc|cc|cc|}
        \hline
        \hline
         Name & Bank & $m_1 {\scriptstyle( \rm M_\odot)}$ & $m_2 {\scriptstyle(\rm M_\odot)}$ & $\chi_{\rm eff}$ & $z$ & $\ln\mathcal{L}_{\rm max}$
         & $\rho_{\rm H}^2$ & $\rho_{\rm L}^2$ & IFAR (yr)\footnote{The inverse false alarm rates (IFARs) are computed within each bank and are given in terms of years based on a total analysis time of 106 days for Hanford--Livingston coincidence.} & $p_{\rm astro}$  \\
         \hline
         \hline
         \rowcolor{gray!30} GW190707\_083226 & \texttt{BBH\_4} & $52_{-12}^{+17}$ & $32_{-11}^{+12}$ & $-0.2_{-0.6}^{+0.5}$ & $0.6_{-0.3}^{+0.4}$ & $43.9$ & $37.0$ & $31.5$ & $23.2$ & $0.94$ \\%(4, 2)
GW190711\_030756 & \texttt{BBH\_3} & $80_{-40}^{+50}$ & $18_{-7}^{+11}$ & $0.2_{-0.7}^{+0.3}$ & $0.41_{-0.16}^{+0.24}$ & $49.5$ & $19.8$ & $60.7$ & $11.2$ & $0.93$ \\%(3, 1)
\rowcolor{gray!30} GW190818\_232544 & \texttt{BBH\_4} & $67_{-19}^{+23}$ & $38_{-15}^{+17}$ & $0.7_{-0.3}^{+0.2}$ & $1.0_{-0.4}^{+0.6}$ & $40.5^*$ & $33.0$ & $32.0$ & $3.4$ & $0.81$ \\%(4, 3)
GW190704\_104834 & \texttt{BBH\_0} & $7_{-2}^{+6}$ & $3.2_{-1.1}^{+1.2}$ & $0.20_{-0.14}^{+0.27}$ & $0.10_{-0.03}^{+0.03}$ & $48.7^*$ & $47.0$ & $32.1$ & $2.8$ & $0.81$ \\%(0, 0)
\rowcolor{gray!30} GW190906\_054335 & \texttt{BBH\_3} & $37_{-8}^{+12}$ & $24_{-8}^{+8}$ & $0.1_{-0.5}^{+0.4}$ & $0.9_{-0.3}^{+0.4}$ & $34.1^*$ & $23.6$ & $38.1$ & $0.73$ & $0.61$ \\%(3, 1)
GW190821\_124821 & \texttt{BBH\_1} & $7.6_{-1.7}^{+3.9}$ & $4.0_{-1.1}^{+1.0}$ & $-0.45_{-0.17}^{+0.33}$ & $0.17_{-0.06}^{+0.06}$ & $48.5^*$ & $28.1$ & $49.4$ & $0.71$ & $0.60$ \\%(1, 0)
\rowcolor{gray!30} GW190814\_192009 & \texttt{BBH\_5} & $68_{-19}^{+28}$ & $48_{-18}^{+21}$ & $0.5_{-0.6}^{+0.4}$ & $1.5_{-0.7}^{+0.8}$ & $25.2$ & $29.9$ & $33.4$ & $0.65$ & $0.64$ \\%(5, 4)
GW190910\_012619 & \texttt{BBH\_1} & $34_{-3}^{+3}$ & $2.9_{-0.2}^{+0.3}$ & $-0.87_{-0.11}^{+0.19}$ & $0.16_{-0.04}^{+0.04}$ & $40.2^*$ & $35.7$ & $32.1$ & $0.65$ & $0.58$ \\%(1, 1)
\rowcolor{gray!30} GW190920\_113516 & \texttt{BBH\_0} & $6.0_{-1.5}^{+3.3}$ & $3.2_{-1.0}^{+0.9}$ & $0.60_{-0.07}^{+0.26}$ & $0.13_{-0.05}^{+0.05}$ & $40.7$ & $26.4$ & $48.0$ & $0.56$ & $0.57$ \\%(0, 0)
GW190718\_160159 & \texttt{BBH\_1} & $10.0_{-1.8}^{+4.5}$ & $6.8_{-2.1}^{+1.4}$ & $0.73_{-0.17}^{+0.10}$ & $0.28_{-0.09}^{+0.10}$ & $41.1^*$ & $23.5$ & $47.6$ & $0.48$ & $0.53$ \\%(1, 0)
         \hline
         \hline
    \end{tabular}
    \label{tab:signalsFound}
\end{table*}
\egroup

%%%%%%%%%%%%%%%%%%%%%%%%%%%%%%%%%%%%%%%%%%%%%%%%%%%%%%%%%
%%%%%%%%%%%%%%%%%%%%%%%%%%%%%%%%%%%%%%%%%%%%%%%%%%%%%%%%%
%%%%%%%%%%%%%%%%%%%%%%%%%%%%%%%%%%%%%%%%%%%%%%%%%%%%%%%%%
%%%%%%%%%%%%%%%%%%%%%%%%%%%%%%%%%%%%%%%%%%%%%%%%%%%%%%%%%

\section{Newly Reported BBH mergers}
\label{sec:events}

Table~\ref{tab:signalsFound} summarizes the basic properties of the newly reported events: their parameters (source-frame masses, effective spin, and redshift), inverse false alarm rate (IFAR), and estimated $p_{\rm astro}$ (computed using the procedure described in Appendix~\ref{Appendix:pastro}). Appendix~\ref{Appendix:posteriors} contains intrinsic parameter and redshift posteriors for all the new events, and PE samples are publicly available at \url{https://github.com/seth-olsen/new_BBH_mergers_O3a_IAS_pipeline}. \changed{Interestingly, some of the new events near the detection threshold have properties unlike those of louder signals. At first sight, this may seem odd because one might expect that only about $\mathcal{O}(10\%)$ of any one kind of astrophysical source are detected near threshold (depending on the astrophysical distribution of source distances, as well as the pipeline's noise background distribution). This suggests that it is less likely for the first detection of any one kind of event to be marginal. The explanation for the population outliers among our marginal events might be a combination of occasional fluctuations (expected since there are many detections and many ways to be considered an outlier), plus some contamination from background triggers.} The sampling prior is uniform in detector-frame constituent masses, effective spin, and comoving volume-time ($VT$), with other extrinsic parameters drawn from standard geometric priors (i.e., isotropic orientation angles and locally uniform coalescence time). Redshifts are computed using a $\Lambda$ cold dark matter ($\Lambda$CDM) cosmology with Planck15 results \cite{cosmology_planck2015}.

More detail on the priors for intrinsic parameters can be found in \citet{flat_chieff_prior_o1event1216_formation_channels2019}, and a comparison of the flat effective spin prior with the isotropic spin prior used in other catalogs \cite{lvc_o3a_deep_gwtc2_1_update_2021, nitz_o3a_3ogc_catalog_2021} is given in Section II of \citet{ias_gw190521_prd2021olsen}. One observation that can be made about several events in Table~\ref{tab:signalsFound} is that the $\ln \mathcal{L}_{\rm max}$ achieved in PE was significantly larger than half the sum of the pipeline's squared SNR in Hanford and Livingston. Analytically, the maximum (coherent) network squared SNR is equivalent to twice the log of the maximum likelihood ratio for that same model and data, so the difference evidently comes from the additional information incorporated in the PE that does not enter into the pipeline SNR: Virgo data (when available) and a waveform model that includes the effects of higher-order multipole modes and spin precession (IMRPhenomXPHM \cite{xphm_pratten2020}). This suggests that future searches incorporating these effects might find those detections to be substantially more secure. We cannot yet precisely quantify the statistical significance of this difference \changed{because it will depend on the change in expected SNR of background (noise) triggers under the full waveform model}, but it motivates the development of such search algorithms. \changed{Beyond the O3a data, an efficient method for coherently integrating a score over three detectors and including the effects of higher harmonics and precession would extend detection sensitivity into regions of parameter space where current searches have low effectualness. In addition to the possibility of improving the significance of events already in the catalogs, these developments could uncover additional events in the least-explored subspaces of the BBH source parameter manifold.} In the remainder of this section, we briefly comment on the properties of each of the new source binaries. 
%\changed{Note that these results depend on the choice of prior, and we have added to the public GitHub repository PE posteriors sampled under the isotropic spin prior used in the GWTC \cite{lvc_o3a_gwtc2_catalog_2021, lvc_o3a_deep_gwtc2_1_update_2021} and OGC \cite{nitz_o3a_3ogc_catalog_2021} catalogs. A population analysis will also involve recomputing $\pastro$ estimates under different models (see, e.g., Ref.~\cite{ias_popO2_Roulet_2020}), and the results will be most sensitive to priors for triggers just above and just below the detection threshold. Under the prior used in this work we can count $\pastro$ values and estimate that roughly three of the ten new events have non-astrophysical sources, but under a different prior (such as one based on the population of secure detections, which would boost triggers from templates with small effective spins and penalize the highly spinning ones) this sum could become smaller or larger, as could the total number of triggers above and below the detection bar. We encourage anyone who would like to follow up on these results with an analysis under other priors to contact us with any questions about using the public repository (\url{https://github.com/seth-olsen/new_BBH_mergers_O3a_IAS_pipeline}).}

\subsection{High-mass sources}

\paragraph{\textbf{GW190707\_083226}}
This event is our most secure new detection, with $\pastro = 0.94$. The primary BH mass posterior extends to the UMG, although $m_{1} = 52^{+17}_{-12}\,\msun$ does not place it confidently above $45 \msun$ (see Fig.~\ref{fig:GW190707_083226_corner_plot}). The effective spin is consistent with zero and there is no preference for precessing signals in the posterior. The maximum likelihood sample has non-negligible contribution from higher-order multipole modes, with the whitened $(\ell, |m|) = (2, 1)$ amplitude becoming comparable to the fundamental mode near 100\,Hz in Livingston. The $(\ell, |m|) = (3, 3)$ gives the dominant contribution for $f \in [150, 200]$\,Hz, and above 200\,Hz the (4, 4) mode is the leading order amplitude. The presence of higher modes is expected due to a combination of the high total mass ($93\,\msun$), unequal masses ($m_2 / m_1 \approx 0.23$), and an inclination that does not favor the fundamental mode ($\iota \approx 1.2$ rad).

\begin{figure*}
    \centering
    \subfloat[Comoving Posteriors]{
    \includegraphics[width=.333\linewidth]{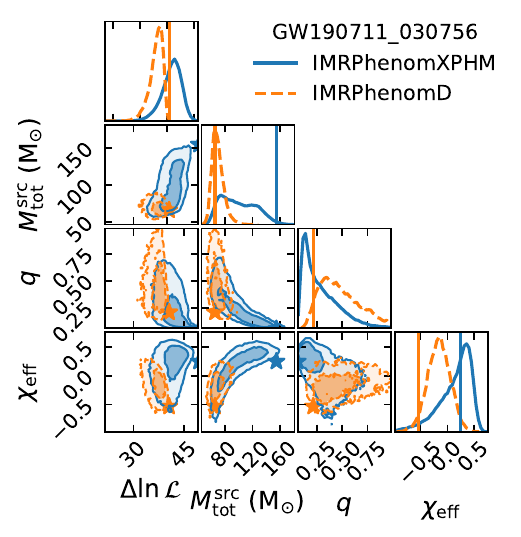}\label{fig:gw190711_XPHMvPhenomD}}
    \subfloat[Multipole Mode Content]{
    \includegraphics[width=.333\linewidth]{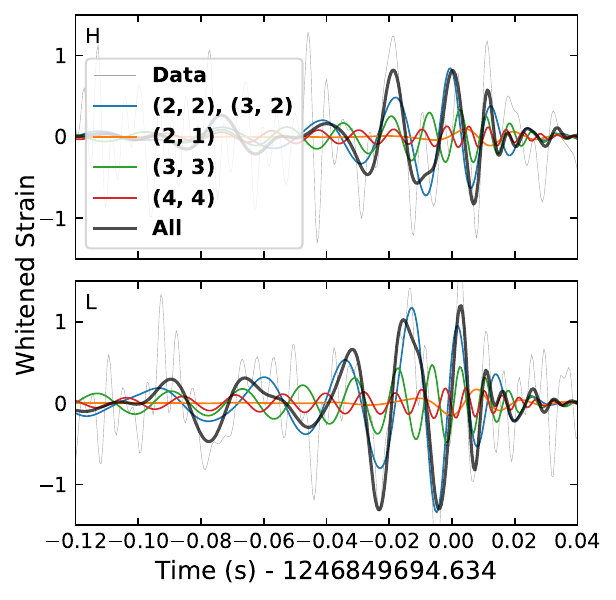}\label{fig:gw190711_lnLmax_wf_HM}}
    \subfloat[In-Plane Spin]{
    \includegraphics[width=.333\linewidth]{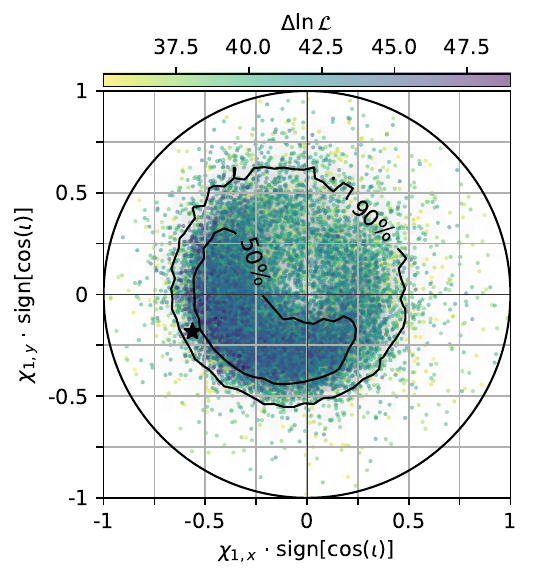}\label{fig:gw190711_inplane_spin}}
    \caption{A closer look at the posteriors for GW190711\_030756. Panel (a) shows the differences in constituent masses and effective spin between the parameter estimation with IMRPhenomXPHM \cite{xphm_pratten2020}, which has higher harmonics and generic spins, versus IMRPhenomD \cite{IMRPhenomD_approximant_Khan2016}, which has only the fundamental mode and aligned spins. The contours indicate $50\%$ and $90\%$ confidence intervals, and the stars indicate the maximum likelihood sample in each posterior. In panel (b) we plot the whitened waveform of the maximum likelihood sample broken down by multipole modes sharing the same value of $m$ (this is what determines a mode's frequency clocking). The maximum likelihood sample has a primary BH mass of $147 \,\msun$ and a secondary mass of $8\, \msun$ at redshift $0.26$, with an effective spin of $0.25$ and a strongly precessing primary tilted at $63^{\circ}$ from the orbital angular momentum and having dimensionless in-plane spin magnitude $\sqrt{\chi_{1,x}^2 + \chi_{1,y}^2} = 0.59$. The whitened detector strain is labeled `Data' and is low-passed before plotting to contain frequencies below 256 Hz for ease of visualization (note that the template contains no weight above 200 Hz). In panel (c) we plot the in-plane spin posterior for the primary BH, which is clearly pulled away from the prior (uniform within each disk of constant $\chi_{j, z}$) by a coherent peak in the likelihood manifold at nonzero tilt away from alignment. The $50\%$ and $90\%$ contours represent the full posterior, but for visualization we scatter only every eighth sample with $\mathcal{L} / \mathcal{L}_{\rm max} > e^{-15}$ (the black star indicates the maximum likelihood sample). }
    \label{fig:newevent_GW190711_030756_3panel}
\end{figure*}

%%%%%%%%%%%%%%%%%%%%%%%%%%%%%%%%%%%%%%%%%

\paragraph{\textbf{GW190711\_030756}}
This event, with a high $\pastro = 0.93$, presents a mass ratio significantly different from unity, a mildly positive effective spin, and primary mass that is most likely in the UMG (see Fig.~\ref{fig:GW190711_030756_corner_plot}). In comparing the PE results with the search we find $2 \ln \mathcal{L}_{\rm max} - \rho_{\rm H}^2 - \rho_{\rm L}^2 \approx 18$, which indicates that this event would likely be even more secure in future searches with templates that include higher modes and precession. This is supported by the result that PE with IMRPhenomD \cite{IMRPhenomD_approximant_Khan2016}--the same aligned-spin fundamental mode approximant used in the search--does not produce posteriors covering the higher likelihood region at extreme mass ratio (see Fig.~\ref{fig:gw190711_XPHMvPhenomD}). There is some preference for precessing waveforms in the posterior (see Fig.~\ref{fig:gw190711_inplane_spin}), and there is evidently a contribution from higher harmonics in the extreme mass ratio solution. In the maximum likelihood sample (see Fig.~\ref{fig:gw190711_lnLmax_wf_HM}), the amplitudes of the $(\ell, |m|) = (3, 3)$ and $(4, 4)$ modes overtake the fundamental mode at frequencies above 90 Hz and 100 Hz, respectively. The strength of higher harmonics near the peak of the likelihood makes this a good candidate for quasinormal mode analysis similar to that of \citet{quasinormal_ringdown_190521_nitzCapano2021} in their study of the GW190521 ringdown. The primary BH mass posterior's 90\% confidence interval extends beyond $125\, \msun$, meaning that the extreme mass ratio solution consists of an IMBH merging with a stellar mass BH.

%%%%%%%%%%%%%%%%%%%%%%%%%%%%%%%%%%%%%%%%%

\paragraph{\textbf{GW190818\_232544}}
This event, with $\pastro = 0.81$, has similar masses to GW190707\_083226 but has a very large and positive effective spin at high confidence: $\chieff = 0.7^{+0.2}_{-0.3}$ (see Fig.~\ref{fig:GW190818_232544_corner_plot}). The inferred mass of $67^{+23}_{-19}\, \msun$ puts the primary BH in the UMG, while the secondary is fairly heavy but can easily avoid the UMG. This source joins a pileup of events with total mass near $100 \,\msun$ and positive effective spin (see Fig.~\ref{fig:population}). There is no indication of precession and the maximum likelihood waveform has a similar $(\ell, |m|) = (3, 3)$ and $(4, 4)$ contribution to GW190707\_083226, but with the $(2, 1)$ mode losing significance.

\paragraph{\textbf{GW190814\_192009}} \label{sec:new190814}
This event, with $p_{\rm astro} = 0.64$, is not the most marginal in the statistical sense, yet it poses challenges in PE due to its low coherent network SNR. Both bank searches and likelihood maximization methods can find higher likelihood solutions at lower masses, but the increase in SNR is not enough to outweigh the look-elsewhere penalty we apply to low-mass candidates due to the large numbers of templates in that region of parameter space. More importantly, the coherence between detectors is weak in the sense that the coherent score with bank templates (no higher modes, aligned spins) and the likelihood maximization with IMRPhenomXPHM (higher modes, generic spins) both converge on two-detector coherent results which are significantly lower than the sum of the same maximization methods performed on individual detectors. The overall result is that the two-detector likelihood manifold has comparable peaks throughout a vast region of intrinsic parameter space, which means that priors may have a heavy hand in determining the inferred parameters. For this reason we cannot be confident that the inferred redshift of $z = 1.5^{+.8}_{-.7}$ indeed makes this the farthest ever detected GW signal (see Fig.~\ref{fig:GW190814_192009_corner_plot}). If real, however, this may be the most distant source to date. Note that, despite its considerably higher SNR, GW190521 also posed a formidable parameter estimation challenge \cite{lvc_event_GW190521, ias_gw190521_prd2021olsen, nitz2020_gw190521}, and hence this lack of a robust solution may not be surprising given the small number of cycles in the sensitive band.

\paragraph{\textbf{GW190906\_054335}}
This event, with $\pastro = 0.61$, is at the heavy end of the stellar collapse BH regime but does not pose issues for the UMG, with inferred masses of $37^{+12}_{-8}\,\msun$ and $24^{+8}_{-8}\,\msun$ (see Fig.~\ref{fig:GW190906_054335_corner_plot}). This is approaching the sweet spot in the total mass and mass ratio plane where the detector's sensitive volume is optimized: the binary is light enough to have a long signal with the fundamental mode's merger frequency within the detector's sensitive band, but heavy enough to be loud and with mass ratio near unity allowing the intrinsic luminosity distance to move toward optimality. The exceptional detectability of this mass configuration explains the fact that this source is among the farthest yet found, with a redshift of $0.9_{-0.3}^{+0.4}$. The effective spin of GW190906\_054335 is consistent with zero, and it shows no clear evidence for precession.

\begin{figure*}
    \centering
    \includegraphics[width=\linewidth]{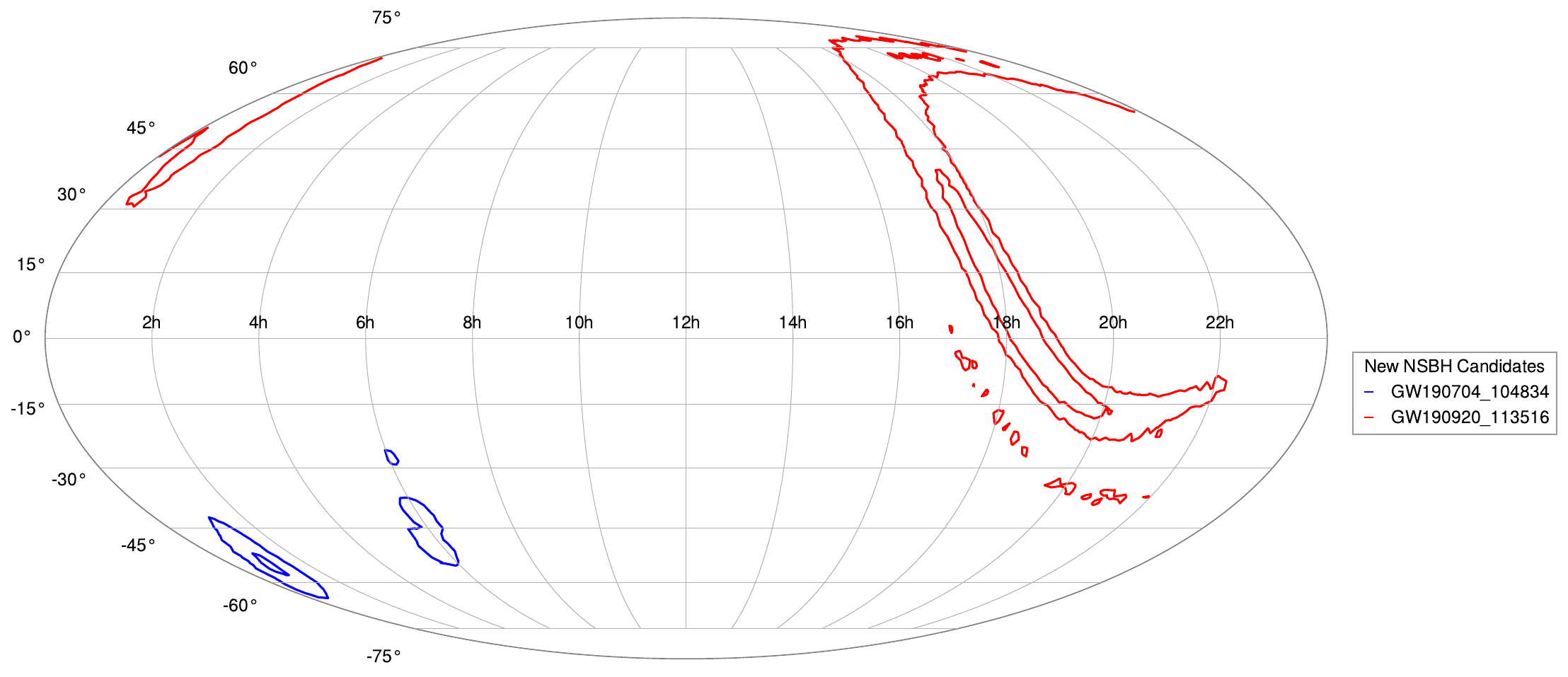}
    \caption{Sky localization for the possible NSBH candidates from the events new to this work, with priors that are uniform in detector-frame constituent masses, effective spin, and comoving $VT$. Two-dimensional 50\% and 90\% contours are drawn. The $x$-axis represents right ascension in hours, and the $y$-axis represents declination in degrees.} \label{fig:new_nsbh_skylocs}
\end{figure*}

%%%%%%%%%%%%%%%%%%%%%%%%%%%%%%%%%%%%%%%%%%%%%%%%%%%%%%%%%%%%%%%%%%
%%%%%%%%%%%%%%%%%%%%%%%%%%%%%%%%%%%%%%%%%%%%%%%%%%%%%%%%%%%%%%%%%%
\subsection{Low-mass sources}

\paragraph{\textbf{GW190704\_104834}}
This event is one of the more confident detections, with $p_{\rm astro} = 0.81$. The secondary BH, with an inferred mass of $3.2^{+1.2}_{-1.1}\,\msun$, may be a BH in the LMG or a heavy NS (see Fig.~\ref{fig:GW190704_104834_corner_plot}). The LMG solution has a small positive $\chieff$, and as the mass ratio becomes more extreme the effective spin increases to roughly 0.5 for the NSBH solution. A catalogue search for an EM counterpart of a NSBH merger at the time and direction of this event may prove fruitful. The sky localization is well constrained and is presented in Fig.~\ref{fig:new_nsbh_skylocs}.

\paragraph{\textbf{GW190821\_124821}}
This event has $p_{\rm astro} = 0.61$ in our search, but PE results suggest that its significance could improve in future searches that use Virgo data and templates with higher modes and precession. 
The source's effective spin is almost surely negative, with $\chi_{\rm eff} = -0.45_{-0.17}^{+0.33}$ (see Fig.~\ref{fig:GW190821_124821_corner_plot}), indicative of a dynamical formation channel \cite{need_dynamical_to_misalign_orbit_and_spin_Rodriguez2016, iasGW151226_prd2022chia}. The secondary BH, with an inferred mass mass of $4^{+1}_{-1.1}\,\msun$, is confidently in the LMG. The direction of the $\chi_{\rm eff}$--$q$ degeneracy is such that the non-spinning solution is the one with lowest $m_2 \approx 3\,\msun$. This event will improve rate measurements for systems containing BHs in the LMG.

\paragraph{\textbf{GW190910\_012619}}
This event, with $\pastro = 0.58$ is intriguing because it has a very well measured and extreme mass ratio of $m_2 / m_1 = 0.087^{+0.012}_{-0.012}$, and a large negative effective spin which is also well measured at $\chieff = -0.87_{-0.11}^{+0.19}$ (see Fig.~\ref{fig:GW190910_012619_corner_plot}). 
Such a large and negative effective spin has never been measured for a GW candidate before. 
The secondary BH falls in the LMG at high confidence with a mass of $2.9^{+0.3}_{-0.2}\,\msun$. 
This event also shows some evidence of precession, with Bayesian evidence ratio of $e^5$ in favor of precession when comparing the evidence computed by \texttt{PyMultinest} for the same waveform model and priors but with a likelihood model that takes in-plane spin components to be zero. 
We suspect future searches with precessing templates could improve the significance of this detection.

\paragraph{\textbf{GW190920\_113516}}
This event, with $p_{\rm astro} = 0.58$ is a possible NSBH, with $m_2 = 3.2^{+0.9}_{-1.0}\,\msun$ making the secondary constituent either a heavy NS or a BH in the LMG (see Fig.~\ref{fig:GW190920_113516_corner_plot}). Due to its low total mass and high effective spin ($\chieff \sim 0.6$), this source would be an excellent candidate for observing an EM counterpart associated to a merging NSBH. In Fig.~\ref{fig:new_nsbh_skylocs} we present the sky localization for the two NSBH candidates. Although the sky position of GW190920\_113516 is poorly constrained, we encourage a follow-up search for an EM counterpart wherever possible. This event shows no evidence of significant higher mode content or precession.

\paragraph{\textbf{GW190718\_160159}}
This event, with $p_{\rm astro} = 0.53$, is the most marginal in the set. The total mass is low despite both constituents avoiding the LMG, and the source presents a confidently large positive effective spin, $\chi_{\rm eff} = 0.73^{+0.1}_{-0.17}$ (see Fig.~\ref{fig:GW190718_160159_corner_plot}). In combination with the roughly equal masses, this configuration is quite rare under an isotropic spin distribution. Therefore, this event may help constrain BBH formation channel rates. The likelihood shows no preference for precessing waveforms, and near the peak there is no significant contribution from harmonics beyond the fundamental mode.

\section{Comparison to previously reported catalogues}
\label{sec:previous_events}

%%%%%%%%%%%%%%%%%%%%%%%%%%%%%%%%%%%%%%%%%%%%%%  LVC EVENTS
\bgroup
\def\arraystretch{1.25}
\begin{table*}
    \centering
    \caption{Hanford--Livingston coincident events already reported in the GWTC-2.1 catalog \cite{lvc_o3a_deep_gwtc2_1_update_2021} and the 3-OGC catalog \cite{nitz_o3a_3ogc_catalog_2021} as detected by our pipeline. Three events found by LVC in Hanford--Livingston coincidence were vetoed in our search: GW190521, GW190924\_021846, and GW190403\_051519. The following events are not included in this table because they were detected in Livingston--Virgo or Hanford--Virgo coincidence, or single detector search, all of which we have yet to run: GW190910\_112807, GW190925\_232845, GW190620\_030421, GW190630\_185205, GW190708\_232457, and GW190814. The inverse false alarm rate (IFAR) values in the GWTC-2.1 column were taken from the highest (most recent) version number of each event in the GWOSC catalog (\url{https://www.gw-openscience.org/eventapi/json/GWTC-2.1-confident/}), which corresponds to whichever LVC pipeline achieved the highest astrophysical probability for that event in the GWTC-2.1 analysis. The 3-OGC column was taken from the catalog summary data on the \texttt{GitHub} listed in the publication (\url{https://raw.githubusercontent.com/gwastro/3-ogc/master/3-OGC\_top.txt}).}
    \begin{tabular}{|c|c|cc|c|ccc|}%cccc|}
        %\hline
        \hline
        \multirow{2}{*}{Event Name} & \multirow{2}{*}{Bank} & \multirow{2}{*}{$\rho_{\rm H}^2$} & \multirow{2}{*}{$\rho_{\rm L}^2$} & \multirow{2}{*}{$p_{\rm astro}$} & \multicolumn{3}{c|}{IFAR (yr)} \\
        %& \multirow{2}{*}{$m_1 {\scriptstyle( \rm M_\odot)}$} & \multirow{2}{*}{$m_2 {\scriptstyle(\rm M_\odot)}$} & \multirow{2}{*}{$\chi_{\rm eff}$} & \multirow{2}{*}{$z$}
         \cline{6-8}
         & & & & &
         IAS\footnote{The IFARs are computed within each bank\changed{, and we do not include any additional trials factor}.} & $|$ GWTC-2.1 $|$ & 3-OGC \\
         %& & & & \\
         \hline
         \rowcolor{gray!30} GW190403\_051519 & \texttt{BBH\_4} & $23.1$ & $29.7$ & --- & Veto & $0.13$ & --- \\%(4, 0)
GW190408\_181802 & \texttt{BBH\_3} & $95.4$ & $109.2$ & $1.00$ & $> 1000$ & $> 1000$ & $> 1000$ \\%(3, 0)
\rowcolor{gray!30} GW190412\_053044 & \texttt{BBH\_2} & $76.2$ & $245.5$ & $1.00$ & $> 1000$ & $> 1000$ & $> 1000$ \\%(2, 0)
GW190413\_052954 & \texttt{BBH\_4} & $26.7$ & $50.5$ & $0.83$ & $4.2$ & $1.2$ & $1.4$ \\%(4, 0)
\rowcolor{gray!30} GW190413\_134308 & \texttt{BBH\_4} & $30.1$ & $62.3$ & $1.00$ & $> 1000$ & $2.9$ & $6.4$ \\%(4, 2)
GW190421\_213856 & \texttt{BBH\_4} & $68.0$ & $42.0$ & $1.00$ & $> 1000$ & $71.4$ & $> 1000$ \\%(4, 1)
\rowcolor{gray!30} GW190426\_190642 & \texttt{BBH\_5} & $24.1$ & $42.7$ & $0.33$ & $0.19$ & $0.24$ & --- \\%(5, 5)
GW190503\_185404 & \texttt{BBH\_3} & $83.2$ & $57.7$ & $1.00$ & $> 1000$ & $> 1000$ & $> 1000$ \\%(3, 1)
\rowcolor{gray!30} GW190512\_180714 & \texttt{BBH\_2} & $39.4$ & $119.4$ & $1.00$ & $> 1000$ & $> 1000$ & $> 1000$ \\%(2, 0)
GW190513\_205428 & \texttt{BBH\_3} & $78.0$ & $66.0$ & $1.00$ & $> 1000$ & $> 1000$ & $> 1000$ \\%(3, 0)
\rowcolor{gray!30} GW190514\_065416 & \texttt{BBH\_4} & $38.9$ & $31.7$ & $0.98$ & $290$ & $0.36$ & $0.19$ \\%(4, 1)
GW190517\_055101 & \texttt{BBH\_3} & $48.7$ & $58.5$ & $1.00$ & $> 1000$ & $9.1$ & $66.1$ \\%(3, 1)
\rowcolor{gray!30} GW190519\_153544 & \texttt{BBH\_4} & $81.6$ & $128.7$ & $1.00$ & $> 1000$ & $> 1000$ & $> 1000$ \\%(4, 2)
GW190521\_030229 & \texttt{BBH\_5} & $65.0$ & $129.8$ & --- & Veto & $769$ & $805$ \\%(5, 5)
\rowcolor{gray!30} GW190521\_074359 & \texttt{BBH\_3} & $142.3$ & $431.3$ & $1.00$ & $> 1000$ & $> 1000$ & $> 1000$ \\%(3, 1)
GW190527\_092055 & \texttt{BBH\_3} & $27.4$ & $46.9$ & $0.92$ & $10.8$ & $4.3$ & $0.37$ \\%(3, 1)
\rowcolor{gray!30} GW190602\_175927 & \texttt{BBH\_4} & $41.9$ & $111.6$ & $1.00$ & $> 1000$ & $> 1000$ & $391$ \\%(4, 3)
GW190701\_203306 & \texttt{BBH\_2} & $25.1$ & $53.8$ & $0.23$ & $0.084$ & $1.8$ & $0.13$ \\%(2, 2)
\rowcolor{gray!30} GW190706\_222641 & \texttt{BBH\_4} & $91.3$ & $79.2$ & $1.00$ & $> 1000$ & $2.9$ & $> 1000$ \\%(4, 3)
GW190707\_093326 & \texttt{BBH\_1} & $63.7$ & $97.5$ & $1.00$ & $> 1000$ & $> 1000$ & $> 1000$ \\%(1, 0)
\rowcolor{gray!30} GW190719\_215514 & \texttt{BBH\_3} & $37.0$ & $33.2$ & $0.90$ & $8.5$ & $1.6$ & $0.25$ \\%(3, 1)
GW190720\_000836 & \texttt{BBH\_1} & $44.7$ & $62.3$ & $1.00$ & $> 1000$ & $10.6$ & $559$ \\%(1, 0)
\rowcolor{gray!30} GW190725\_174728 & \texttt{BBH\_1} & $31.3$ & $59.1$ & $0.96$ & $34.2$ & $2.2$ & $0.41$ \\%(1, 0)
GW190727\_060333 & \texttt{BBH\_4} & $76.0$ & $61.3$ & $1.00$ & $> 1000$ & $> 1000$ & $> 1000$ \\%(4, 0)
\rowcolor{gray!30} GW190728\_064510 & \texttt{BBH\_1} & $58.4$ & $110.1$ & $1.00$ & $> 1000$ & $> 1000$ & $> 1000$ \\%(1, 0)
GW190731\_140936 & \texttt{BBH\_3} & $28.9$ & $39.6$ & $0.76$ & $2.1$ & $0.53$ & $0.43$ \\%(3, 1)
\rowcolor{gray!30} GW190803\_022701 & \texttt{BBH\_3} & $30.6$ & $43.7$ & $0.94$ & $15.7$ & $2.6$ & $2.4$ \\%(3, 1)
GW190805\_211137 & \texttt{BBH\_4} & $18.8$ & $54.8$ & $0.81$ & $3.3$ & $1.6$ & --- \\%(4, 1)
\rowcolor{gray!30} GW190828\_063405 & \texttt{BBH\_3} & $112.6$ & $142.3$ & $1.00$ & $> 1000$ & $> 1000$ & $> 1000$ \\%(3, 1)
GW190828\_065509 & \texttt{BBH\_2} & $54.5$ & $53.6$ & $1.00$ & $> 1000$ & $> 1000$ & $> 1000$ \\%(2, 1)
\rowcolor{gray!30} GW190909\_114149\footnote{The LVC reduced GW190909\_114149 to the marginal candidate list between GWTC-2 \cite{lvc_o3a_gwtc2_catalog_2021} and GWTC-2.1 \cite{lvc_o3a_deep_gwtc2_1_update_2021}, so we include it here since our recovery with $\pastro = 0.52$ is not the first detection. We leave other marginal LVC candidates (such as GW190531\_023648 and GW190426\_152155) off this list since they were sub-threshold in our analysis as well.} & \texttt{BBH\_3} & $31.3$ & $32.4$ & $0.52$ & $0.45$ & $0.010$ & --- \\%(3, 1)
GW190915\_235702 & \texttt{BBH\_3} & $92.4$ & $71.1$ & $1.00$ & $> 1000$ & $> 1000$ & $> 1000$ \\%(3, 0)
\rowcolor{gray!30} GW190916\_200658 & \texttt{BBH\_4} & $27.1$ & $36.5$ & $0.93$ & $20.7$ & $< 0.001$ & $0.22$ \\%(4, 2)
GW190917\_114629 & \texttt{BBH\_0} & $26.8$ & $40.6$ & $0.35$ & $0.17$ & $1.5$ & --- \\%(0, 0)
\rowcolor{gray!30} GW190924\_021846 & \texttt{BBH\_1} & $31.9$ & $94.9$ & --- & Veto & $> 1000$ & $> 1000$ \\%(1, 0)
GW190926\_050336 & \texttt{BBH\_3} & $45.3$ & $31.4$ & $0.96$ & $25.3$ & $0.91$ & $0.27$ \\%(3, 1)
\rowcolor{gray!30} GW190929\_012149 & \texttt{BBH\_5} & $40.2$ & $51.2$ & $1.00$ & $> 1000$ & $6.2$ & $3.1$ \\%(5, 2)
GW190930\_133541 & \texttt{BBH\_1} & $41.1$ & $55.6$ & $1.00$ & $> 1000$ & $55.6$ & $295$ \\%(1, 0)
         \hline
         %\hline
    \end{tabular}
    \label{tab:lvc events}
\end{table*}
\egroup

Table~\ref{tab:lvc events} summarizes our pipeline's results for the O3a Hanford--Livingston coincident events published by the LVC in the GWTC-2.1 catalog \cite{lvc_o3a_deep_gwtc2_1_update_2021}. We also include the significance reported by the 3-OGC catalog \cite{nitz_o3a_3ogc_catalog_2021}, which was the first to report four of the eight new events that LVC added between the original GWTC-2 catalog \cite{lvc_o3a_gwtc2_catalog_2021} and the refined results presented in GWTC-2.1. We restrict the focus of this section to events declared as confident in GWTC-2.1, but for completeness we note here that the previously declared \cite{lvc_o3a_gwtc2_catalog_2021} and subsequently revoked \cite{lvc_o3a_deep_gwtc2_1_update_2021} event GW190909\_114149 was detected with $\pastro = 0.52$ in our pipeline. That event is included in the population presented in Fig.~\ref{fig:population}, whereas LVC sub-threshold candidates that were also below the threshold in our analysis (such as GW190531\_023648 and GW190426\_152155) are not.

In the remainder of the section we briefly summarize the differences in significance and mention the event space excluded from our search. Note that the O3b data was released and updated catalogs have been produced \cite{lvc_gwtc3_o3_ab_catalog_2021, nitz_4ogc_o3_ab_catalog_2021} (along with population analysis \cite{LVC_pop_o3_ab_2021lvk}), but we do not discuss data beyond O3a here. \changed{One important distinction to keep in mind is between the estimated $\pastro$, which is based on the distribution of all O3a foreground and background triggers, and the IFAR, which is computed independently for each template bank. The astrophysical probability is the statistic used to determine whether a signal is declared as a detection, whereas the false alarm rate tells us how often the detector noise produces a trigger of a given SNR peaking in the same frequency band as that template (\texttt{BBH~0-4} are naturally separated by central merger frequency since they are delineated by chirp mass).}

\paragraph{\textbf{Confidently recovered events}}
Our analysis retains all previously reported Hanford--Livingston (HL) coincident BBH triggers except for the three candidates which were vetoed (indicated by the word ``Veto" in the IFAR column). Another three events (GW190701\_203306, GW190917\_114630, and GW190426\_190642) fall below the $\pastro = 0.5$ threshold to be declarable in our analysis. All the other events were detected with confidence comparable to or better than the LVC catalog. The inferred parameters from our analysis are largely consistent with the GWTC-2.1 and 3-OGC analyses, in all cases having overlap in the $90\%$ confidence intervals of constituent BH masses, effective spin, and redshift despite the difference in spin prior.

%\changed{\sout{Three of the events were given IFARs under 100 years by all pipelines in the GWTC-2.1 analysis but were improved to IFARs over 1000 years in our pipeline: GW190421\_213856, GW190930\_133541, and GW190720\_000836. Another four events which saturated the IFAR in our analysis were given comparable significance by at least one of the pipelines in the GWTC-2.1 analysis but achieved their maximum $\pastro$ in pipelines with much lower significance: GW190517\_055101, GW190929\_012149, GW190706\_222641, and GW190413\_134308. Three of the new events reported by Nitz et al. were given false alarm rates of $\gtrsim 1$ per year by LVC but achieved much higher significance in our pipeline with orders of magnitude smaller false alarm rates: GW190725\_174728, GW190926\_050336, and GW190916\_200658. Note that our IFAR computations do not include an additional trials factor for the number of template banks used in our analysis, but we only highlight improvements here that are more than an order of magnitude so that the presence or absence of such a penalty would not change the comparison.}}

\changed{For a detailed study of the effects that various choices in signal processing and statistical methodology have on the sensitivity of detection pipelines, collaboration between analysis groups is essential. It is important to note that neither IFARs nor $\pastro$ values should be directly compared between our results and the LVC catalogs, because there are a number of ways in which the analyses differ. Two such differences are the spin prior and the method for aggregating results. Although the spin priors used in the various LVC pipelines are closer to our flat effective spin prior than they are to a population-informed prior (with PyCBC inheriting the uniform template prior from their hybrid geometric-stochastic placement and GstLAL being uniform in the orbit-aligned spin components), there are still differences that must be accounted for in any rigorous comparison. Moreover, our IFARs are computed within smaller template banks and then ranking scores are combined over our six banks to compute $\pastro$ for a single pipeline result, whereas the LVC IFARs are computed over the whole search space of each pipeline and $\pastro$ is then chosen by maximizing over five pipelines.
Since our six template banks (\texttt{BBH\_0-5}) are delineated to minimize overlap, whereas the five LVC pipelines (cWB \cite{cWB}, MBTA \cite{mbta_o3a_pastro_andres2022}, GstLAL \cite{gstlal}, PyCBC and PyCBC\_BBH \cite{PYCBCPipeline}) have similar search spaces covering a larger region than our banks, our method of assigning IFARs accounts for less of the look-elsewhere effects that are penalizing the IFARs in the LVC catalog.}

\changed{This means that the only clear improvements are where IFARs change by orders of magnitude for triggers with comparable template prior. We expect this to be the case for some of the confirmed detections because our methodology accepts a small increase in the false negative rate for very loud triggers (which we assume the LVC will detect) in order to gain sensitivity near the detection bar. This naturally results in a loss of some very loud events and a gain of some marginal events, as well as the improvement of some events which were previously marginal. Adopting the approach of the GWTC catalogs \cite{gwtc1_o2catalog_LVC2018,lvc_o3a_gwtc2_catalog_2021,lvc_o3a_deep_gwtc2_1_update_2021,lvc_gwtc3_o3_ab_catalog_2021}, which maximize over pipelines rather than comparing pipeline-specific catalogs, our results are not only adding new detections but also making previous detections more secure. In the cases where we increase the significance of detections that were near-threshold in all previous analyses, we extend the list that can be used in population studies which choose to include only very secure events.} 
These improvements are in large part due to our aggressive vetoes, which also have some probability of rejecting high-SNR triggers that might otherwise be declared as confident events.

\paragraph{\textbf{Vetoed candidates}}

Vetoes are automated checks that improve sensitivity by rejecting noise transients (glitches) through a series of signal consistency tests. However, a template bank's incompleteness in the physical parameter space, along with noise limitations, may cause astrophysical signals to be vetoed. While it is important to respect the determination of the veto procedure in order to uphold the integrity of the reported IFARs, we also know a priori that incompleteness arises (in all existing end-to-end matched-filter pipelines \cite{lvc_o3a_deep_gwtc2_1_update_2021, nitz_o3a_3ogc_catalog_2021}) from the use of waveform approximants that force in-plane spin components to be zero and neglect multipole modes beyond the fundamental harmonic \cite{IMRPhenomD_approximant_Khan2016}. Three of the events that LVC reported as astrophysical were vetoed by our pipeline, the most notable of which is GW190521. This event and GW190403\_051519 were vetoed because of trigger checks called split tests, where we require that the accumulation of SNR in the matched-filter with the data is consistent with the accumulation of SNR in the template's self-overlap (i.e., the expected SNR in noiseless data). GW190924\_021846 was vetoed due to an excess sine Gaussian power test in the \SI{60}{\hertz} band.

\paragraph{\textbf{Marginally recovered candidates}}

There were also three confident detections by the LVC which are neither vetoed nor significant in our pipeline: GW190701\_203306, GW190917\_114630, and GW190426\_190642. GW190701\_203306 was an event for which Virgo contained close to the same SNR as Hanford, so it is likely that the significance of this event will increase substantially when Virgo data is included in our coincident detection. GW190917\_114630 is an event which was not recovered by 3-OGC or any of the PyCBC-based searches in GWTC-2.1, but GstLAL recovered it with a network SNR of 9.5 including Virgo, so again we expect improvement upon incorporating Virgo into the coincident search. GW190426\_152155 was unfortunately not covered by our banks due to the upper limit we placed on the secondary mass, so the closest template was still a relatively poor match despite reaching a moderate SNR. This will be addressed by improvements to \texttt{BBH 5} in our upcoming analysis of the O3b data (which has already been released and analyzed by other pipelines \cite{lvc_gwtc3_o3_ab_catalog_2021, nitz_4ogc_o3_ab_catalog_2021}).

\paragraph{\textbf{Search space excluded from this work}}

We did not perform a BNS search or a dedicated NSBH search (and therefore we do not provide results on the BNS event GW190425). We leave those to future analyses. In \cite{fishingIAS19_prd2021} we have noted that when the response of the operating detectors is very disparate, a focused analysis is required in order to achieve robust results. The results of this analysis will be reported in a separate publication. In this upcoming analysis we will cover the six events in GWTC-2.1 (as well as new detections) which occurred at times when either Hanford or Livingston were offline or unusable (there have been no Virgo-only detections): GW190620\_030421 (LV), GW190630\_185205 (LV), GW190910\_112807 (LV), GW190925\_232845 (HV), GW190708\_232457 (LV), GW190814 (LV). Note the exclusion of the Livingston-only event GW190424\_180648, which appears in GWTC-2 \cite{lvc_o3a_gwtc2_catalog_2021} but was reduced to a subthreshold candidate in the GWTC-2.1 update \cite{lvc_o3a_deep_gwtc2_1_update_2021}. The O3b data was recently released along with catalogs \cite{lvc_gwtc3_o3_ab_catalog_2021, nitz_4ogc_o3_ab_catalog_2021} and a population analysis \cite{LVC_pop_o3_ab_2021lvk}, but we do not address data beyond O3a in this work.

%%%%%%%%%%%%%%%%%%%%%%%%%%%%%%%%%%%%%%%%%%%%%%%%%%%%%%%%%
%%%%%%%%%%%%%%%%%%%%%%%%%%%%%%%%%%%%%%%%%%%%%%%%%%%%%%%%%
%%%%%%%%%%%%%%%%%%%%%%%%%%%%%%%%%%%%%%%%%%%%%%%%%%%%%%%%%
%%%%%%%%%%%%%%%%%%%%%%%%%%%%%%%%%%%%%%%%%%%%%%%%%%%%%%%%%
\section{Discussion}
\label{sec:conclusions}

We defer a quantitative population analysis to future work, but here we offer a brief qualitative discussion of the ways in which our new events might be significant in furthering an empirical understanding of the astrophysical population of merging BBH. We conclude with a summary of our O3a results and a note on the planned updates for the O3b analysis. 

\subsection{Astrophysical implications of the new events} \label{sec:astro_implications}

\paragraph{\textbf{The lower mass gap (LMG)}}

In modelling the mass distribution of BHs in the progenitors of NSBH mergers based on EM observations of several low-mass X-ray binaries (sample sizes varying from 6 to 16), studies over the years have found some evidence of a gap between the minimal stellar BH mass and the maximal NS mass \cite{lmg_OBS_EM_stellarBH_Mdist_from_few_xrayNSBH_Bailyn1998, lmg_OBS_EM_BHdistInfer_from_16xrayNSBH_ozel2010, lmg_OBS_EM_BHdist_from_15xrayNSBHrocheOver_Farr2011ilya}. Although some formation channel models exist which could produce such a gap \cite{lmg_or_not_sim3methods_differ_Patton2021eldridge, lmg_from_CCSNe_simulations_Liu2020-21sun}, it is unclear whether the inference of the gap from the X-ray binary samples is primarily astrophysical, or if instead the leading order factors are observational limitations \cite{lmg_model_3methods_1gap_fastSNtime_2012, lmg_NO_theoretical_support_Fryer1999kalogera} and/or systematic errors in analysis \cite{lmg_NOgap_OBS_EM_BHdistInfer_from_xrayNSBH_kreidberg2012farr_bailyn}.

A simple argument for why this apparent mass gap may be driven in large part by observational sensitivity as opposed to being a significant feature in the astrophysical distribution is that more and more BHs under $5\, \msun$ are detected as the state-of-the-art sensitivity increases, both through inference of dark companions to giant stars in EM data \cite{lmg_BH_EMobs_science2019, lmg_unicorn_close_3msunBH_EMobs_Jayasinghe2021, lmg_confirmingRV_unicorn_close_3msunBH_EMobs_Masuda2021}, and through GW signals from BBH mergers \cite{GW190814ApJ2020_LVC, lvc_o3a_deep_gwtc2_1_update_2021}. On the GW side, \citet{lmg_requires_pop_feature_Fishbach_2020holz} found that the lack of LMG mergers detected up through GWTC-1 \cite{gwtc1_o2catalog_LVC2018} indicated a gap-like feature in the LMG region; but in a follow-up using GWTC-2, \citet{lmg_bridging_need_NSBHfeature_Farah2021fishbach_holz} found that the empirical evidence for the LMG was not as strong. This change is driven almost exclusively by GW190814, the single example in GWTC-2 of a BBH constituent mass confidently below $5 \,\msun$ \cite{lvc_o3a_population_properties_2021, ias_o3a_population_analysis_prd2021roulet}.

In this work we present four events with a secondary mass $\lesssim 5\, \msun$ at $90\%$ confidence. GW190910\_012619 is a system with similar masses to GW190814 but with a large negative effective spin, and it could arise from the kind of dynamics channels proposed as possibilities for producing GW190814 (see, e.g., \citet{lmg_model190814_BNSremnant_lu2020bonnerot} and \citet{lmg_from_AGN_model190814_Yang2020bartos}). Although it is possible for GW190814 to have come from an isolated binary evolution channel (e.g., if the model allows Hertzsprung-gap donors to survive common-envelope evolution \cite{lmg_model190814_CEcrossHertzsprung_mandel2020ilya}), this origin is unlikely for GW190910\_012619 due to its anti-alignment. GW190821\_124821 is another event with negative effective spin and a secondary mass of $4^{+1.0}_{-1.1}$ confidently in the LMG, but its primary BH is only $\sim 2$ times as massive (in contrast to the extreme mass ratios of GW190910\_012619 and GW190814). Parameter estimation of GW190821\_124821 with an aligned-spin model using only the fundamental mode preferred an NSBH solution with a small positive effective spin, but with higher modes and precession we uncover an anti-aligned maximum likelihood solution that is strongly precessing with likelihood ratio $\sim e^{12}$ over the NSBH solution. 

The other two events, GW190920\_113516 and GW190704\_104834, have secondary constituents which might be BHs in the LMG or heavy NSs. The latter may be a more astrophysically interesting scenario to probe, especially if accompanied by an EM counterpart (see Fig.~\ref{fig:new_nsbh_skylocs}), and if one imposes the LMG through the mass prior then the NSBH solution is sure to be weighted much more heavily in the posteriors. Under the current mass prior, however, which is the same uniform-in-detector-masses prior used in other catalogs \cite{nitz_o3a_3ogc_catalog_2021, lvc_o3a_deep_gwtc2_1_update_2021}, the BBH solution with a secondary in the LMG and a mass ratio of $\sim 0.5$ is favored over the NSBH solution with mass ratio $\sim 0.2$. In either case the effective spin is positive, which (along with the masses) makes these systems feasible to produce in standard isolated binary evolution models \cite{isotropic_vs_aligned_spin_subpops_prefer_iso_Farr2017ilya, chieff_isolated_field_binary_matias2018, bimodal_spin_subpops_formation_channels_Galaudage2021ilya}. Since the maximum NS mass is uncertain (according to \citet{max_NS_mass_est_from_EOS_space_alsing2018berti} it can be constrained to $[2, 2.6]\, \msun$), determining whether these are BBH or NSBH mergers may be sensitive to prior assumptions. With or without these two additional examples, the new low-mass events presented here represent a substantial increase in our sample size of LMG mergers and can be expected to impact population inference of the astrophysical mass distribution.

\paragraph{\textbf{The upper mass gap (UMG)}}

It is difficult to populate the BH mass range of roughly $\sim$\,$45 \msun$ to $\sim$\,$135 \msun$ with stellar collapse because of pulsational pair instability and pair instability supernova ~\cite{umg_pre_pairinstability_early_fowler1964ApJS, umg_Barkat1967PairInstability, umg_another_old_pairinstability_bond1984ApJ, umg_HegerWoosley2002massgap, umg_Woosley2007pairinstabilitySN, umg_Woosley2017ppisn, umg_Farmer2019LowerEdgeBHMassGap, umg_in_collapse_simulations_chen2014, umg_yoshida2016_PPI_simulations, umg_pair_instability_mass_loss_Belczynski2016}.
An inferred BH mass in this range could indicate a hierarchical merger scenario \cite{hierarchical_7merger_scenario2020b, hierarchical_evidence_inGWTC2_assume_dynamical_Kimball2020b, hierarchical_from_dynamical_in_any_star_cluster2020b, hierarchical_from_triples2021, hierarchical_mergerFamily_dynamical_mass_dist_matters2021, hierarchical_mergers_agn_kocsis2019, hierarchical_multiples_2g_190521_190814_Liu2020a, hierarchical_rate_sensitive_to_natal_spins_Fragione2021kocsis, rate_hierarchical_cluster_spinningLO_nospinHI_rodriguez2019, cluster_hierarchical_metallicity_spin_mapelli2021} (though this can be reasonably excluded if the BH has low spin \cite{umg_hierarchical_must_have_BHspin_Gerosa2021, review_hierarchical_BBH_channels_rates_problems_Gerosa2021fishbach}), 
or stars with spin and metallicity conditions tuned to allow gravitational collapse to a BH larger than $45\, \msun$ \cite{umg_Mapelli2020rotationEffectOnMassGap, umg_pop3isolatedBinaryEvol_rate_simulations_lowMetallicity_Tanikawa2021, umg_collapsars_fromAbove_spinHe_Siegel2021} (this can push the bottom of the UMG up to $\gtrsim 85 \,\msun$ in the most fine-tuned stellar environments \cite{low_metallicity_1gBHs_to85m_shell_interactions_mass_gap_Farrell2021}), or possibly even sustained and highly efficient accretion \cite{accretion_for_stellar_mass_case_Rice2021, umg_gw190521_from_relativistic_accretion_Cruz-Osorio2021}.

Unlike the LMG, the boundary of the UMG is a regime of high sensitivity for current LVC detectors, with a pair of $\sim 50 \msun$ BHs being capable of producing $\sim 40$ times the squared SNR of a pair of $\sim 5 \msun$ BHs at the same distance with a typical O3a PSD. Therefore, despite 4 of the 38 BBH events from GWTC-2 having primary mass posteriors confidently above $50\,\msun$ \cite{lvc_o3a_gwtc2_catalog_2021}, population analyses have consistently found a significant die-off feature in the BH mass distribution around the lower edge of the UMG \cite{lvc_o3a_population_properties_2021, ias_o3a_population_analysis_prd2021roulet}. However, beyond the upper edge of the UMG we have severely limited sensitivity because we begin to lose the merger and late inspiral frequency range of the fundamental mode to the low-frequency noise wall in the detector PSD, which ramps up below $\sim$\,\SI{60}{\hertz} and by \SI{20}{\hertz} has risen by a factor of $\gtrsim$\,1000 compared to the optimally sensitive band (roughly $[100, 200]$\,Hz for typical O3a detector sensitivity). This makes it unclear whether we should expect a population model that includes an explicit UMG to do much better at describing the population than a power law with a peak feature, as used by LVC after GWTC-2 \cite{lvc_o3a_population_properties_2021}. Follow-up studies targeting the UMG found that a power law with a peak was a sufficient model to describe the observed mass distribution \cite{umg_from_population_baxter2021, umg_poking_holes_mass_gap_from_pop_edelman2021}, and these conclusions appear robust to whether or not one uses the kind of ``leave-one-out" analyses used by LVC \cite{pop_mass_address_outliers_but190521and190412_NOToutliers_Essick2021fishbach_holz_thrane}.

An additional two events with BHs in the UMG were reported in GWTC-2.1 \cite{lvc_o3a_deep_gwtc2_1_update_2021}, but they were not recovered by our pipeline or in 3-OGC \cite{nitz_o3a_3ogc_catalog_2021}: GW190426\_190642, which has a secondary mass that is redshifted to well over $100\,\msun$ in the detector frame; and GW190403\_051519, which (like GW190521) was vetoed by our pipeline. It remains to be seen whether one or both of the two heavy events that we vetoed will be recovered in our reanalysis of O3a and O3b, at which point we will have revisited whether all of the vetoes used in this analysis are indeed optimally applicable to the new high-mass bank (see Section~\ref{sec:changes}). At this point, however, we retain only three of the four UMG events from GWTC-2 and neither of the two additions in GWTC-2.1 (see Table~\ref{tab:lvc events}). From our new detections we add two mergers with a primary mass in the UMG at $90\%$ confidence (see Table~\ref{tab:signalsFound}), and a third whose posterior is bimodal with the possibility of the primary being either a precessing IMBH or a $\sim 45 \,\msun$ BH with poorly-measured spin at a larger distance (see Figures~\ref{fig:gw190711_inplane_spin} and \ref{fig:GW190711_030756_corner_plot}).

The least believable of these UMG violations is GW190814\_192009, which is not the lowest $\pastro$ in our catalog but has hints of falling into the false alarm bin due to its issues in parameter estimation (\changed{see Section \ref{sec:new190814}}). We also have more reason to mistrust our $\pastro$ estimation at those high masses due to the small sample size, as noted by \citet{lvc_o3a_deep_gwtc2_1_update_2021}. If we are to believe the inferred redshift of $1.5^{+0.8}_{-0.7}$, making it the most distant detection to date, then we also run the risk that the standard distance prior (uniform in comoving $VT$) is not accurately representing the cosmological rate evolution, which will have a significant impact on these distant sources \cite{umg_distance_and_mass_of_pop_shifts_distribution_ezquiaga2021}.

Even including GW190814\_192009, this gives our catalog only one more confident UMG detection than GWTC-2 \cite{lvc_o3a_gwtc2_catalog_2021} and 3-OGC \cite{nitz_o3a_3ogc_catalog_2021}, and one fewer than GWTC-2.1 \cite{lvc_o3a_deep_gwtc2_1_update_2021}, so overall we do not have much to offer beyond what has already been done to constrain the UMG in astrophysical populations. GW190711\_030756 will add some statistical value to these constraints, because although its primary BH mass $90\%$ confidence interval extends below $45\, \msun$, the likelihood has a clear preference for the extreme mass ratio solution with the primary as an IMBH (mass above $100 \,\msun$) and the secondary with mass below $20\, \msun$ (see Fig.~\ref{fig:gw190711_XPHMvPhenomD}). Like the other two UMG sources new to this work, the IMBH solution of GW190711\_030756 has a substantial positive effective spin, and this seems to be a trend in the population masses and spins shown in Fig.~\ref{fig:population}: there is an apparent build-up of high-mass events at $\chieff \gtrsim 0.5$, to which we now turn.

\paragraph{\textbf{Effective spin}}

The total energy radiated by a merger sets an intrinsic luminosity which, averaged over detector and BBH orientations, allows sources in some regions of parameter space to be observable from larger distances than others, leading to an advantage in total detection rate given a fixed astrophysical rate density. Thus when we look at Fig.~\ref{fig:population} and see many more events above a total mass of $\sim 40 \msun$ than below it, we must account for the significant difference between the sensitive volumes of these regions before inferring their relative astrophysical rates. It has long been understood that effective spin is another parameter which is positively correlated to total radiated energy (and therefore loudness) due to the so-called orbital hang-up effect \cite{orbital_hangup_Campanelli2006}, and this effect is even more pronounced in heavier systems \cite{BBHpop_MchiCorr_Roulet_2019, intrinsic_obs_dist_by_chieff_mtot_mehta2021observing}. Thus one might imagine that the relative abundance of high mass sources with positive $\chieff$ compared to negative (seen in Fig.~\ref{fig:population}) could be entirely explained by the dependence of sensitive volume on intrinsic parameters. This dependence is discussed in Section IV.B of Reference~\cite{ias_gw190521_prd2021olsen} where a $VT_{\rm{max}}$ is estimated as a function of intrinsic parameters, and similarly (but independent of cosmology) we can define some maximum observable luminosity distance $D_{\rm{max}}$ for a fiducial SNR. The dependence of $D_{\rm{max}}$ on intrinsic parameters makes the detected population a biased sample of the astrophysical distribution, so one must correct for this selection effect before attempting to infer the parameter dependence of astrophysical rates from detection catalogs.

The mass distribution is coupled to the spin distribution in many formation channel models (see, e.g., discussions in \citet{signatures_of_dynamical_mapelli2020}), especially near mass gap edges \cite{umg_Mapelli2020rotationEffectOnMassGap}. Population models which include correlation between the mass and spin dimensions are significantly better at fitting the population than models which do not \cite{who_ordered_that-better_models_have_q_chieff_corr-Callister2021farr}, as are models which allow for independently modeled sub-populations \cite{ias_o3a_population_analysis_prd2021roulet, bimodal_spin_subpops_formation_channels_Galaudage2021ilya}. Whether trends in the data reflect the rate distributions predicted by formation channel models is a question that yields different answers depending on the sample of events and the population modeling method \cite{BBHpop_MchiCorr_Roulet_2019, mass_chieff_trends_pop_farr2020Safarzadeh, who_ordered_that-better_models_have_q_chieff_corr-Callister2021farr, bimodal_spin_subpops_formation_channels_Galaudage2021ilya}. Moreover, though we can make some theoretically robust predictions associating effective spin characteristics to formation channels \cite{need_dynamical_to_misalign_orbit_and_spin_Rodriguez2016, isotropic_vs_aligned_spin_subpops_prefer_iso_Farr2017ilya, chieff_isolated_field_binary_matias2018, field_binary_no_chieff_pop2020}, the predicted distributions and the relative rates between channels can be sensitive to uncertain priors like progenitor metallicity and natal BH mass and spin distributions \cite{rate_hierarchical_cluster_spinningLO_nospinHI_rodriguez2019, hierarchical_mergerFamily_dynamical_mass_dist_matters2021, star_cluster_rates_eccentricity_rigriguez2018samsing, cluster_hierarchical_metallicity_spin_mapelli2021, metallicity_effects_mapelli2017, star_cluster_dynamical_v_isolated_rates_mapelli2020}.

On the observational side, some constraints on population spin inference can be obtained in a prior-agnostic way \cite{pop_spin_isolated_field_vs_dynamical_NOprior_Talbot2017thrane} but it is impossible to completely remove the effects of assumptions about the astrophysical population on modeled searches and PE, and the choice of priors used for individual events can impact the results of population inference \cite{impact_of_priors_Vitale2017}. Without knowing the true astrophysical distribution, one can maximize the role of the likelihood (i.e., the data) in determining the posterior by using priors that are uniform in the best-measured parameters \cite{ias_gw190521_prd2021olsen}. This motivates us to use an intrinsic prior that is uniform in effective spin \cite{flat_chieff_prior_o1event1216_formation_channels2019} instead of the isotropic spin prior used in other catalogs \cite{lvc_o3a_deep_gwtc2_1_update_2021, nitz_o3a_3ogc_catalog_2021}, which was motivated by the predicted distribution in dynamical formation channels in which constituent BH spins are randomly oriented with respect to the orbital angular momentum.

Of the ten new events reported in this work, six had confidently nonzero $\chieff$ under this uniform effective spin prior, with four in the positive (aligned) direction and two in the negative (anti-aligned) direction (see Table~\ref{tab:signalsFound}). GW190704\_104834 has a small but positive effective spin with a tail extending to higher values; GW190818\_232544, GW190920\_113516, and GW190718\_160159 all have $\chieff \gtrsim 0.5$ at high confidence. Notably, all of these results are robust to reweighting \cite{reweighting_importance_sampling_thrane2019} from the uniform $\chieff$ prior to the isotropic prior which suppresses large effective spin magnitudes. These events cover the mass range all the way from the LMG to the UMG and their addition to the catalog could lend support to the type of bi-modal effective spin distribution used by \citet{bimodal_spin_subpops_formation_channels_Galaudage2021ilya}, which may help constrain rate contributions from isolated binary evolution channels \cite{chieff_isolated_field_binary_matias2018} even after accounting for the effects of $\partial D_{\rm{max}} / \partial \chieff$ and $\partial D_{\rm{max}} / \partial M$.

Dynamical channels, on the other hand, are expected to be responsible for producing negative effective spins \cite{need_dynamical_to_misalign_orbit_and_spin_Rodriguez2016, signatures_of_dynamical_mapelli2020}, which have never been observed at high confidence under the isotropic spin prior. Here we report two events with confidently negative effective spin: GW190821\_124821 and GW190910\_012619. The more secure event is GW190821\_124821, which has a more moderately negative effective spin that becomes consistent with zero at the $10\%$ level under the isotropic spin prior. GW190910\_012619, with $\chieff = -0.87^{+0.19}_{-0.11}$ under the uniform $\chieff$ prior, remains confidently negative even under the isotropic prior expected to describe dynamical channels, with $\chieff = -0.78^{+0.17}_{-0.12}$ after reweighting. This makes GW190910\_012619 the first detection of BBH anti-alignment measured under an isotropic spin prior. One possible concern is that the most confidently large effective spin magnitude measurements are associated to the least secure events, i.e., an apparent trend of $\partial \pastro / \partial |\chieff| < 0$. While GW190818\_232544 is quite secure with $\pastro > 0.8$, the other three extreme effective spin events are the least secure of the new detections, all with $\pastro < 0.6$. We do expect that some fraction of the declared events near the detection threshold are in fact noise transients, and $\pastro$ has higher variance in these underpopulated regions, but their collective statistical presence will be helpful in improving the ongoing empirical investigation of effective spin the observed BBH population \cite{lvc_o3a_population_properties_2021, ias_o3a_population_analysis_prd2021roulet, bimodal_spin_subpops_formation_channels_Galaudage2021ilya}.

\subsection{Concluding remarks}

We have reported ten new BBH merger events, declared based on the criteria that $\pastro > 0.5$, following \citet{lvc_o3a_deep_gwtc2_1_update_2021} and \citet{nitz_o3a_3ogc_catalog_2021}. Our computation of the ranking score and $\pastro$ are given in Appendices~\ref{Appendix:collection} and \ref{Appendix:pastro}, respectively. Notable detections include GW190910\_012619: the first reported event with well-measured negative effective spin at high confidence under the isotropic spin prior (which describes the kind of dynamical channels that can produce anti-aligned mergers \cite{need_dynamical_to_misalign_orbit_and_spin_Rodriguez2016}); and GW190704\_104834: a possible NSBH candidate that is well-localized on the sky (see Fig.~\ref{fig:new_nsbh_skylocs}), with the NSBH solution having a confidently positive effective spin that makes it a good candidate for an EM counterpart search. The collection of new events will have a statistically interesting impact on future population inference of the effective spin distribution, providing a number of detections in sparsely populated regions of the $M$--$\chieff$ plane (see Fig.~\ref{fig:population}). These outlying examples will also inform the investigation of the BH mass spectrum, with four detections confidently in the lower mass gap and two detections confidently in the upper mass gap, as well as GW190711\_030756: a multi-modal likelihood that favors a solution with a precessing IMBH ($m_1 \sim 120 \msun$) at extreme mass ratio ($q \sim 0.1$).

By simply summing the complements of the $\pastro$ values in Table~\ref{tab:signalsFound}, we can estimate that roughly three of the new events are noise transients rather than astrophysical signals. \changed{Estimates of $\pastro$ and source parameters depend on the choice of prior, and results become more sensitive to the prior as SNR decreases. The information needed for using different astrophysical models to estimate $p_{\rm astro}$ (see, e.g., Ref.~\cite{ias_popO2_Roulet_2020}) and reweight posterior samples (see, e.g., Ref.~\cite{reweighting_importance_sampling_thrane2019}) is available to the public at \url{https://github.com/seth-olsen/new_BBH_mergers_O3a_IAS_pipeline}. The flat effective spin prior used in this work differs from the isotropic spin prior used for PE in the GWTC \cite{lvc_o3a_gwtc2_catalog_2021, lvc_o3a_deep_gwtc2_1_update_2021} and OGC \cite{nitz_o3a_3ogc_catalog_2021} catalogs in that it does not penalize solutions with large effective spins (for a more detailed comparison of these priors, see \cite{ias_gw190521_prd2021olsen}). Our public GitHub of results includes PE posteriors sampled under the isotropic spin prior, but a full population analysis will also require new $\pastro$ estimates using the full list of triggers  (\textit{IAS\_O3a\_triggers.hdf}). In the case of priors that favor sources with small effective spin, triggers from highly spinning templates will be down-weighted and templates near zero effective spin will be boosted. It will be interesting to see how the total number of events and the source parameter distributions change under the priors implied by various astrophysical channels, and we encourage anyone interested in population studies to contact us with any questions about analyzing the publicly available triggers.}

We 
%\changed{\sout{saw a significant increase in IFARs for many of the events}} 
confirm or improve the significance of detections previously reported by other pipelines in Hanford--Livingston coincidence, \changed{retaining all such GWTC-2.1 events} except for three vetoes (GW190521, GW190403\_051519, and GW190924\_021846) and three previously declared events which dropped below $\pastro = 0.5$ in our analysis (GW190701\_203306, GW190917\_114630, and GW190426\_190642).
We also bring back to significance (albeit with a $\pastro$ of only $0.52$) the previously declared event GW190909\_114149 from the GWTC-2 catalog \cite{lvc_o3a_gwtc2_catalog_2021}, which was reduced to a sub-threshold in the GWTC-2.1 update \cite{lvc_o3a_deep_gwtc2_1_update_2021}.
From the O3a data, this amounts to a total of 42 BBH detections by our pipeline's Hanford--Livingston coincident search (see Tables~\ref{tab:signalsFound} and \ref{tab:lvc events}). This will soon be expanded with results from our disparate detector response search, which also includes data having coincidence in only Hanford--Virgo or Livingston--Virgo.

In our upcoming unified analysis of O3a and O3b, we aim to implement several pipeline improvements such as the use of Virgo data in the coincident search, the expansion of our heaviest template coverage to higher masses (possibly restructuring the amplitude categorization of templates to be organized by total mass rather than chirp mass for the heavier banks), and bank-dependent updates to the veto procedure so that the small number of in-band cycles for the heaviest events does not lead to over-aggressive vetoing in the presence of template bank incompleteness. We are also exploring methods for incorporating the effects of higher-order multipole modes and spin precession in the ranking statistic. While implementing these improvements, we intend to complete an analysis of the O3b data using the same version of the pipeline as this work, which will be released on a similar timescale to our disparate detector search. The result of adding these IAS pipeline detections into analyses of catalogs combining results from other pipelines will be to refine our understanding of both astrophysical populations and fundamental physics.

%%%%%%%%%%%%%%%%%%%%%%%%%%%%%%%%%%%%%%%%%%%%%%%%%%%%%%%%%%%%%%%%%%%%%%%%%%%%
\section*{Acknowledgements}
%%%%%%%%%%%%%%%%%%%%%%%%%%%%%%%%%%%%%%%%%%%%%%%%%%%%%%%%%%%%%%%%%%%%%%%%%%%%

This research has made use of data, software and/or web tools obtained from the Gravitational Wave Open Science Center (\url{https://www.gw-openscience.org/}), a service of LIGO Laboratory, the LIGO Scientific Collaboration and the Virgo Collaboration. LIGO Laboratory and Advanced LIGO are funded by the United States National Science Foundation (NSF) as well as the Science and Technology Facilities Council (STFC) of the United Kingdom, the Max-Planck-Society (MPS), and the State of Niedersachsen/Germany for support of the construction of Advanced LIGO and construction and operation of the GEO600 detector. Additional support for Advanced LIGO was provided by the Australian Research Council. Virgo is funded, through the European Gravitational Observatory (EGO), by the French Centre National de Recherche Scientifique (CNRS), the Italian Istituto Nazionale di Fisica Nucleare (INFN) and the Dutch Nikhef, with contributions by institutions from Belgium, Germany, Greece, Hungary, Ireland, Japan, Monaco, Poland, Portugal, Spain.

SO acknowledges support as an NSF Graduate Research Fellow under Grant No. DGE-2039656. Any opinions, findings, and conclusions or recommendations expressed in this material are those of the authors and do not necessarily reflect the views of the National Science Foundation.
TV acknowledges support by the National Science Foundation under Grant No. 2012086. 
JR is supported by a grant to the KITP from the Simons Foundation (\#216179).
BZ is supported by a research grant from the Center for New Scientists at the Weizmann Institute of Science and a research grant from the Ruth and Herman Albert Scholarship Program for New Scientists.
MZ is supported by the Canadian Institute for Advanced Research (CIFAR) program on Gravity and the Extreme Universe and the Simons Foundation Modern Inflationary Cosmology initiative.

%%%%%%%%%%%%%%%%%%%%%%%%%%%%%%%%%%%%%%%%%%%%%%%%%%%%%%%%%
%%%%%%%%%%%%%%%%%%%%%%%%%%%%%%%%%%%%%%%%%%%%%%%%%%%%%%%%%
\appendix

%%%%%%%%%%%%%%%%%%%%%%%%%%%%%%%%%%%%%%%%%%%%%%%%%%%%%%%%%
%%%%%%%%%%%%%%%%%%%%%%%%%%%%%%%%%%%%%%%%%%%%%%%%%%%%%%%%%
\section{Posteriors for the new events}
\label{Appendix:posteriors}

Here we present the parameter estimation posteriors for the new events reported in this work under priors that are uniform in detector-frame constituent masses (as in GWTC-2.1 and 3-OGC), effective spin (for more details on the flat effective spin prior, see \cite{flat_chieff_prior_o1event1216_formation_channels2019} or \cite{ias_gw190521_prd2021olsen}), and comoving $VT$ (using a $\Lambda$CDM cosmology with Planck15 results \cite{cosmology_planck2015}). The samples are publicly available at \url{https://github.com/seth-olsen/new_BBH_mergers_O3a_IAS_pipeline}. We use a new parameter estimation software called \texttt{cogwheel}, created by the authors of this paper and 
%to be released for public use in a forthcoming publication.
\changed{recently released for public use (accessible at \url{https://github.com/jroulet/cogwheel}, described in Ref.~\cite{IAScogwheel_roulet2022prd})}. The parameter estimation software uses the \texttt{PyMultinest} sampler \cite{pymultinest} (based on the \texttt{MultiNest} importance nested sampling library \cite{multinest_orig_Feroz2007, multinest_orig_Feroz2007, multinest_Feroz2013}) and these posteriors were generated with a log-evidence tolerance value of $0.1$ and 8192 live points. 

Waveforms are generated with the IMRPhenomXPHM approximant \cite{xphm_pratten2020}. Likelihoods are computed using a relative binning method similar to that of \citet{relative_binning} but adapted so that multipole modes contributions are computed in groups of modes with with common values of $m$ for improved efficiency \cite{relative_binning_higher_modes}. The sampling coordinates, which \changed{are} 
%will be 
described in detail in the \texttt{cogwheel} release paper \cite{IAScogwheel_roulet2022prd}, are carefully designed to minimize correlations throughout the intrinsic and extrinsic parameter space. These coordinate choices improve the sampling efficiency and reduce the risk of pathological convergence. While the new coordinates do improve spin measurements, the effective spin remains the only consistently well-measured spin variable, so we do not include any additional spin parameters in these corner plots. Notably, there is not a reliable way to quantify precession in the population, although \citet{generalized_chi_precession2021} have developed generalized precession parameters to this end which may prove informative as measurements improve in other spin dimensions. As seen in Fig.~\ref{fig:gw190711_inplane_spin}, our preferred visualization method for examining precession in an individual event is a likelihood mapping of the in-plane spin posterior for the primary BH. We provide code to produce these plots in the public repository with the samples: \url{https://github.com/seth-olsen/new_BBH_mergers_O3a_IAS_pipeline}.

\begin{figure}[H]
    \centering
    \includegraphics[width=\linewidth]{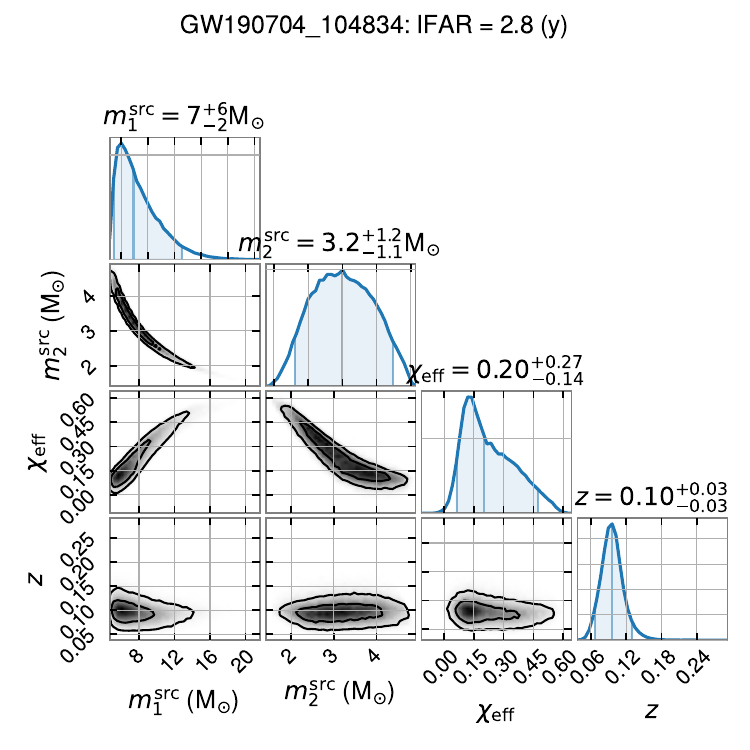}
    \caption{GW190704\_104834 has $p_{\rm astro} = 0.81$ and a secondary which may be a BH in the LMG or a heavy NS. This source should be targeted by searches for EM counterparts of NSBH mergers (sky localization, which is well-constrained, is presented in Fig.~\ref{fig:new_nsbh_skylocs}).}
    \label{fig:GW190704_104834_corner_plot}
\end{figure}

\begin{figure}[H]
    \centering
    \includegraphics[width=\linewidth]{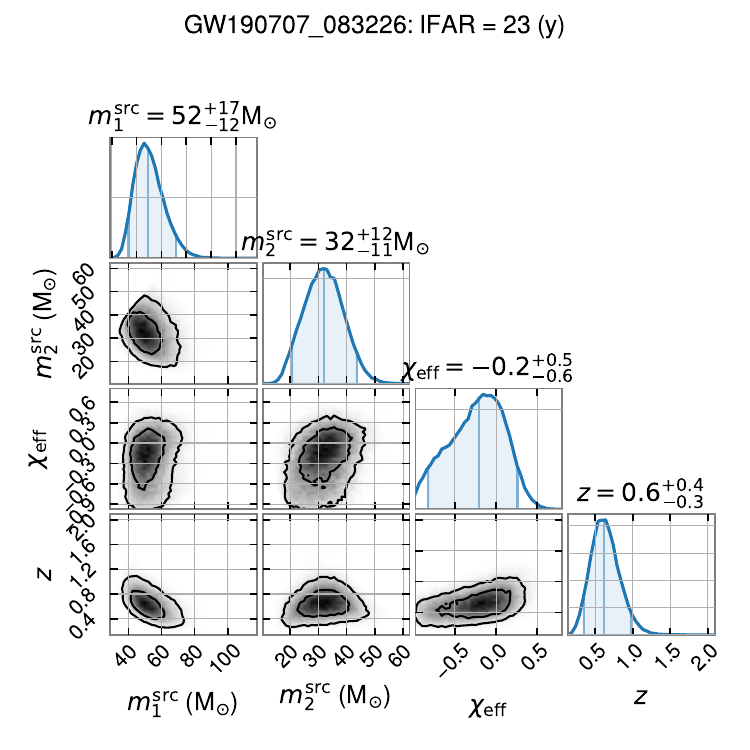}
    \caption{GW190707\_083226 has $\pastro = 0.94$ and the maximum likelihood region has a significant contribution from higher harmonics.}
    \label{fig:GW190707_083226_corner_plot}
\end{figure}

\begin{figure}[H]
    \centering
    \includegraphics[width=\linewidth]{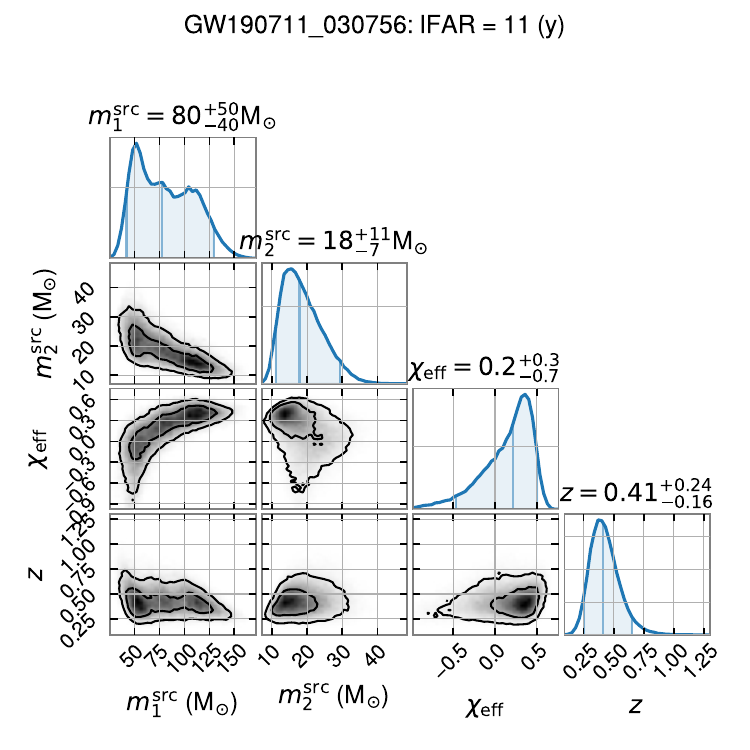}
    \caption{GW190711\_030756 has $\pastro = 0.93$ and the likelihood favors an extreme mass ratio solution whose primary is a precessing IMBH (see Figures~\ref{fig:gw190711_XPHMvPhenomD}, \ref{fig:gw190711_lnLmax_wf_HM}, \ref{fig:gw190711_inplane_spin})}.
    \label{fig:GW190711_030756_corner_plot}
\end{figure}

\begin{figure}[H]
    \centering
    \includegraphics[width=\linewidth]{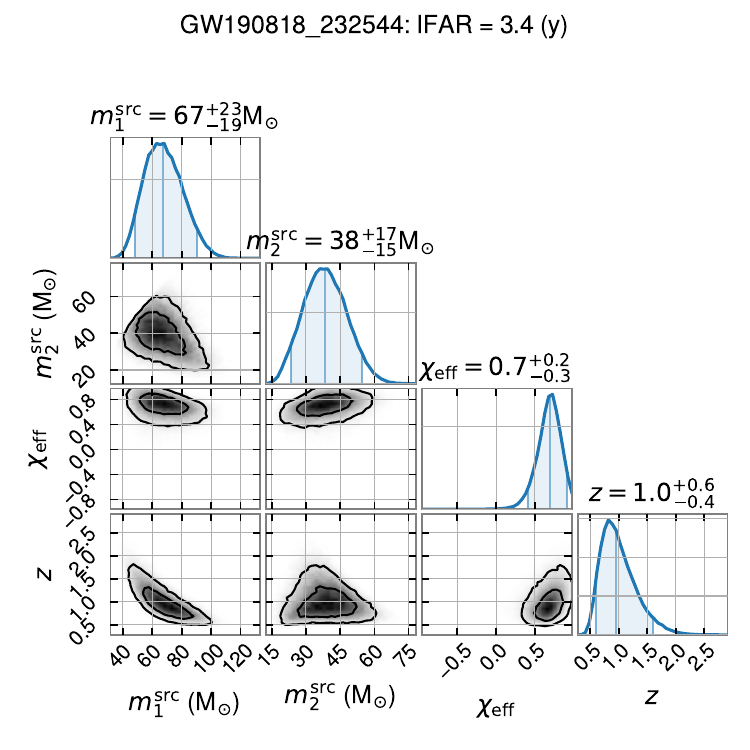}
    \caption{GW190818\_232544 has $\pastro = 0.81$, a primary BH in the UMG, and a very large and positive effective spin at high confidence.}
    \label{fig:GW190818_232544_corner_plot}
\end{figure}

\begin{figure}[H]
    \centering
    \includegraphics[width=\linewidth]{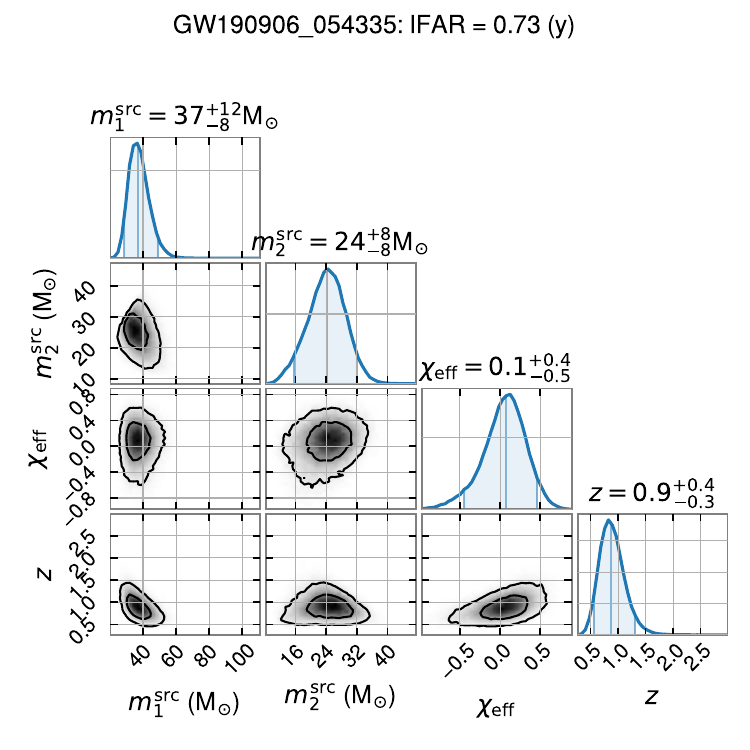}
    \caption{GW190906\_054335 has $\pastro = 0.61$ and the BH masses are favorable for optimizing the intrinsic luminosity, which helps to explain why the inferred redshift places it farther than every event in GWTC-2.1 \cite{lvc_o3a_deep_gwtc2_1_update_2021} except for GW190403\_051519.}
    \label{fig:GW190906_054335_corner_plot}
\end{figure}

\begin{figure}[H]
    \centering
    \includegraphics[width=\linewidth]{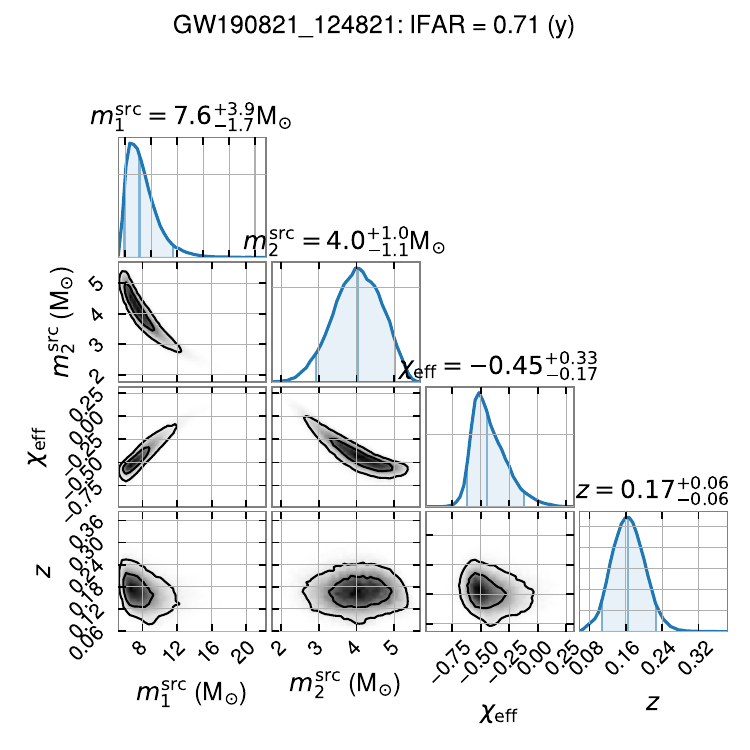}
    \caption{GW190821\_124821 $p_{\rm astro} = 0.61$ and might be considered more secure in future searches where we incorporate Virgo data in the ranking score and use templates with higher harmonics and precession. The source's effective spin is negative and the secondary BH is in the LMG, both at $90\%$ confidence.}
    \label{fig:GW190821_124821_corner_plot}
\end{figure}

\begin{figure}[H]
    \centering
    \includegraphics[width=\linewidth]{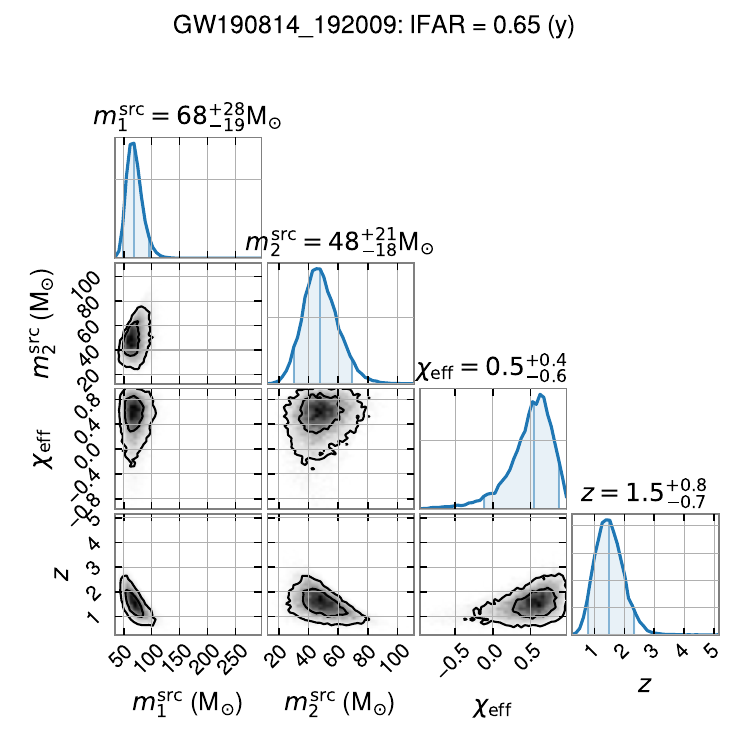}
    \caption{GW190814\_192009 has $p_{\rm astro} = 0.64$ and causes problems in PE because the coherent two-detector solution does not achieve squared SNR comparable to the sum of the single-detector maxima. Both template bank searches and likelihood maximization methods find higher likelihoods at lower chirp masses, but not so much higher as to overcome the large look-elsewhere effect of the low-mass templates, meaning that that source must be heavy in order to achieve a ranking score above the detection threshold. The likelihood manifold has many comparable peaks throughout intrinsic parameter space, which means that priors may have a heavy hand in determining the inferred parameters. The inferred redshift of $z = 1.5^{+.8}_{-.7}$ would make this the farthest ever detected GW signal if real. Note that, despite its dramatically higher SNR, GW190521 also posed issues in PE \cite{lvc_event_GW190521, ias_gw190521_prd2021olsen, nitz2020_gw190521} due to the small number of waveform cycles in the sensitive band, which is a factor here as well.}
    \label{fig:GW190814_192009_corner_plot}
\end{figure}

\begin{figure}[H]
    \centering
    \includegraphics[width=\linewidth]{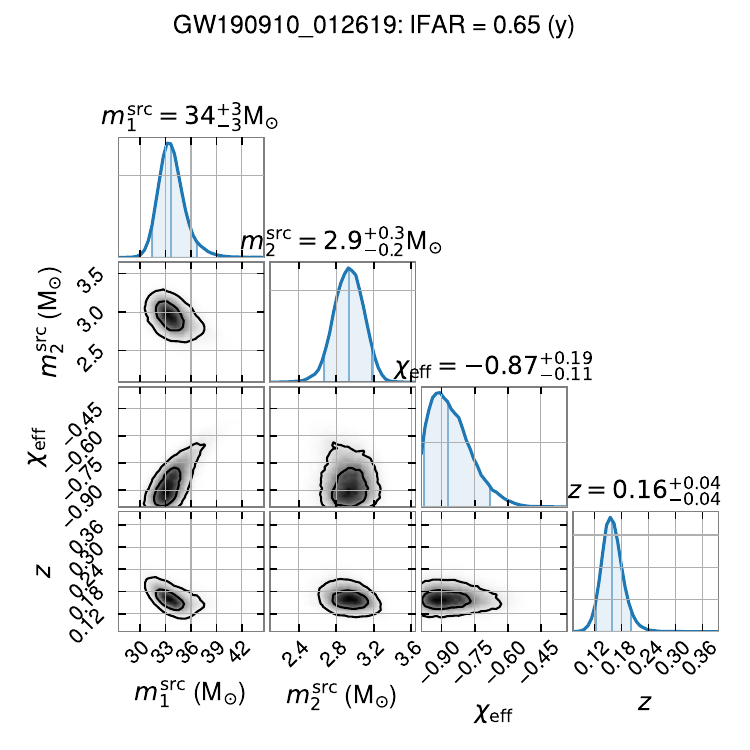}
    \caption{GW190910\_012619 has $\pastro = 0.58$, a well-measured extreme mass ratio of $q = 0.087^{+0.012}_{-0.012}$, and a well-measured (and large) negative effective spin which is robust to changes in the spin prior. No previously detected BBH merger has been confidently measured to have negative effective spin. The secondary BH falls in the LMG at high confidence and there is evidence of precession, with Bayesian evidence ratio $e^5$ in favor of precession when comparing the evidence computed by \texttt{PyMultinest} for the same waveform model and priors but with a likelihood model that internally sets in-plane spin components to zero. We expect future searches with precessing templates and higher harmonics to find this event even more significant.}
    \label{fig:GW190910_012619_corner_plot}
\end{figure}

\begin{figure}[H]
    \centering
    \includegraphics[width=\linewidth]{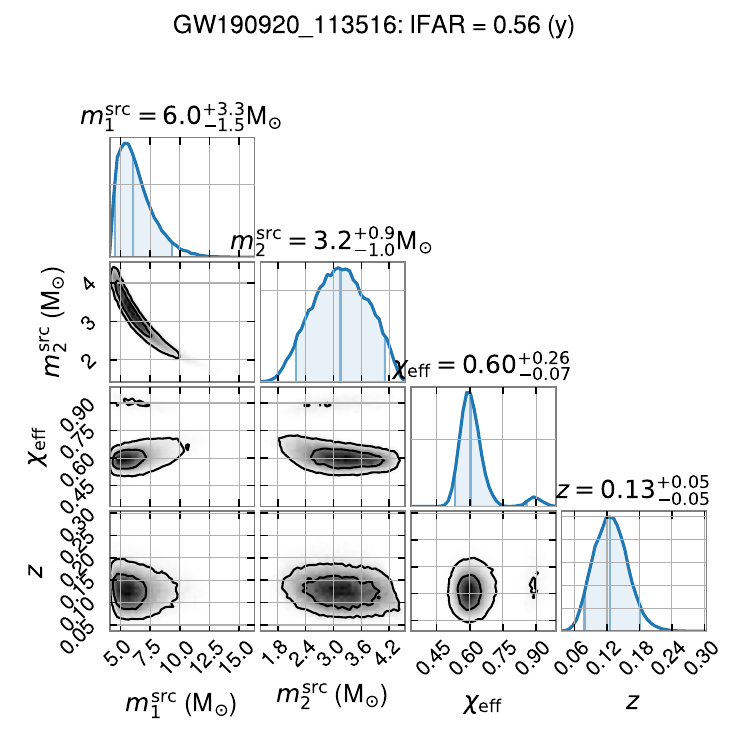}
    \caption{GW190920\_113516 has $p_{\rm astro} = 0.58$ and a secondary constituent which is either a heavy NS or a BH in the LMG. Despite poorly-measured sky localization (see Fig.~\ref{fig:new_nsbh_skylocs}), the large effective spin gives this event some hope as a candidate for an NSBH merger association with an EM counterpart.}
    \label{fig:GW190920_113516_corner_plot}
\end{figure}

\begin{figure}[H]
    \centering
    \includegraphics[width=\linewidth]{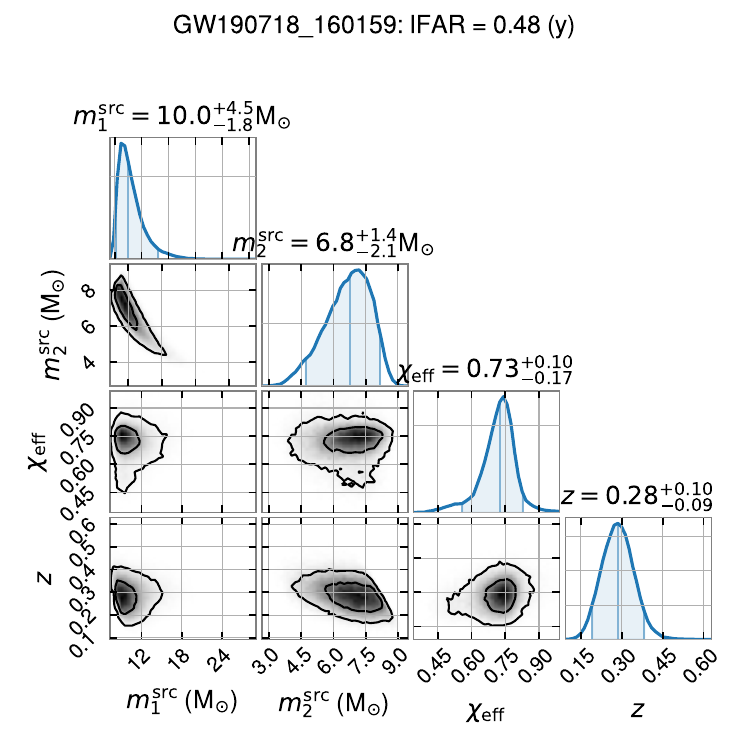}
    \caption{GW190718\_160159 has $p_{\rm astro} = 0.53$ and a large positive effective spin at high confidence.}
    \label{fig:GW190718_160159_corner_plot}
\end{figure}

%%%%%%%%%%%%%%%%%%%%%%%%%%%%%%%%%%%%%%%%%%%%%%%%%%%%%%%%%
%%%%%%%%%%%%%%%%%%%%%%%%%%%%%%%%%%%%%%%%%%%%%%%%%%%%%%%%%
%%%%%%%%%%%%%%%%%%%%%%%%%%%%%%%%%%%%%%%%%%%%%%%%%%%%%%%%%
%%%%%%%%%%%%%%%%%%%%%%%%%%%%%%%%%%%%%%%%%%%%%%%%%%%%%%%%%

\section{Computing Astrophysical Probabilities}
\label{Appendix:pastro}

The probability that an event is astrophysical, $\pastro$, is defined as the ratio of the data's Bayesian evidence under the signal detection hypothesis ($\althyp$: the data contains a GW signal from a BBH merger) to the sum of this signal evidence and the data's evidence under the noise hypothesis ($\nullhyp$: the data does not contain a BBH merger signal). Both hypotheses pose issues in evidence computations. The evidence under $\nullhyp$ (and likelihood computation more generally) suffers from the failure of the stationary noise assumption, since not all non-stationarity can be removed in data processing. Our pipeline takes steps to mitigate this such as correcting likelihood computations for a linear PSD drift and in-painting bad data segments \cite{psd_drift}. We veto glitches with the same methods as in the previous IAS catalog \cite{ias_o2_pipeline_new_events_prd2020}.

The evidence under $\althyp$ requires an astrophysical prior in order to reflect the differences between detectable merger rates in different regions of physical parameter space, but this prior's dependence on intrinsic parameters is unknown. In an attempt to remain as agnostic as possible about the astrophysical population, we do not introduce additional prior information beyond what already goes into the computation of the ranking score (see \S \ref{Appendix:collection} and \S \ref{Appendix:templateprior}). We aim to devise a method that is as simple as possible so it is straightforward for the reader to identify where their own choice of astrophysical prior could enter. More specifically, we would like to measure $\pastro$ directly from the distribution of triggers, which includes an additional 2000 O3a runtimes worth of noise realizations generated from the O3a data using timeslides (for triggering details, see \cite{ias_pipeline_o1_catalog_new_search_prd2019} and \cite{ias_o2_pipeline_new_events_prd2020}). To this end, we estimate the densities of triggers appearing in Equation~\ref{eq:pastrodef}, where we express the astrophysical probability as a function of our ranking score ($\Sigma$):
\begin{align}
    p_{\rm astro}(\Sigma)
     = \frac{{\rm d} N/{\rm d}\Sigma{(\Sigma \mid \althyp)}}
    {{\rm d} N/{\rm d}\Sigma{(\Sigma \mid \nullhyp)} + {\rm d} N/{\rm d}\Sigma{(\Sigma \mid \althyp)}}.
\end{align}

\begin{figure}[H]
    \centering
    \includegraphics[width=\linewidth]{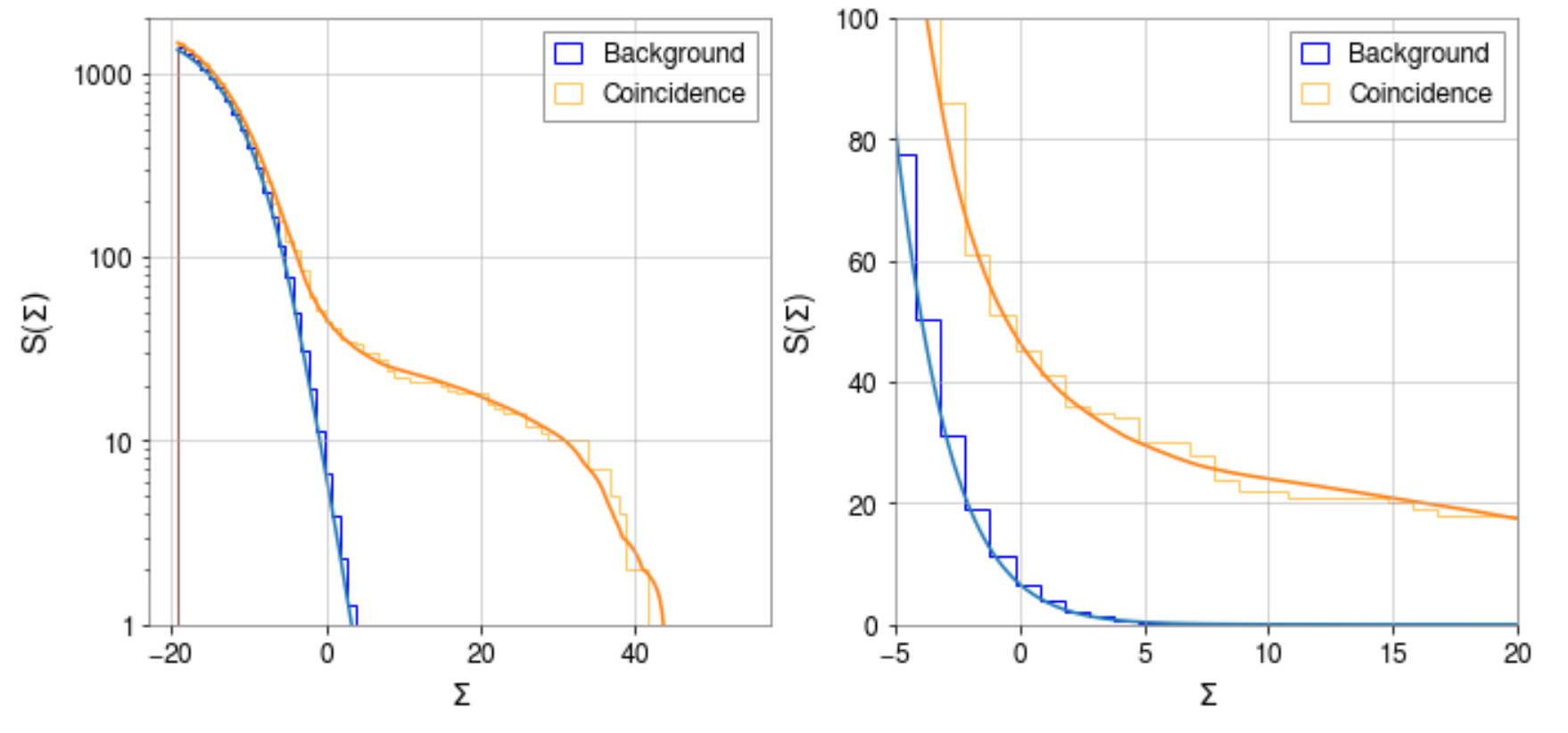}%
    \caption{Survival function for scores $\Sigma$ as defined in equation \ref{eq:surv-function}. The right panel shows a zoomed in version in the range relevant for estimating $p_{\rm astro}$ near $p_{\rm astro} \sim 0.5$. The lines show the fit from our simple analytical model (equation \ref{eq:simple-model}).}
    \label{fig:distribution-of-scores}
\end{figure}

\begin{figure}[H]
    \centering
    \includegraphics[width=\linewidth]{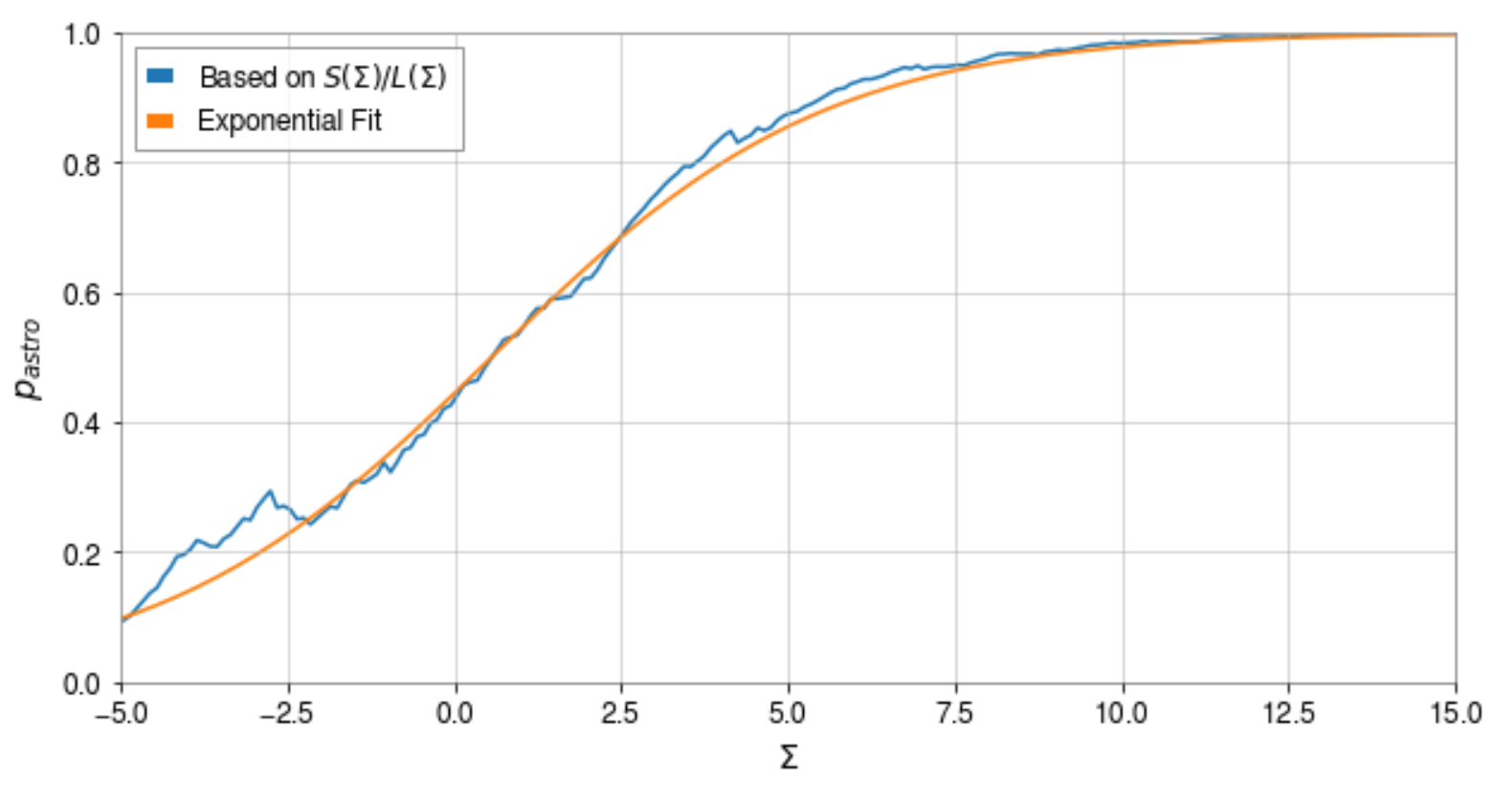}%
    \caption{$p_{\rm astro}$ as a function of score $\Sigma$ as computed based on our simple analytical model (equation \ref{eq:simple-model}). The smooth line shows the fit of equation \ref{eq:pastro-model}.} 
    \label{fig:pastro-vs-scores}
\end{figure}

Figure \ref{fig:distribution-of-scores} shows the distribution of scores in our search. For the purpose of determining $p_{\rm astro}$ we combined all our banks together. To bring the scores computed for each bank into a common scale we subtract a constant from the scores in each bank such that a score of zero corresponds to an expectation of one trigger during O3a in the background distribution of that bank. Figure \ref{fig:distribution-of-scores} shows the survival function, defined as:
\begin{equation}
    S(\Sigma)=\int_\Sigma^{\infty} \frac{{\rm d} N}{{\rm d}\Sigma}(\Sigma') {\rm d}\Sigma'.
    \label{eq:surv-function}
\end{equation}
Thus $S(\Sigma)$ quantifies the probability of obtaining a score higher than $\Sigma$.

To obtain $\frac{{\rm d} N}{{\rm d}\Sigma}(\Sigma)$ we could fit a parametrized model to the distribution of events. We will do something simpler and construct a model directly based on $S(\Sigma)$. To do so we need to divide $S(\Sigma)$ by a quantity with units of $\Sigma$ that quantifies the range of scores over which events are being accumulated at a given value of $\Sigma$. Let us introduce: 
\begin{equation}
    L(\Sigma)=S^{-1}(\Sigma)\int_\Sigma^{\infty} (\Sigma'-\Sigma) \frac{{\rm d} N}{{\rm d}\Sigma}(\Sigma') {\rm d}\Sigma' 
\end{equation}
Both  $S(\Sigma)$ and $L(\Sigma)$ can easily be estimated from the data. It turns out that their ratio can be used to construct a good model for the probability distribution functions.

It is useful to consider some simple probability distributions and work out the relation between $\frac{{\rm d} N}{{\rm d}\Sigma}(\Sigma)$, $S(\Sigma)$ and $L(\Sigma)$. Two examples that are relevant for the distribution of background and astrophysical events in our search are the exponential and power law distributions. In those cases one gets:
\begin{align}
    \frac{{\rm d} N}{{\rm d}\Sigma}(\Sigma) = A \exp^{-\gamma \Sigma} & \rightarrow \frac{{\rm d} N}{{\rm d}\Sigma}(\Sigma)= S(\Sigma)/L(\Sigma)\nonumber \\
    \frac{{\rm d} N}{{\rm d}\Sigma}(\Sigma) = A \Sigma^{-\gamma} & \rightarrow \frac{{\rm d} N}{{\rm d}\Sigma}(\Sigma)= \frac{(\gamma-2)}{(\gamma-1)} S(\Sigma)/L(\Sigma)
    \label{eq:simple-model}
\end{align}
As might be expected for these simple distributions that do not have a preferred scale, $\frac{{\rm d} N}{{\rm d}\Sigma}$ is given by $S(\Sigma)/L(\Sigma)$ times a normalization coefficient. An overall coefficient of one provides a good fit for the background in our search, while for the astrophysical distribution a coefficient of 0.5 provides a better fit\footnote{Figure \ref{fig:distribution-of-scores} shows that at low scores the distribution of triggers in our search turns over. This is the result of incompleteness in our search that stems from the fact that we only collect triggers above a hard cut-off in the incoherent scores of each detector. This has the effect of making the coefficients slightly score-dependent at low scores. We model this by allowing the normalization coefficient to evolve linearly with the score at low scores. This is a small complication that does not affect the range of the distributions relevant for the calculation of $p_{\rm astro}$ in the range of interest.}. Figure \ref{fig:distribution-of-scores} shows the cumulative distribution obtained by integrating our simple models for the probability distributions. The agreement is very good.

Figure \ref{fig:pastro-vs-scores} shows the $p_{\rm astro}$ obtained by using our models for the distribution functions in Eq.~\eqref{eq:pastrodef}. We note that, because the background of our search is very well approximated by an exponential distribution, while the astrophysical distribution is better fit by a power law which is approximately constant over the range of scores where $p_{\rm astro}$ transitions between $p_{\rm astro} \ll 1$ and $p_{\rm astro}\sim 1$, one can obtain a very simple fitting formula for $p_{\rm astro}$:
\begin{equation}
    p_{\rm astro}(\Sigma) \approx \frac{1}{1+a \, e^{-\gamma \Sigma}} \ , \ \ (a,\gamma)=(1.25,0.4),
    \label{eq:pastro-model}
\end{equation}
which provides a good fit over the relevant range. We use this simple formula in the main text. \changed{Note that this result is similar to the astrophysical probability analysis of the O3a triggers from the Multi-Band Template Analysis (MBTA) pipeline, described in Section 6 of \citet{mbta_o3a_pastro_andres2022}.}

%%%%%%%%%%%%%%%%%%%%%%%%%%%%%%%%%%%%%%%%%%%%%%%%%%%%%%%%%
%%%%%%%%%%%%%%%%%%%%%%%%%%%%%%%%%%%%%%%%%%%%%%%%%%%%%%%%%
%%%%%%%%%%%%%%%%%%%%%%%%%%%%%%%%%%%%%%%%%%%%%%%%%%%%%%%%%
%%%%%%%%%%%%%%%%%%%%%%%%%%%%%%%%%%%%%%%%%%%%%%%%%%%%%%%%%
\section{Template prior}
\label{Appendix:templateprior}

In previous works (\citet{ias_o2_pipeline_new_events_prd2020, ias_pipeline_o1_catalog_new_search_prd2019}), we used a template prior that was uniform in the geometric coordinates within each chirp mass bank (see \cite{ias_template_bank_PSD_roulet2019} for details on geometric placement). This assumed that, within each bank, there is approximate proportionality between phase space densities in physical parameters space and in the template grid (with coordinates $c_{\alpha}(m_1, m_2, \chi_{1,z}, \chi_{2,z})$ corresponding to waveform phase components). In this work, we refine this assumption by assigning a prior probability density to each template based on its physical parameters. The prior is uniform in the constituent masses and the effective spin $\chieff$, defined in Eq.~\eqref{eq:chieff_def}. We remain agnostic about the relative prior probabilities between different chirp-mass banks. This is similar to the prior implemented by \citet{CoherentScore} (section IV therein). \changed{Note that this prior serves only to keep as much of parameter space as possible recoverable under another set of priors, since the optimal template prior is the distribution of the astrophysical population \cite{optimal_template_prior_popdist_dent2014} and this is not known.}

For each bank $B$, we draw $N_B^{\rm samples} = 10^{7}$ sets of physical parameters $(m_1, m_2, \chi_{1,z}, \chi_{2,z})$. First the detector-frame masses are drawn uniformly for each bank by rejection-sampling, under the constraint that the chirp mass $\mathcal{M}$ and mass-ratio are within the bank's ranges (see Fig.~\ref{fig:banks}). Then we draw effective spins uniformly from the interval $(-0.99, 0.99)$. The effective spin values and the condition $|\chi_{1,2}| < 0.99$ provide conditional ranges for the complementary spin parameter $\chi_a \equiv \chi_{1, z} - \chi_{2,z}$, from which values are drawn uniformly. These parameters are then associated to their best-fit templates based on the best match (with the same PSD used in bank generation), which are denoted by the chirp mass bank number ($B$), the index of the sub-bank ($i$) giving a reference amplitude profile, and the grid coordinates ($c_{\alpha}$) specifying the waveform phase (refer to \citet{ias_template_bank_PSD_roulet2019} for a detailed description of template bank construction).

For each sub-bank $i$, we use the $N^{\rm samples}_{B,i}$ samples to create a multi-dimensional histogram ${\rm hist}_{B,i}(c_{\alpha})$ counting the number of samples falling into each bin of the geometric space, which must then be normalized. This histogram provides us with an estimate of the prior: 
\begin{equation}
    \Pi_{{\rm BBH}(B,i)}(c_{\alpha}) {\rm d} V_{(B,i)} \approx  \frac
    {{\rm hist}_{(B,i)}(c_{\alpha})}
    {N_{B}^{\rm samples}}
\end{equation}
where ${\rm d} V_{(B,i)}$ is the product of the grid spacing $\Delta c_{\alpha}$ over all the dimensions $\alpha$ in \texttt{BBH}($B, i$). This is normalized to integrate to one over the search grid within each chirp mass bank.

Some practical modifications are necessary to protect our histogramming method from numerical pathologies. First, the finite sample size results in some low-probability template regions being under-sampled. In particular, we do not want to allow stochastic fluctuations to take the prior to zero where there should have been a few points. Thus, in order to prevent from rejecting any physical templates a priori, we add a single count to each empty bin in the histogram. To further mitigate the effects of under-sampling, we limit our resolution in each dimension to $\Delta c_\alpha = 0.5$ and we marginalize over dimensions that have less than two physical grid points. For histogram dimensions with more than 100 bins, we decrease the resolution so that 100 bins covers the full extent.

The marginalization, which reduces most histograms to two or less dimensions, also helps reduce gradients in the prior map. We must handle large gradients with care, both at the edges of the bank and near sharp features within the physical grid, because the presence of noise can shift the best-fit $c_{\alpha}$ from their true values by $\delta c \sim {\rm SNR}^{-1}$. This stochastic misplacement can easily degrade the estimated prior's accuracy in the vicinity of sharp features. To address this effect at the edges of the physical grid, we expand the $c_{\alpha}$ extents and demand that the prior map be smooth over this larger region. We enforce smoothness throughout the prior map with an iterative filtering procedure.

\begin{figure}[H]
    \centering
    \includegraphics[width=\linewidth]{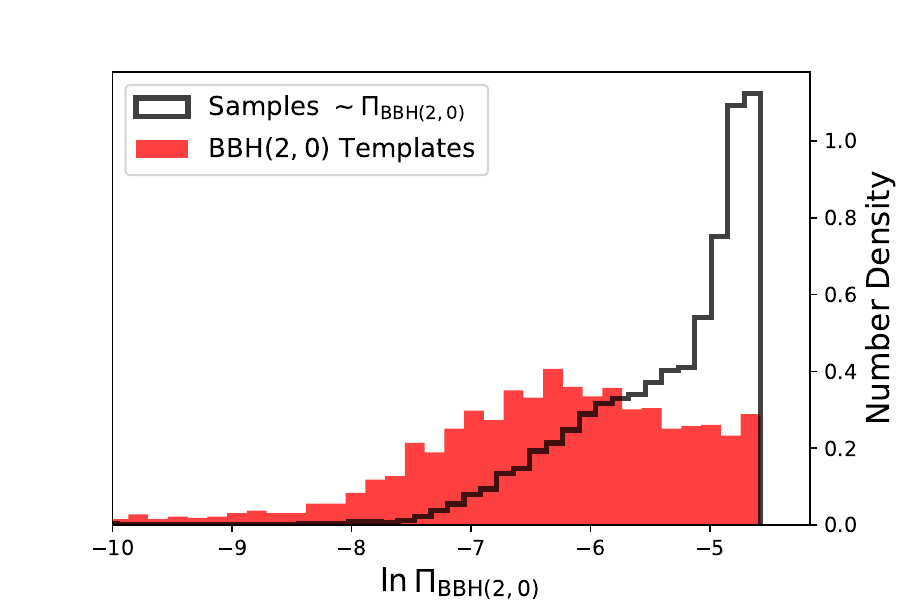}
    \caption{Histograms of the (log) prior assigned to the search templates in \texttt{BBH (2,0)} (red histogram whose samples are the red points in Fig.~\ref{fig:calpha_prior_map}) and of that assigned to $10^4$ draws from the analytic prior in physical parameter space (black histogram whose samples are generated from uniform distributions in detector frame constituent masses and effective spin). The relative broadening of the search grid's tail compared to the prior indicates that the ratio of grid coordinate volume to physical parameter space volume is large in the vicinity of some fraction of the search templates in this sub-bank. Therefore a prior that is uniform in the geometric coordinates will over-represent regions of small prior volume in physical parameter space. Correcting for this brings our ranking score closer to optimality.
    }
    \label{fig:histograms_calpha_vs_prior_samples}
\end{figure}

\begin{figure}[H]
    \centering
    \includegraphics[width=\linewidth]{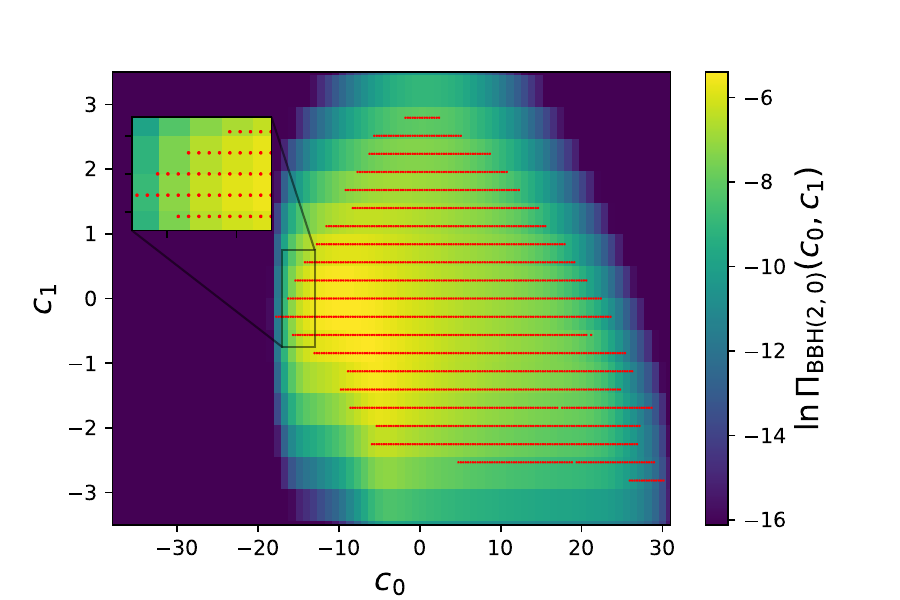}
    \caption{The prior map for the leading dimensions of \texttt{BBH (2,0)}, produced by the histogramming and smoothing procedure described in the text. Red grid points indicate the coordinates of search templates, where the uniform spacing in these geometric coordinates represents some optimal (and generically nonuniform) coverage of the physical parameter space covered by this sub-bank (which itself is delineated in an optimal way described in \citet{ias_template_bank_PSD_roulet2019}).
    }
    \label{fig:calpha_prior_map}
\end{figure}

We begin by filtering the histogram with a Gaussian smoothing kernel of standard deviation $\sigma_c$ in each dimension. Then we draw another $5\times 10^{4}$ samples as a test set for each chirp mass bank. This amounts to $\mathcal{O}(10^4)$ random samples falling into each sub-bank. Next we create another set of samples from the first by adding Gaussian noise with scale $\delta c$ in each dimension. We use the centers of histogram bins to interpolate linearly the prior value for the original samples ($\Pi_1$) and for the noisy samples ($\Pi_2$). If $\left|\log(\Pi_1/\Pi_2)\right| > 0.5$  for any of the samples (equivalent to a change of $\sim$1 in ${\rm SNR}^2$), we repeat the process with a kernel of larger $\sigma_c$. The filtering process ended for all sub-banks with kernel width less than unity. An example of a smoothed prior map can be in Fig.~\ref{fig:calpha_prior_map}. We test convergence by repeating the process with four smaller values of $N_{B}^{\rm samples}$, ranging from $10^{5}$ to $5\times10^{6}$. In each case we evaluate the prior ratio between the computations at this value of $N_{B}^{\rm samples}$ and the high-resolution version ($N_{B}^{\rm samples} = 10^7$) for $10^{4}$ random samples per chirp mass bank. From the computation at $5\times10^{6}$ to the high resolution version the probability density change at physical grid templates is negligible.

We can see from Fig.~\ref{fig:calpha_prior_map} that the prior varies significantly within a sub-bank. Figure~\ref{fig:histograms_calpha_vs_prior_samples} demonstrates that uniform $c_{\alpha}$ grid spacing over-represents regions of low prior relative to its coverage of the regions where physical templates are most likely. The template prior therefore decreases the impact of the look-elsewhere effect on the most physically probable regions of geometric grid space by raising the effective detection bar for templates in regions of grid space where the ratio of grid coordinate volume to physical parameter space volume is too large.

\changed{It is important to note that `physically probable' is to some degree about phase space volume being physical or non-physical, but it also reflects our assumptions about the empirical distribution of astrophysical systems (which may differ drastically from what is physically \textit{possible}). Therefore making results portable depends critically on using an initial template prior which avoids suppressing as much of parameter space as possible, so that results can be reweighted a broad range of different astrophysical priors using a methodology such as that described by \citet{ias_popO2_Roulet_2020}. This is the main motivation for choosing a template prior that is flat in the effective spin (see Ref.~\cite{ias_gw190521_prd2021olsen} for a discussion of the advantages of this spin prior in data analysis with minimal assumptions about the astrophysical distribution).}

%%%%%%%%%%%%%%%%%%%%%%%%%%%%%%%%%%%%%%%%%%%%%%%%%%%%%%%%%
%%%%%%%%%%%%%%%%%%%%%%%%%%%%%%%%%%%%%%%%%%%%%%%%%%%%%%%%%
%%%%%%%%%%%%%%%%%%%%%%%%%%%%%%%%%%%%%%%%%%%%%%%%%%%%%%%%%
%%%%%%%%%%%%%%%%%%%%%%%%%%%%%%%%%%%%%%%%%%%%%%%%%%%%%%%%%

\onecolumngrid

\section{Coherent Multi-detector Detection statistic}
\label{Appendix:collection}

In this section, we will outline a derivation of the detection statistic our pipeline uses in the Gaussian noise case (i.e., before the corrections to account for the presence of glitches kick in). 
Search pipelines analyze streams of data from multiple detectors, and identify triggers as segments of data in which some test-statistics (which are intermediate quantities in the search process) are above some suitable threshold. 
Having collected a set of triggers, a choice to make is the effective statistic according to which the triggers will be ranked (``ranking statistic"). 
Given a choice of ranking statistic, pipelines then model its distribution in the null hypothesis (useful for computing false-alarm rate), and signal hypothesis (useful for computing probability of astrophysical origin, $\pastro$). 
In this sequence of steps, almost any choice of the ranking statistic is admissible (as long as its distribution is modeled properly), but in order to maximize sensitivity, we should adopt a statistic that utilizes all the signal information contained in the data.

Previous versions of our pipeline \cite{ias_pipeline_o1_catalog_new_search_prd2019, ias_o2_pipeline_new_events_prd2020}, as well as other currently existing pipelines \cite{gstlal, PYCBCPipeline, nitz_o3a_3ogc_catalog_2021}, use summary statistics for each trigger that are based on the peak-value of the matched-filter score's timeseries in each detector (as well as a few additional statistics that are built directly or indirectly from the full timeseries, designed to down-weight non-Gaussian transients). 
In the new version of our pipeline that we are presenting in this paper, we adopt a statistic that uses the full timeseries of matched-filtering scores in both detectors (Hanford and Livingston), meaning we retain more information that can be used to discriminate between the signal and noise hypotheses (even in the Gaussian noise case).

Following the Neyman-Pearson lemma \cite{neymanpearson}, the optimal detection statistic is the likelihood ratio for the entire data associated with the trigger under the signal and noise hypotheses, marginalized over the free parameters inherent to the signal hypothesis (along with the appropriate priors for these parameters) -- in parameter estimation, this is the Bayesian evidence for the signal. 
In this section, we separate the signal parameters into intrinsic parameters ($\mathcal{I}$, masses and spins, which we iterate over in the template bank), and extrinsic parameters ($\mathcal{E}$, such as sky-location, distance, orbital inclination, etc.), and derive the likelihood ratio marginalized over only the extrinsic parameters $\mathcal{E}$, whose priors are known. 

%\teja{Talk about bank subbank and rank functions.}
%Finally, we use an approximation to additionally integrate over the intrinsic parameters (only used when dec. 

In the rest of this section, we will define and expand upon the detection statistic and outline a method to efficiently compute it.

\subsection{Preliminaries} 

Given a timeseries $x(t)$, we use the DFT with the convention
\begin{align*}
  x( f ) & = \sum_{t} x(t) \,e^{-2 \pi i f t}, {\rm and} \\
  x( t ) & = \frac1N \sum_{f} x(f) \,e^{2 \pi i f t},
\end{align*}
where $N$ is the length of the timeseries. If $x(t)$ is real-valued, its Fourier coefficients satisfy 
\begin{align}
  x(-f) & = x^\ast(f). \label{eq:realseries}
\end{align}
More generally, it is convenient to split timeseries into their positive and negative frequency components:
\begin{align}
  x(t) & = x^{\oplus}(t) + x^{\ominus}(t),
\end{align}
which are composed of the appropriate subsets of Fourier components (note that even if $x$ is real-valued, neither $x^{\oplus}$ nor $x^{\ominus}$ is real-valued). If $x(t)$ is real-valued, these components satisfy $\overline{x^\ominus} = x^\oplus$. For a real-valued timeseries $x_c(t)$ (the `cosine' timeseries), we can define its real-valued `sine' counterpart, which satisfies
\begin{align}
  x_s^\oplus (f) & = i x_c^\oplus (f). \label{eq:sinewaveform}
\end{align}
Using this, we can also define the ``complexified" form of the timeseries, as 
\begin{align}
  \mathbb{x}(t) = x_c(t) - i x_s(t). \label{eq:complexified}
\end{align}
We can use Eq.~\eqref{eq:sinewaveform}, along with Eq.~\eqref{eq:realseries} (due to the real-valued nature of $x_c$ and $x_s$), to see that 
\begin{align}
  \mathbb{x} & = 2 x_c^\oplus, \label{eq:compl-positive}
\end{align}
i.e., the complexified series $\mathbb{x}$ consists of only the positive-frequency parts of $x_c$ (multiplied by a factor of 2). Conversely, from Eqs.~\eqref{eq:compl-positive} and \eqref{eq:complexified}, we see that if we have a positive-frequency timeseries, its real and imaginary parts behave like cosine and sine waveforms.

In the absence of a signal, the data is assumed to be stationary Gaussian random noise with one-sided power spectrum $\cn(f) = \cn(\vert f \vert)$. Given a timeseries $x(t)$, we define the whitened timeseries $x_w(t)$ through convolution with the whitening filter $w$ corresponding to the power spectrum $\cn$, i.e., 
\begin{align}
  x_w(t) & = \left( x \conv w \right) (t) \\
  & = \frac1N \sum_f x(f) w(f) e^{2 \pi i f t} \\
  & = \frac{\sqrt{2 \Delta t}}{N} \sum_f \frac{x(f)}{\sqrt{\cn(f)}} e^{2 \pi i f t}, \, {\rm i.e.,} \\
  x_w(f) & = \sqrt{\frac{2 \Delta t}{\cn(f)}} x(f).
\end{align}
The noise-weighted inner product between two timeseries $a(t)$ and $b(t)$ is
\begin{align}
  \langle a \vert b \rangle & = \sum_t a_w^\ast(t) b_w(t) \label{eq:innerproduct-td}\\
  & = \frac1N \sum_f a_w^\ast(f) b_w(f) \\
  & = \frac{2 \Delta t}{N} \sum_f \frac{a^\ast(f) b(f)}{\cn(f)} \label{eq:innerproduct-fd}. 
\end{align}
Depending on context, we call this the overlap of the whitened timeseries $a_w(t)$ and $b_w(t)$.

%\begin{align}
%  \langle s_w(t) \rangle^2 & = \frac{\Delta t}{N}  \sum_f \frac12 N(f) \vert w(f) \vert^2 \\
%  & = \frac12 \lambda^2 \Delta t
%\end{align}

\subsection{Signal Model}

In the search, we assume the signals are from binaries on circular orbits, radiating in their dominant quadrupole mode. We can divide the set of parameters describing a merger into three classes: 
\begin{enumerate}
  \item Intrinsic parameters $\mcal{I}$, i.e., masses and spins, which do not depend on the geometry of the system relative to us (in reality, we measure the detector-frame, or redshifted, masses rather than the source-frame values, so we treat the detector-frame masses as intrinsic parameters).
  \item Extrinsic parameters $\mcal{E}_m$ that describe the location of the observer on the merger's sky --- these are the inclination $\iota$ and orbital phase $\phi$, here defined in an equatorial coordinate system relative to the merger's orbital plane (note that the inclination is sometimes defined as the supplement of this value, i.e.\  w.r.t the line of sight from the observer); and
  \item Extrinsic parameters $\mcal{E}_d$ that describe the location of the merger relative to the detector --- these are the right-ascension $\alpha$, declination $\delta$, polarization angle $\psi$ (roll-angle of the major axis of the orbit's projected ellipse on the observer's sky), luminosity distance $D$, and the merger time, $\tau$.
\end{enumerate}
The space of all detectable signals for a fixed set $\mcal{I}$ is two-dimensional in nature. To get the basis elements, we follow the notation in \S 2 of Ref.~\cite{Sphharmconventions}. 

At any point on a merger's sky, the plus (`+') and cross (`$\times$') polarizations are the strains measured by a pair of hypothetical detectors that lie on the tangent space (i.e., are oriented with the merger on their zenith), and with their arms along the $\mathbf{e}_\theta$ and $\mathbf{e}_\phi$ directions, and rotated by $\pi/4$ around the radial direction w.r.t the first one, respectively. 
These polarizations satisfy
\begin{align}
  h_+(\mcal{I}, \mcal{E}_m, t) - i h_\times(\mcal{I}, \mcal{E}_m, t) & = \sum_m h_{2 m}(\mcal{I}, t) \, {}_{-2}Y^{2m}(\iota, \phi), \label{eq:sphharm}
\end{align}
where $\iota$ and $\phi$ represent the location of the detector on the merger's sky, the ${}_{-2}Y^{2m}(\iota, \phi)$ are spin-weighted spherical harmonics, and the $h_{2m}(\mcal{I}, t)$ are related to the evolution of the quadrupole source term\footnote{In the non-spinning case, the source term is proportional to an integration of the mass distribution against the appropriate spherical harmonics, but it has a more general meaning.}; the amplitude of the signals in Eq.~\eqref{eq:sphharm} are defined at some fiducial distance which we will call $D_0$. 
We keep only the dominant harmonics ($m = \pm 2$), and use the relation $h_{2 -2}(\mcal{I}, t) = h_{22}^\ast(\mcal{I}, t)$ between the moments (valid under assumptions that hold for mergers with aligned spins), to get
\begin{align}
  %~~~~ & \!\!\!\!
  h_+(\mcal{I}, \mcal{E}_m, t) - i h_\times(\mcal{I}, \mcal{E}_m, t) & = h_{2 2}(\mcal{I}, t) \, {}_{-2}Y^{2 2}(\iota, \phi) + h_{2 2}^\ast(\mcal{I}, t) \, {}_{-2}Y^{2 -2}(\iota, \phi) \\
  & = A \left[ \left(1 + \mu \right)^2 e^{2 i \phi} h_{2 2}(\mcal{I}, t) + \left(1 - \mu \right)^2 e^{-2 i \phi} h_{2 2}^\ast(\mcal{I}, t) \right] \\
  & = 4 A \Bigl[ \frac{1 + \mu^2}{2} {\rm Re} \left\{ e^{2 i \phi} h_{2 2}(\mcal{I}, t) \right\} + i \mu \, {\rm Im} \left\{ e^{2 i \phi} h_{2 2}(\mcal{I}, t) \right\} \Bigr],
\end{align}
where $\mu = \cos{\iota}$ and $A$ is a numerical constant. We equate real and imaginary parts of the LHS and RHS, and absorb the numerical constants into the amplitude of $h_{22}$ to obtain
\begin{align}
  h_+(\mcal{I}, \mcal{E}_m, t) & = \frac{1 + \mu^2}{2} {\rm Re} \left\{ e^{2 i \phi} h_{2 2}(\mcal{I}, t) \right\} \\
  & = \frac{1 + \mu^2}{2} {\rm Re} \left\{ e^{2 i \phi} \left[ h_{2 2}^\oplus(\mcal{I}, t) + h_{2 2}^\ominus(\mcal{I}, t) \right] \right\} \\
  & = \frac{1 + \mu^2}{4} \left[ e^{2 i \phi} \left\{ h_{2 2}^\oplus(\mcal{I}, t) + h_{2 2}^\ominus(\mcal{I}, t) \right\} + e^{-2 i \phi} \left\{ \overline{h_{2 2}^\oplus}(\mcal{I}, t) + \overline{h_{2 2}^\ominus}(\mcal{I}, t) \right\} \right] \, {\rm , and} \\
  h_\times(\mcal{I}, \mcal{E}_m, t) & = - \mu {\rm Im} \left\{ e^{2 i \phi} h_{2 2}(\mcal{I}, t) \right\} \\
  & = - \frac{\mu}{2 i} \left[ e^{2 i \phi} \left\{ h_{2 2}^\oplus(\mcal{I}, t) + h_{2 2}^\ominus(\mcal{I}, t) \right\} - e^{-2 i \phi} \left\{ \overline{h_{2 2}^\oplus}(\mcal{I}, t) + \overline{h_{2 2}^\ominus}(\mcal{I}, t) \right\} \right].
\end{align}
Specializing to the positive frequency parts, we have
\begin{align}
  h_+^\oplus(\mcal{I}, \mcal{E}_m, t) & = \frac{1 + \mu^2}{4} \left[ e^{2 i \phi} h_{2 2}^\oplus(\mcal{I}, t_R) + e^{-2 i \phi} \overline{h_{2 2}^\ominus}(\mcal{I}, t_R) \right], \, {\rm and} \\
  h_\times^\oplus(\mcal{I}, \mcal{E}_m, t) & = \frac{\mu}{2} i \left[ e^{2 i \phi} h_{2 2}^\oplus(\mcal{I}, t) - e^{-2 i \phi} \overline{h_{2 2}^\ominus}(\mcal{I}, t) \right].
\end{align}
For non-precessing systems, the approximants assume that the quadrupole source term $h_{22}(\mcal{I}, t)$ only consists of negative frequency terms, i.e., $h_{22} = h_{22}^\ominus$ (or exclusively positive ones, depending on DFT convention, but not both). In this case, we have
\begin{subequations}
\label{eq:hpol}
  \begin{align}
    h_+^\oplus(\mcal{I}, \mcal{E}_m, t) & = \frac{1 + \mu^2}{2} e^{-2 i \phi} \overline{h_{2 2}^\ominus}(\mcal{I}, t) = \frac{1 + \mu^2}{2} e^{-2 i \phi} \overline{h_{2 2}}(\mcal{I}, t), \, {\rm and} \label{eq:hpluspositive} \\
    h_\times^\oplus(\mcal{I}, \mcal{E}_m, t) & = - \mu i e^{-2 i \phi} \overline{h_{2 2}^\ominus}(\mcal{I}, t) = - \mu i e^{-2 i \phi} \overline{h_{2 2}}(\mcal{I}, t), \label{eq:hminuspositive}
\end{align}
\end{subequations}
where we absorbed a factor of 2 into the amplitude of $h_{22}^\ominus$. From the definition of the sine and cosine waveforms in Eq.~\eqref{eq:sinewaveform}, we can see that the `$\times$' polarization is a scaled version of the `sine' counterpart of the `+' polarization at the same location, and at a fixed time, the `+' polarization itself rotates in phase as a function of the angle $\phi$ (with twice the rate). 

In general the detector is not oriented in a particular way relative to the $\mathbf{e}_\theta$ and $\mathbf{e}_\phi$ directions, and hence, is sensitive to a combination of the two polarizations; moreover, the detector's zenith is not precisely pointed toward the merger, and hence, the response to the different polarizations is not equal in strength. 
We retain the definition of $h_+$ and $h_\times$ in Eq.~\eqref{eq:hpol}, but allow the detector to have a general distance $D$, and a merger time $\tau$ -- we can quantify this by picking a significant moment in the merger's waveform, setting the time where it is achieved in Eqs.~\eqref{eq:hpol} to zero, and using $\tau$ to label when it is achieved in the detected signal\footnote{There is also the additional subtlety of cosmological redshift, but the effect of redshift can be captured by a change in intrinsic parameters $\cal{I}$.}. The detected waveform is
\begin{align}
  h(\mcal{I}, \mcal{E}_m, \mcal{E}_d, t) &= \frac{D_0}{D} \left[ F_+(\alpha, \delta, \psi) h_+(\mcal{I}, \mcal{E}_m, t - \tau) + F_\times(\alpha, \delta, \psi) h_\times(\mcal{I}, \mcal{E}_m, t - \tau) \right], \label{eq:superposition}
\end{align}
where $F_+$ and $F_\times$ are the detector's polarization response functions (taken to be constant over the time scale of the waveform). We restrict to the positive-frequency part of both sides of Eq.~\eqref{eq:superposition}, and use Eqs.~\eqref{eq:hpluspositive} and \eqref{eq:hminuspositive} to obtain
\begin{align}
  h^\oplus(\mcal{I}, \mcal{E}_m, \mcal{E}_d, t) & = \frac{D_0}{D} \left[ F_+(\alpha, \delta, \psi) h_+^\oplus(\mcal{I}, \mcal{E}_m, t - \tau) + F_\times(\alpha, \delta, \psi) h_\times^\oplus(\mcal{I}, \mcal{E}_m, t - \tau) \right] \\
  & = \frac{D_0}{D} \left[ F_+(\alpha, \delta, \psi) \frac{1 + \mu^2}{2} - i F_\times(\alpha, \delta, \psi) \mu \right] e^{-2 i \phi} \overline{h_{2 2}}(\mcal{I}, t - \tau). \label{eq:detwfpositive}
\end{align}
If we want to give a physical meaning to the term $\overline{h_{22}}(\mcal{I}, t)$ on the RHS, consider a detector located on the north pole of the merger's sky at a distance $D_0$, facing the merger, with one arm parallel to the line joining the masses at a fiducial frequency (which defines the origin for the phase $\phi$): $\overline{h_{22}}(\mcal{I}, t)$ is the positive frequency part of the signal that this detector sees.

\subsection{Detection Statistic}

In this section, we will omit the arguments $\mcal{I}, \mcal{E}_{m}, \mcal{E}_{d}$ for brevity, unless we explicitly need to show that quantities only depend on subsets of the parameters. In our search, we use the waveform $h_T(\mcal{I}, t) = 2 {\rm Re}\left( \overline{h_{22}}(\mcal{I}, t) \right)$ as a `cosine' template -- the complexified waveform $\mathbb{h}_T(t) = 2 \overline{h_{22}}(t)$. %We use a fiducial distance $D_0$ (common to all detectors) to define $h_T(t)$. 

We have several detectors, indexed by $k$; the total number of detectors is $N_d$. In general, the arrival times for the signal at different detectors are different, so we can define a common merger time $\tau_c$, e.g., as the merger time seen by a fiducial detector, or a hypothetical detector at the center of the earth. In general, the merger time seen in detector $k$, $\tau_k \neq \tau_c$, and $\tau_k$ depends on $\mcal{E}_d$ through $\tau_c$ and the angular parameters $\alpha$ and $\delta$:
\begin{align}
  \tau_k(\tau_c, \alpha, \delta) & = \tau_c + \Delta \tau_k(\alpha, \delta). \label{eq:dtdef}
\end{align}
From Eq.~\eqref{eq:detwfpositive} and Eq.~\eqref{eq:compl-positive}, we see that the predicted waveform $h_k(t)$ of the signal in detector $k$ satisfies
\begin{align}\label{eq:predictedwf}
  \mathbb{h}_k(t) & = \frac{D_0}{D} \left[ F_{+, k} \frac{1 + \mu^2}{2} - i F_{\times, k} \mu \right] e^{-2 i \phi} \mathbb{h}_T(t - \tau_k) \\
  & \, \equiv Y \, R_k \, \mathbb{h}_T(t - \tau_k),
\end{align}
where
\begin{align}
  Y = y e^{-2 i \phi} \, \equiv& \, \frac{D_0}{D} e^{-2 i \phi}, \, \, {\rm and} \label{eq:ydef} \\
  R_k \, = \, F_{+, k} \frac{1 + \mu^2}{2} \, &- \, i F_{\times, k} \mu.
\end{align}
The complex number $Y$ is independent of the detector, while the response factor $R_k$ varies from detector to detector. 
In the search, we compute overlaps of the data with a ``unit" template that is rescaled according to the detector sensitivity. For a detector $k$, the unit template is the waveform at a distance $D_k$, which satisfies
\begin{align}
  h_{T, k}(t) & = \frac{D_0}{D_k} h_T(t),  \\
  \langle h_{T, k} \vert h_{T, k} \rangle_k & = 1,
\end{align}
where the inner product is defined with respect to the detector PSD, as in Eq.~\eqref{eq:innerproduct-fd}. The merger time $\tau_k$ at detector $k$ is unknown and to be marginalized over, hence the search returns timeseries of complex overlaps for all times $\tau$. The overlaps are inner products of the complexified unit template $\mathbb{h}_{T, k}$ and the data (as in Eq.~\eqref{eq:innerproduct-td}):
\begin{align}
  Z_k(\mcal{I}, \tau) & =  \, \langle \mathbb{h}_{T, k}(\mcal{I}, \tau, t) \vert d(t) \rangle_k \,  \equiv \langle \mathbb{h}_{T, k}(\mcal{I}, t - \tau) \vert d(t) \rangle_k . \label{eq:complexoverlap}
\end{align}
For a given set of parameters $\tau_c, \alpha, \, {\rm and} \, \delta$, it is convenient to group the overlaps in the detectors into a complex vector as
\begin{align}
  \mathbf{Z}(\mcal{I}, \tau_c, \alpha, \delta) & = \begin{pmatrix}
%    Z_0(\mcal{I}, \tau_0(\mcal{E}_d)) \\ 
%    Z_1(\mcal{I}, \tau_1(\mcal{E}_d)) \\
    \vdots \\
    Z_k(\mcal{I}, \tau_k(\tau_c, \alpha, \delta)) \\
    \vdots \\
  \end{pmatrix}. \label{eq:complexz}
\end{align}
Note that the overlap in this vector for each detector is picked at the respective merger time, which is a different local time due to the differing location of the detectors.

If the data were composed of just the waveform $h_k$ (without any noise), the predicted overlap is
\begin{align}
  z_k(\mcal{I}, \mcal{E}_m, \mcal{E}_d/\tau_c) & = \langle \mathbb{h}_{T, k} \vert h_k \rangle_k \\
  & = Y R_k \langle \mathbb{h}_{T, k} \vert h_T \rangle_k \\
  & = Y R_k \frac{D_k}{D_0} \langle \mathbb{h}_{T, k} \vert h_{T, k} \rangle_k \\
  & = Y R_k \frac{D_k}{D_0} \\
  & = Y z_{0, k}(\mcal{I}, \mcal{E}_m/\phi, \mcal{E}_d/\{\tau_c, D\}) \label{eq:zk0}. 
  %= Y(D, \phi) z_{0, k}(\mcal{I}, \iota, \alpha, \delta, \psi)
\end{align}
Here, $\mcal{E}_d/\tau_c$ indicates a quantity that depends on the parameters in $\mcal{E}_d$ apart from the merger time $\tau_c$. In going from the first line to the second, we used Eq.~\eqref{eq:predictedwf}, along with the fact that the complexified waveform $\mathbb{h}_k$ only has positive frequencies. The quantity $z_{0, k}$ can be interpreted as the complex overlap that would be attained for a signal from a merger with phase $\phi = 0$ and at distance $D_0$ in detector $k$, in the noiseless case with the merger time `aligned' between the waveform and the template. Similar to the vector in Eq.~\eqref{eq:complexz}, we can also define vectors $\mathbf{z}(\mcal{I}, \mcal{E}_m, \mcal{E}_d/\tau_c)$ and $\mathbf{z}_0(\mcal{I}, \mcal{E}_m/\phi, \mcal{E}_d/\{\tau, D\})$:
\begin{align}
  \mathbf{z}(\mcal{I}, \mcal{E}_m, \mcal{E}_d/\tau_c) & = \begin{pmatrix}
%    z_0(\mcal{I}, \mcal{E}_m, \mcal{E}_d/\tau) \\ 
%    z_1(\mcal{I}, \mcal{E}_m, \mcal{E}_d/\tau) \\
    \vdots \\
    z_k(\mcal{I}, \mcal{E}_m, \mcal{E}_d/\tau_c) \\
    \vdots \\
  \end{pmatrix}, \, {\rm and} \\
  \mathbf{z}_0(\mcal{I}, \mcal{E}_m/\phi, \mcal{E}_d/\{D, \tau_c \}) & = \begin{pmatrix}
%    z_0(\mcal{I}, \mcal{E}_m, \mcal{E}_d/\tau) \\ 
%    z_1(\mcal{I}, \mcal{E}_m, \mcal{E}_d/\tau) \\
    \vdots \\
    z_{0, k}(\mcal{I}, \mcal{E}_m/\phi, \mcal{E}_d/\{D, \tau_c\}) \\
    \vdots \\
  \end{pmatrix}.
\end{align}
Given two complex vectors $\mathbf{a}$ and $\mathbf{b}$, we can define their inner product as $\mathbf{a} \cdot \mathbf{b} = \sum_k a_k^\ast b_k$.

The quantity of interest for us is the inner product of the data in detector $k$ with the predicted physical waveform (and not the complexified template). We can derive the relation between this and the quantity in Eq.~\eqref{eq:complexoverlap} as follows:
\begin{align}
  \langle h_k \vert d \rangle_k & = {\rm Re} \left\{ \langle \mathbb{h}_k \vert d_k \rangle_k \right\} \\
  & = {\rm Re} \left\{ Y^\ast R_k^\ast \langle \mathbb{h}_T \vert d_k \rangle_k \right\} \\
  & = {\rm Re} \left\{ Y^\ast R_k^\ast \frac{D_k}{D_0} \langle \mathbb{h}_{T, k} \vert d_k \rangle_k \right\} \\
  & = {\rm Re} \left\{ Y^\ast R_k^\ast \frac{D_k}{D_0} Z_k \right\} \\
  & = {\rm Re} \left\{ z_k^\ast Z_k \right\}.
\end{align}
In going from the first line to the second, we used Eq.~\eqref{eq:predictedwf}, along with the fact that the complexified waveform $\mathbb{h}_k$ only has positive frequencies. The inner product of the predicted waveform with itself is
\begin{align}
  \langle h_k \vert h_k \rangle_k & = \frac12 \langle \mathbb{h}_k \vert \mathbb{h}_k \rangle_k \\
  & = \frac12 y^2 \abs{R_k}^2 \langle \mathbb{h}_T \vert \mathbb{h}_T \rangle_k \\
  & = \frac12 y^2 \abs{R_k}^2 \frac{D_k^2}{D_0^2} \langle \mathbb{h}_{T, k} \vert \mathbb{h}_{T, k} \rangle_k \\
  & = y^2 \abs{R_k}^2 \frac{D_k^2}{D_0^2} \\
  & = \abs{z_k}^2. 
\end{align}
The log-likelihood for the data, given a set of parameters, is
\begin{align}
  2 \ln L(\mcal{I}, \mcal{E}_m, \mcal{E}_d)
  & = - \sum_{k \in {\rm detectors}} \langle d_k - h_k \vert d_k - h_k \rangle_k \\
  & = - \sum_k  \left[ \langle d_k \vert d_k \rangle_k + \langle h_k \vert h_k \rangle_k - 2 \langle h_k \vert d_k \rangle_k \right] \\
  & = - \sum_k  \left[ \langle d_k \vert d_k \rangle_k + \abs{z_{k}}^2 - 2 \langle h_k \vert d_k \rangle_k \right] \\
%  & = - \sum_k \langle d \vert d \rangle_k - \sum_k \left[ \abs{z_{p, k} - z_k}^2 - \abs{z_k}^2 \right] \\
%  & = - \sum_k \langle d \vert d \rangle_k - \sum_k \left[ \abs{Y R_k \frac{D_k}{D_0} - z_k}^2 - \abs{z_k}^2 \right] \\
  & = - \sum_k \langle d_k \vert d_k \rangle_k - \sum_k \left[ \abs{Y}^2 \abs{ z_{0, k} }^2 - 2 {\rm Re}\left( Y^\ast z_{0, k}^\ast Z_k \right) \right] \\
  & = - \sum_k \langle d_k \vert d_k \rangle_k - \left[ \abs{Y}^2 \vabs{\mathbf{z}_0}^2 - 2 {\rm Re}\left( Y^\ast \mathbf{z}_0 \cdot \mathbf{Z} \right) \right] \\
%  & = - \sum_k \langle d \vert d \rangle_k - \left[ \abs{Y}^2 \sum_k \abs{z_{0, k}}^2 - 2 {\rm Re}\left( Y^\ast \sum_k z_{0, k}^\ast Z_k \right) \right] \\
  & = - \sum_k \langle d_k \vert d_k \rangle_k - \vabs{\mathbf{z}_0}^2 \left[ \abs{Y}^2  - 2 {\rm Re}\left( Y^\ast \frac{\mathbf{z}_0 \cdot \mathbf{Z}}{\vabs{\mathbf{z}_0}^2} \right) \right] \\  
  & = - \sum_k \langle d_k \vert d_k \rangle_k - \vabs{\mathbf{z}_0}^2 \left[ \abs{Y - \frac{\mathbf{z}_0 \cdot \mathbf{Z}}{\vabs{\mathbf{z}_0}^2} }^2 - \abs{\frac{\mathbf{z}_0 \cdot \mathbf{Z}}{\vabs{\mathbf{z}_0}^2}}^2 \right] \\
  & = - \sum_k \langle d_k \vert d_k \rangle_k + \frac{\abs{\mathbf{z}_0 \cdot \mathbf{Z}}^2}{\vabs{\mathbf{z}_0}^2} - \vabs{\mathbf{z}_0}^2 \abs{Y - \frac{\mathbf{z}_0 \cdot \mathbf{Z}}{\vabs{\mathbf{z}_0}^2}}^2. \label{eq:lnL_simpl}
\end{align}
We can see that in the final form, we have separated out the dependence of the terms on the various parts of $\mcal{I}, \mcal{E}_m, \mcal{E}_d$: $\mathbf{z}_0$ depends on $\mcal{I}, \mcal{E}_m/\phi, \, {\rm and} \, \mcal{E}_d/\{\tau_c, D\}$ (see Eq.~\eqref{eq:zk0}), $\mathbf{Z}$ depends on $\mcal{I}$ and $\mcal{E}_d$ (see Eq.~\eqref{eq:complexoverlap}), and the term $Y$ depends on the distance $D$ and phase $\phi$. 

If we multiply the likelihood $L$ (derived from the log-likelihood in Eq.~\eqref{eq:lnL_simpl}) with a prior $\Pi$ over the various extrinsic parameters in $\mcal{E}_{d/m}$ and integrate, we get the evidence integral for fixed values of intrinsic parameters, which is formally the probability of seeing the data that we do under the hypothesis that it contains a gravitational wave signal with known intrinsic parameters $\mcal{I}$. 
\begin{align}
  p(d \vert \mcal{I}) & = \int d\Pi (\mcal{E}_m, \mcal{E}_d) L(\mcal{I}, \mcal{E}_m, \mcal{E}_d).
\end{align}
If we have the alternative hypothesis that the data is composed of pure Gaussian random noise, the probability of the data in that case is
\begin{align}
  p(d \vert {\rm noise}) & = \exp{\left( -\frac12 \sum_k \langle d_k \vert d_k \rangle_k \right)}.
\end{align}
According to the Neyman-Pearson Lemma, the optimal detection statistic is the ratio of the two probabilities
\begin{align}
  \frac{p(d \vert \mcal{I})}{p(d \vert {\rm noise})} & = \int {\rm d}\Pi(\mcal{E}_m, \mcal{E}_d) \exp{\left( \frac12 \left[ \frac{\abs{\mathbf{z}_0 \cdot \mathbf{Z}}^2}{\vabs{\mathbf{z}_0}^2} - \vabs{\mathbf{z}_0}^2 \abs{Y - \frac{\mathbf{z}_0 \cdot \mathbf{Z}}{\vabs{\mathbf{z}_0}^2}}^2 \right] \right)}. \label{eq:dstatistic}
\end{align}
The argument of the exponential in the integrand depends on distance $D$ and phase $\phi$ only through the parameter $Y$, so let us keep the other parameters in $\mcal{I}, \mcal{E}_{d/m}$ fixed and simplify the integral over $D$ and $\phi$. Suppose the prior on distance and RA, DEC, is 
\begin{align}
  {\rm d}\Pi = {\rm d}D \, {\rm d}\alpha \, {\rm d}\delta \sin{\lt \delta \rt} \, D^2 \Pi(D),
\end{align}
which is normalized, so that $4\pi \int {\rm d}D \, D^2 \Pi(D) = 1$, and the prior on the phase $\phi$ is uniform. The relevant integral is
\begin{align}
  ~~~ & \!\!\!\!
  G(\mcal{I}, \mcal{E}_m/\phi, \mcal{E}_d/\{ \tau_c, D\}) \notag \\
  & = \int_0^{\infty} {\rm d}D D^2 \Pi(D) \int_0^{2\pi} \frac{{\rm d}\phi}{2\pi} \exp{\left( - \frac{\vabs{\mathbf{z}_0}^2}{2} \abs{Y - \frac{\mathbf{z}_0 \cdot \mathbf{Z}}{\vabs{\mathbf{z}_0}^2}}^2 \right)} \\
  & = D_0^3 \int_0^\infty \frac{{\rm d}y}{y^4} \Pi\left( \frac{D_0}{y} \right) \int_0^{2\pi} \frac{{\rm d}\phi}{2\pi} \exp{\left( - \frac{\vabs{\mathbf{z}_0}^2}{2} \abs{y \exp{(2 i \phi)} - \frac{\mathbf{z}_0 \cdot \mathbf{Z}}{\vabs{\mathbf{z}_0}^2}}^2 \right)} \\
  & = D_0^3 \int_0^\infty \frac{{\rm d}y}{y^4} \Pi\left( \frac{D_0}{y} \right) \exp{\left( - \frac{\vabs{\mathbf{z}_0}^2}{2} \left[ y^2 + \frac{\vert \mathbf{z}_0 \cdot \mathbf{Z} \vert^2}{\vabs{\mathbf{z}_0}^4} \right] \right)} \int_0^{2\pi} \frac{{\rm d}\phi}{2\pi} \exp{\left[ {\rm Re} \left( y^\ast \mathbf{z}_0 \cdot \mathbf{Z} \exp{(- 2 i \phi)} \right) \right]} \\
  & = D_0^3 \int_0^\infty \frac{{\rm d}y}{y^4} \Pi\left( \frac{D_0}{y} \right) \exp{\left( - \frac{\vabs{\mathbf{z}_0}^2}{2} \left[ y^2 + \frac{\vert \mathbf{z}_0 \cdot \mathbf{Z} \vert^2}{\vabs{\mathbf{z}_0}^4} \right] \right)} I_0 \left( y \vert \mathbf{z}_0 \cdot \mathbf{Z} \vert \right) \\
  & = D_0^3 \exp{\left( - \frac12 \frac{\vert \mathbf{z}_0 \cdot \mathbf{Z} \vert^2}{\vabs{\mathbf{z}_0}^2} \right)} \int_0^\infty \frac{{\rm d}y}{y^4} \Pi\left( \frac{D_0}{y} \right) \exp{\left( - \frac{\vabs{\mathbf{z}_0}^2 y^2}{2} \right)} I_0 \left( y \vert \mathbf{z}_0 \cdot \mathbf{Z} \vert \right) \\
  & = D_0^3 \vabs{\mathbf{z}_0}^3 \exp{\left( - \frac12 \vert \hat{\mathbf{z}}_0 \cdot \mathbf{Z} \vert^2 \right)} \int_0^\infty \frac{{\rm d}a}{a^4} \Pi\left( \frac{D_0 \vabs{\mathbf{z}_0}}{a} \right) \exp{\left( - \frac{a^2}{2} \right)} I_0 \left( a \vert \hat{\mathbf{z}}_0 \cdot \mathbf{Z} \vert \right) \label{eq:Idphi}
\end{align}
where $I_0$ is the modified Bessel function of the first kind, and in the last equation, the dummy variable $a = y \vabs{\mathbf{z}_0}$ [the physical interpretation of $a$ is the `network SNR' (i.e., norm of the complex overlaps in all detectors with the whitening filters applied) of the signal from a merger with phase $\phi = 0$ and distance $D$ in the absence of noise]. 

\begin{figure}[htbp]
\begin{center}
  \includegraphics[width=10cm]{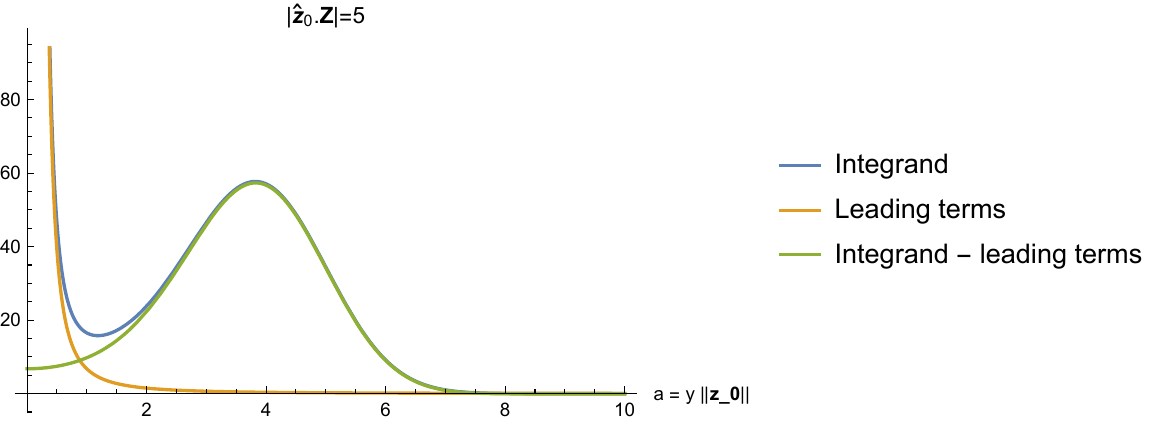}
\caption{Structure of the integrand for the marginalization over distance (without the prior), with and without the leading terms in the expansion near large distances. The coordinate $a$ varies inversely with distance, and can be interpreted as the `network SNR' (norm of the complex overlaps in all detectors with the whitening filter applied) for the signal from a merger with phase $\phi = 0$ and distance $D$ in the absence of noise.}
\label{fig:regulated}
\end{center}
\end{figure}

Consider the terms in the integrand in Eq.~\eqref{eq:Idphi} other than the prior $\Pi$. 
These terms blow up at small values of $a$ (or large distances), with leading terms diverging as $1/{a^4}$ and $1/{a^2}$:
\begin{align}
  \frac{1}{a^4} \exp{\left( - \frac{a^2}{2} \right)} I_0 \left( a \vert \hat{\mathbf{z}}_0 \cdot \mathbf{Z} \vert \right) & = \frac{1}{a^4} \left[ 1 - \frac{a^2}{2} + \cdots \right] \left[ 1 + \frac14 a^2 \vert \hat{\mathbf{z}}_0 \cdot \mathbf{Z} \vert^2 + \cdots \right] \\
  & = \frac{1}{a^4} + \left( \frac14 \vert \hat{\mathbf{z}}_0 \cdot \mathbf{Z} \vert^2 - \frac12 \right) \frac{1}{a^2} + O(a^0). \label{eq:intgtaylor}
\end{align}
Figure \ref{fig:regulated} illustrates the behavior of the integrand in Eq.~\eqref{eq:Idphi} without the prior as a function of $a$. The leading terms are those in Eq.~\eqref{eq:intgtaylor}.

In practice, the behavior of the prior $\Pi$ at large distances regulates the integral; a common choice of distance prior has $\Pi(D) = {\rm constant} = 3 / (4\pi D_{\rm max}^3)$ up to some upper limit $D_{\rm max}$ that is chosen by hand (i.e., uniform in luminosity volume; different priors, e.g., uniform in comoving volume, involve different forms of $\Pi(D)$). 
An upper cutoff in distance corresponds to cutting the lower limits of the integrand in Eq.~\eqref{eq:Idphi} (i.e., that in Fig.~\ref{fig:regulated}) at some small but nonzero value of $a$. 
Even without an artificial cutoff, astrophysically motivated priors can tail off at large distances (or redshifts). 

Motivated by this, we split the integrand into two parts -- one with the leading terms in Eq.~\eqref{eq:intgtaylor} subtracted, and one with the leading terms.
\begin{align}
  ~~ & \!\!\! 
  \int_0^\infty \frac{{\rm d}a}{a^4} \Pi\left( \frac{D_0 \vabs{\mathbf{z}_0}}{a} \right) \exp{\left( - \frac{a^2}{2} \right)} I_0 \left( a \vert \hat{\mathbf{z}}_0 \cdot \mathbf{Z} \vert \right) \notag \\
   & = \int_0^\infty {\rm d}a \, \Pi\left( \frac{D_0 \vabs{\mathbf{z}_0}}{a} \right) \left[ \frac{1}{a^4} \exp{\left( - \frac{a^2}{2} \right)} I_0 \left( a \vert \hat{\mathbf{z}}_0 \cdot \mathbf{Z} \vert \right) - \frac{1}{a^4} - \left( \frac14 \vert \hat{\mathbf{z}}_0 \cdot \mathbf{Z} \vert^2 - \frac12 \right) \frac{1}{a^2} \right] + \notag \\
   & \hspace{10pt} \int_0^\infty {\rm d}a \, \Pi\left( \frac{D_0 \vabs{\mathbf{z}_0}}{a} \right) \left[ \frac{1}{a^4} + \left( \frac14 \vert \hat{\mathbf{z}}_0 \cdot \mathbf{Z} \vert^2 - \frac12 \right) \frac{1}{a^2} \right]. \label{eq:splitintegrand}
\end{align}
The first and second terms in the RHS of Eq.~\eqref{eq:splitintegrand}, respectively, pick up most of their weight at large and small values of $a$ (see Fig.~\ref{fig:regulated}) -- these are, respectively, the relatively local and very distant universe (recall the physical interpretation of $a$). 
If we adopt a prior that is locally uniform in luminosity volume, and cuts off in some regulated manner at large distances, then we can replace $\Pi$ with its local value $\Pi_0$ in the first term. 
The physical interpretation of $\Pi_0$ is the local number of mergers per unit volume, if we set the total number of mergers in the universe (regardless of detectability) to one and the location of this merger is distributed according to the prior. 
Under this simplifying assumption, Eq.~\eqref{eq:splitintegrand} simplifies to
\begin{align}
 ~~ & \!\!\!
  \int_0^\infty \frac{{\rm d}a}{a^4} \Pi\left( \frac{D_0 \vabs{\mathbf{z}_0}}{a} \right) \exp{\left( - \frac{a^2}{2} \right)} I_0 \left( a \vert \hat{\mathbf{z}}_0 \cdot \mathbf{Z} \vert \right) \\
  \notag
  & \approx \Pi_0 \left\{ \int_0^\infty {\rm d}a \, \left[ \frac{1}{a^4} \exp{\left\{ - \frac{a^2}{2} \right\}} I_0 \left( a \vert \hat{\mathbf{z}}_0 \cdot \mathbf{Z} \vert \right) - \frac{1}{a^4} - \left( \frac14 \vert \hat{\mathbf{z}}_0 \cdot \mathbf{Z} \vert^2 - \frac12 \right) \frac{1}{a^2} \right] + f_1 + f_2 \vert \hat{\mathbf{z}}_0 \cdot \mathbf{Z} \vert^2 \right\} \\
  & = \Pi_0 \left\{ \frac{1}{18} \sqrt{\frac{\pi}{2}} \exp{\left( \frac{\vert \hat{\mathbf{z}}_0 \cdot \mathbf{Z} \vert^2}{4} \right)} \left[ \left( \vert \hat{\mathbf{z}}_0 \cdot \mathbf{Z} \vert^4 - 6 \vert \hat{\mathbf{z}}_0 \cdot \mathbf{Z} \vert^2 + 6 \right) I_0 \left( \frac{\vert \hat{\mathbf{z}}_0 \cdot \mathbf{Z} \vert^2}{4} \right) - \vert \hat{\mathbf{z}}_0 \cdot \mathbf{Z} \vert^2 \left( \vert \hat{\mathbf{z}}_0 \cdot \mathbf{Z} \vert^2 - 4 \right) I_1\left( \frac{\vert \hat{\mathbf{z}}_0 \cdot \mathbf{Z} \vert^2}{4} \right) \right] \right. \notag \\
  & \hspace{30pt} \left. + f_1 + f_2 \vert \hat{\mathbf{z}}_0 \cdot \mathbf{Z} \vert^2 \right\} \label{eq:dmarg}
\end{align}
where $f_1$ and $f_2$ are numerical constants whose value depends on how the distance prior $\Pi$ is cutoff at large distances. Substituting Eq.~\eqref{eq:dmarg} into Eq.~\eqref{eq:Idphi}, we get 
\begin{align}
  G(\mcal{I}, \mcal{E}_m/\phi, \mcal{E}_d/\{ \tau_c, D\}) & = \Pi_0 D_0^3 \vabs{\mathbf{z}_0}^3 g\left( \vert \hat{\mathbf{z}}_0 \cdot \mathbf{Z} \vert, f_1, f_2 \right), \, {\rm where} \label{eq:Gdef}\\
  g\left( \vert \hat{\mathbf{z}}_0 \cdot \mathbf{Z} \vert, f_1, f_2 \right) & = \left\{ \right. \notag \\
  & \!\!\!\!\!\!\!\!\!\!\!\!\!\!\!\!\!\!\!\!\!\!\!\!\!\!\!\!\!\!\!\!\!\!\! \left. \frac{1}{18} \sqrt{\frac{\pi}{2}} \exp{\left( - \frac{\vert \hat{\mathbf{z}}_0 \cdot \mathbf{Z} \vert^2}{4} \right)} \left[ \left( \vert \hat{\mathbf{z}}_0 \cdot \mathbf{Z} \vert^4 - 6 \vert \hat{\mathbf{z}}_0 \cdot \mathbf{Z} \vert^2 + 6 \right) I_0 \left( \frac{\vert \hat{\mathbf{z}}_0 \cdot \mathbf{Z} \vert^2}{4} \right) - \vert \hat{\mathbf{z}}_0 \cdot \mathbf{Z} \vert^2 \left( \vert \hat{\mathbf{z}}_0 \cdot \mathbf{Z} \vert^2 - 4 \right) I_1\left( \frac{\vert \hat{\mathbf{z}}_0 \cdot \mathbf{Z} \vert^2}{4} \right) \right] \right. \notag \\
  & \!\!\!\!\!\!\!\!\!\!\!\!\!\!\!\!\!\!\!\!\!\!\!\!\!\!\!\!\!\!\!\!\!\!\! \left. + f_1 \exp{\left( - \frac12 \vert \hat{\mathbf{z}}_0 \cdot \mathbf{Z} \vert^2 \right)} + f_2 \vert \hat{\mathbf{z}}_0 \cdot \mathbf{Z} \vert^2 \exp{\left( - \frac12 \vert \hat{\mathbf{z}}_0 \cdot \mathbf{Z} \vert^2 \right)} \right\} \label{eq:gdef}
%  G(\mcal{I}, \mcal{E}_m/\phi, \mcal{E}_d/\{ \tau_c, D\}) \notag \\
%  & = \Pi_0 D_0^3 \vabs{\mathbf{z}_0}^3 \left\{ \right. \notag \\
%  & \hspace{20pt} \left. \frac{1}{18} \sqrt{\frac{\pi}{2}} \exp{\left( - \frac{\vert \hat{\mathbf{z}}_0 \cdot \mathbf{Z} \vert^2}{4} \right)} \left[ \left( \vert \hat{\mathbf{z}}_0 \cdot \mathbf{Z} \vert^4 - 6 \vert \hat{\mathbf{z}}_0 \cdot \mathbf{Z} \vert^2 + 6 \right) I_0 \left( \frac{\vert \hat{\mathbf{z}}_0 \cdot \mathbf{Z} \vert^2}{4} \right) - \vert \hat{\mathbf{z}}_0 \cdot \mathbf{Z} \vert^2 \left( \vert \hat{\mathbf{z}}_0 \cdot \mathbf{Z} \vert^2 - 4 \right) I_1\left( \frac{\vert \hat{\mathbf{z}}_0 \cdot \mathbf{Z} \vert^2}{4} \right) \right] \right. \notag \\
%  & \hspace{20pt} \left. + f_1 \exp{\left( - \frac12 \vert \hat{\mathbf{z}}_0 \cdot \mathbf{Z} \vert^2 \right)} + f_2 \vert \hat{\mathbf{z}}_0 \cdot \mathbf{Z} \vert^2 \exp{\left( - \frac12 \vert \hat{\mathbf{z}}_0 \cdot \mathbf{Z} \vert^2 \right)} \right\}
\end{align}
%The function $g(\vert \hat{\mathbf{z}}_0 \cdot \mathbf{Z} \vert)$ we use in the code corresponds to $(3/2) g\left( \vert \hat{\mathbf{z}}_0 \cdot \mathbf{Z} \vert, (2/9)\sqrt{2/\pi}, (2/9)\sqrt{2/\pi} \right)$, and any extra factors depending on the unit convention for $\vabs{\mathbf{z}_0}$. 
For large values of $\vert \hat{\mathbf{z}}_0 \cdot \mathbf{Z} \vert$ ($\gtrsim 6$, say), the function $g$ falls off as $\sim 1/\vert \hat{\mathbf{z}}_0 \cdot \mathbf{Z} \vert^5$. 
Henceforth, we omit the particular values of $f_1$ and $f_2$ in the arguments to $g$ in Eqs.~\eqref{eq:Gdef} and \eqref{eq:gdef} for brevity (in practice, we use the values $f_1 = f_2 = (2/9)\sqrt{2/\pi}$, which were chosen with the arbitrary criterion that the different terms in Eq.~\eqref{eq:gdef} give equal contributions when integrated over $\vert \hat{\mathbf{z}}_0 \cdot \mathbf{Z} \vert$ from $0$ to $\infty$).

We now go back to Eq.~\eqref{eq:dstatistic} and substitute the solution of Eq.~\eqref{eq:Gdef} for the integral in Eq.~\eqref{eq:Idphi}, and write out the dependence on parameters explicitly:
%\begin{align}
%  \frac{p(d \vert \mcal{I})}{p(d \vert {\rm noise})} & = \Pi_0 D_0^3 \int d\Pi(\mcal{E}_m/\phi, \mcal{E}_d/D) \exp{\left( \frac12 \abs{\hat{\mathbf{z}}_0 \cdot \mathbf{Z} }^2 \right)} \vabs{\mathbf{z}_0}^3 g\left( \vert \hat{\mathbf{z}}_0 \cdot \mathbf{Z} \vert \right)
%\end{align}
%We make the dependence on parameters explicit:
\begin{align}
  \frac{p(d \vert \mcal{I})}{p(d \vert {\rm noise})} & = \frac{4\pi \Pi_0 D_0^3}{T} \int \frac{{\rm d}\mu}{2} \frac{{\rm d}\alpha}{2\pi} \frac{{\rm d}\delta \cos{\delta}}{2} \frac{{\rm d}\psi}{2\pi} {\rm d}\tau_c \, \exp{\left( \frac12 \abs{\hat{\mathbf{z}}_0\left( \mu, \alpha, \delta, \psi \right) \cdot \mathbf{Z}\left[ \tau_{\{ k \}}\left(\tau_c, \alpha, \delta\right) \right] }^2 \right)} \vabs{\mathbf{z}_0}^3 g\left( \vert \hat{\mathbf{z}}_0 \cdot \mathbf{Z} \vert \right),
\end{align}
where $T$ is the allowance for the merger time in the length of the data we are analyzing (i.e., in play for the merger to happen within), and the notation $\mathbf{Z}\left[ \tau_k\left(\tau_c, \alpha, \delta\right) \right]$ is shorthand for saying that the vector $\mathbf{Z}$ depends on the parameters $\tau_c, \alpha, \, {\rm and} \, \delta$ only through the set of arrival times $\tau_k$ (we left the dependence on intrinsic parameters $\mcal{I}$ implicit above). Note that the polarization angle $\psi$ could have been restricted to the interval $[0, \pi]$ without changing the answer, though we integrate over $[0, 2 \pi]$ in practice. We can rewrite the probability as
\begin{align}
  \frac{p(d \vert \mcal{I})}{p(d \vert {\rm noise})} & = \frac{4\pi \Pi_0 D_0^3}{T} \prod_{k \in \, {\rm detectors}} \int {\rm d}\tau_k \, \dd(\tau_k - \tau_k(\tau_c, \alpha, \delta)) \, \int \frac{{\rm d}\mu}{2} \frac{{\rm d}\alpha}{2\pi} \frac{{\rm d}\delta \cos{\delta}}{2} \frac{{\rm d}\psi}{2\pi} {\rm d}\tau_{\rm c} \times \, \notag \\
  & \hspace{80pt} \exp{ \left( \frac12 \abs{\hat{\mathbf{z}}_0\left( \mu, \alpha, \delta, \psi \right) \cdot \mathbf{Z}(\tau_{\{k\}}) }^2 \right) } \vabs{\mathbf{z}_0}^3 g\left( \vert \hat{\mathbf{z}}_0 \cdot \mathbf{Z} \vert \right) \\
  & = \frac{4\pi \Pi_0 D_0^3}{T} \prod_{k \in \, {\rm detectors}} \int {\rm d}\tau_k \, \exp{ \left( \frac12 \abs{Z_k(\tau_k)}^2 \right)} \int  \frac{{\rm d}\alpha}{2 \pi} \frac{{\rm d}\delta \cos{\delta}}{2} {\rm d}\tau_c \, \dd(\tau_k - \tau_k(\tau_c, \alpha, \delta)) \times \, \notag \\
  & \hspace{80pt} \int \frac{{\rm d}\mu}{2} \frac{{\rm d}\psi}{2\pi} \, \exp{ \left\{ - \frac12 \left( \vabs{\mathbf{Z}}^2 - \abs{\hat{\mathbf{z}}_0\left( \mu, \alpha, \delta, \psi \right) \cdot \mathbf{Z} }^2 \right) \right\} } \vabs{\mathbf{z}_0}^3 g\left( \vert \hat{\mathbf{z}}_0 \cdot \mathbf{Z} \vert \right), \label{eq:ldphi2}
\end{align}
where $\dd$ denotes the Dirac-delta function. %, and in the second equation, we pulled out a factor of $\exp{\lt 1/2 \vabs{\mathbf{Z}}^2 \rt }$ (compensated inside the integrand). 
Without loss of generality, let us assume the common merger time $\tau_c$ is that seen in the first detector. The integral over $\tau_c$ in Eq.~\eqref{eq:ldphi2} then fixes $\tau_c = \tau_0$. Defining $\Delta \tau_k = \tau_k - \tau_0$ for $k > 0$, we have
\begin{align}
  \frac{p(d \vert \mcal{I})}{p(d \vert {\rm noise})}  & = \frac{4\pi \Pi_0 D_0^3}{T} \prod_{k \in \, {\rm detectors}} \int {\rm d}\tau_k \, \exp{ \left( \frac12 \abs{Z_k(\tau_k)}^2 \right)} \int \frac{{\rm d}\alpha}{2 \pi} \frac{{\rm d}\delta \cos{\delta}}{2} \, \left[ \delta_{k, 0} + (1 - \delta_{k, 0}) \dd(\Delta \tau_k - \Delta \tau_k(\alpha, \delta)) \right] \times \, \notag \\
  & \hspace{80pt} \int \frac{{\rm d}\mu}{2} \frac{{\rm d}\psi}{2\pi} \, \exp{ \left\{ - \frac12 \left( \vabs{\mathbf{Z}}^2 - \abs{\hat{\mathbf{z}}_0\left( \mu, \alpha, \delta, \psi \right) \cdot \mathbf{Z} }^2 \right) \right\} } \vabs{\mathbf{z}_0}^3 g\left( \vert \hat{\mathbf{z}}_0 \cdot \mathbf{Z} \vert \right). \label{eq:ldphi3}
\end{align}
Given the structure of the integral in Eq.~\eqref{eq:ldphi2}, we can use the following sequence of steps to evaluate it using a Monte-Carlo method:
\begin{enumerate}
  \item Discretize the arrival times in each detector, $\tau_k$, on a time grid that we choose to be fine enough to capture any structure in the $Z(\tau_k)$ timeseries. Select a tuple of arrival times $(\cdots, \tau_k, \cdots)$ (i.e., a cell in the $N_d$-dimensional space of times, where $N_d$ is the number of detectors) with each component $\tau_k$ picked according to a probability $\sim \exp{ \left( \frac12 \abs{Z_k(\tau_k)}^2 \right)}$.
  \item In general, not all tuples are physically realizable, for two reasons. 
  The first is simple: physically-realizable delays $\Delta \tau_k$ between detectors $k$ and $0$ are bounded. 
  The second reason is more subtle: given several detectors (number $N_d > 3$), physically realizable arrival times live inside a three-dimensional subspace of the $N_d$ dimensional space of the $\tau_k$ (or two-dimensional for the $\Delta \tau_k$ with $k > 0$), with degrees of freedom corresponding to a free global arrival time, and the two degrees of freedom on the sky.
  \item If the cell corresponding to the given tuple $(\cdots, \tau_k, \cdots)$ does not intersect the physical subspace, the contribution of this sample to the integral is zero (i.e., the argument of the delta function in Eq.~\eqref{eq:ldphi3} never hits zero). If the cell intersects the physical surface, the delta function picks up a factor of the fraction of the total sky-area that produces delays contained within the cell. 
  \item To account for the final terms $\int ({\rm d}\mu/2) \, {\rm d}\psi/(2\pi) \cdots$, we can pick a random point in the sky area within the cell, as well as the range of $\mu$ and $\psi$, and record the sample's contribution to the integral as the integrand multiplied by the fraction of the sky-area with delays contained within the cell, and a normalization constant for the probability distributions over times $\tau_k$.
\end{enumerate}
To facilitate steps 3 and 4 above, we discretize the sky into cells of equal area (to a high enough angular resolution), compute the tuples of delays corresponding to each discretized sky-cell's center, record which delay-cell it lies in, and build up a dictionary that maps a given delay-cell (i.e., cell in $\Delta \tau_k$) to all the sky-cells that produced delays consistent with this delay-cell.

\twocolumngrid

\bibliographystyle{apsrev4-1-etal}
\bibliography{main}
%------------------------------------------------------------------------------

\end{document}